\documentclass[notoc]{JHEP3}

\usepackage{amsmath}
\usepackage{amsfonts}
\usepackage{slashed}

\newcommand{\eps}{\epsilon}
\newcommand{\ord}{\begin{cal}O\end{cal}}

\def\beq{\begin{equation}}
\def\eeq{\end{equation}}
\def\bsp#1\esp{\begin{split}#1\end{split}}

\newenvironment{sloppyequation}[0]{\sloppy\begin{flushleft}\hspace*{0.75cm}\(}{\)\end{flushleft}\fussy}

\newcommand{\beqsloppy}{\begin{sloppyequation}}
\newcommand{\eeqsloppy}{\end{sloppyequation}}

\newcommand{\cF}{\begin{cal}F\end{cal}}

\newcommand{\cM}{\begin{cal}M\end{cal}}
\newcommand{\cN}{\begin{cal}N\end{cal}}

\title{Two-loop splitting amplitudes and the single-real contribution to inclusive Higgs production at N$^3$LO}

\author{Claude Duhr\\
Center for Cosmology, Particle Physics and Phenomenology (CP3),
Universit\'{e} Catholique de Louvain,
Chemin du Cyclotron 2,
B-1348 Louvain-La-Neuve,
Belgium}

\author{Thomas Gehrmann, Matthieu Jaquier\\Physik-Institut, 
Universit\"at Z\"urich, Wintherturerstrasse 190,\\CH-8057 Z\"urich, Switzerland}

\abstract{The factorisation of QCD matrix elements in the limit of two external partons becoming collinear
 is described by process-independent 
splitting amplitudes, which can be expanded systematically in perturbation theory. Working in conventional dimensional 
regularisation, we compute the two-loop splitting amplitudes for all simple collinear splitting processes, including subleading 
terms in the regularisation parameter. Our results are then applied to derive an analytical expression for the two-loop 
single-real contribution to inclusive Higgs boson production in gluon fusion to fourth order (N$^3$LO) in 
perturbative QCD. }

\preprint{CP3-14-72, ZU-TH 38/14}

\begin{document}

\catcode`\@=11
\font\manfnt=manfnt
\def\Watchout{\@ifnextchar [{\W@tchout}{\W@tchout[1]}}
\def\W@tchout[#1]{{\manfnt\@tempcnta#1\relax%
  \@whilenum\@tempcnta>\z@\do{%
    \char"7F\hskip 0.3em\advance\@tempcnta\m@ne}}}
\let\foo\W@tchout
\def\dubious{\@ifnextchar[{\@dubious}{\@dubious[1]}}
\let\enddubious\endlist
\def\@dubious[#1]{%
  \setbox\@tempboxa\hbox{\@W@tchout#1}
  \@tempdima\wd\@tempboxa
  \list{}{\leftmargin\@tempdima}\item[\hbox to 0pt{\hss\@W@tchout#1}]}
\def\@W@tchout#1{\W@tchout[#1]}
\catcode`\@=12

\section{Introduction}
\label{sec:intro}

Scattering amplitudes in massless QCD diverge if one or more of the external momenta become soft or if two or more 
external momenta become collinear. In these infrared limits, one observes process-independent factorisation 
properties: the divergent behaviour of the amplitude is described by universal unresolved factors, multiplying a 
reduced amplitude with lower partonic multiplicity. These unresolved factors are expanded in a two-fold manner in 
perturbative QCD: in the number of unresolved partons and in the number of virtual loops correcting the factors 
at fixed multiplicity. The eikonal factors describing the simple soft behaviour and the collinear splitting amplitudes 
were established already long ago~\cite{Altarelli:1977zs,treesoft}. They are sufficient to understand the 
infrared singular structure of real radiation contributions at next-to-leading order (NLO) in QCD,
and have been instrumental in devising systematic subtraction schemes for NLO 
calculations~\cite{Frixione:1995ms,Catani:1996vz,Kosower:1997zr}.
They also form the starting point for a systematic expansion around unresolved limits to all orders in 
perturbation theory~\cite{Kosower:1999xi,Catani:1998bh}.

At next-to-next-to-leading order (NNLO), tree-level contributions with up to two unresolved particles 
contribute, again displaying universal factorization properties in collinear and soft limits~\cite{Campbell:1997hg,Catani:1998nv}. At the 
same order, one-loop corrections to simple collinear~\cite{Bern:1998sc} and simple soft~\cite{Catani:2000pi} emissions need to 
be accounted for. The behaviour in all these limits is understood in detail, and served as a construction principle for 
subtraction methods at NNLO~\cite{GehrmannDeRidder:2005cm,Catani:2007vq,Czakon:2011ve}. In an actual calculation of 
higher order corrections to a collider observable, the universal soft and collinear factors need to be integrated over their 
appropriate phase spaces, including a regulator for the infrared divergences. Consequently, subleading terms 
in the regulator will yield finite contributions to the final result: simple soft configurations demand two subleading orders, and 
simple collinear configurations one subleading order. 

With advances in multi-loop calculations, calculations of next-to-next-to-next-to-leading order (N$^3$LO) corrections to benchmark observables are now becoming 
feasible. Multiple unresolved limits at zero and one loop order were understood already some time 
ago~\cite{Birthwright:2005ak,Catani:2003vu,Buchta:2014dfa}.  
Two-loop corrections to simple soft and simple collinear limits were derived 
only to finite order in the regulator~\cite{Bern:2004cz,Badger:2004uk} by taking the appropriate limits 
of the two-loop three-parton decay matrix  elements~\cite{Garland:2001tf,Gehrmann:2011aa}. With the two-loop soft gluon 
current recently derived to all orders in the regulator~\cite{Duhr:2013msa,Li:2013lsa}, 
N$^3$LO calculations at the two-loop single 
real level are now missing only the two-loop corrections to simple collinear limits, described by the two-loop splitting amplitudes.
It is the aim of this paper to derive these, based on a new derivation of the two-loop matrix elements for 
three-parton decays, expanded to higher orders in the regulator. 

The most important application of our newly derived results are the N$^3$LO corrections to 
inclusive Higgs production in gluon 
fusion. Following the derivation of the three-loop virtual corrections~\cite{formfactor}, of the required renormalization 
and factorization counterterm contributions~\cite{hfact}, and of expansions of single and multiple real radiation contributions 
around the soft limit~\cite{hn3loall,Duhr:2013msa,Li:2013lsa}, the threshold approximation of this coefficient function was computed 
earlier this year~\cite{Anastasiou:2014vaa}.  We use our results for the two-loop splitting amplitudes to derive a closed 
analytic expression 
for the two-loop single-real contribution to the inclusive N$^3$LO coefficient function for Higgs production. 
Depending on the availability of results for the other channels (one-loop double-real and triple-real), our expression 
can be used to derive further terms in the threshold expansion, or to obtain an exact expression for the full 
coefficient function. 

Our paper is structured as follows. In Section~\ref{sec:collinear_limits}, we establish the notation and discuss the 
behaviour of multi-loop amplitudes in single collinear limits. Section~\ref{sec:splitting_amps} describes the extraction of the 
two-loop splitting amplitudes from the calculation of two-loop matrix elements in three-particle decay kinematics. The analytical 
expressions for the splitting amplitudes are quite lengthy (especially in the term subleading in the 
dimensional regulator), and are collected in Appendix~\ref{app:twoloop}. These results are then applied in
 Section~\ref{sec:VVR_xsecs} to compute the two-loop single-real contribution to Higgs production at N$^3$LO. The
two-loop matrix elements relevant to this process were previously known only for external four-dimensional helicity,
and are re-derived in conventional dimensional regularization in Appendix~\ref{app:HCDR}, their 
contribution to the inclusive coefficient function is documented in Appendix~\ref{app:coeff}. We conclude with an outlook 
in Section~\ref{sec:conclusion}.


\section{Collinear limits of multi-loop QCD amplitudes}
\label{sec:collinear_limits}

We consider an $\ell$-loop QCD amplitude with a certain number of coloured partons in the final state. Our example will be the decay of a heavy colourless state $X$ into $n$ partons with momenta $p_i$, $i=1,\ldots,n$, and all other cases can be obtained by crossing symmetry. In the limit where a pair of partons become collinear, say $i$ and $j$, then the corresponding Mandelstam invariant vanishes, $s_{ij}= 2p_i\cdot p_j\to 0$. The amplitude diverges in the limit, and the divergence is characterised by a simple pole at $s_{ij}=0$. The residue at the pole can be described by the factorisation formula
\beq\bsp
\langle\cM^{(0)}&(\ldots,p_i,p_j,\ldots)|\cM^{(\ell)}(\ldots,p_i,p_j,\ldots)\rangle\\
&\simeq \frac{8\pi\,\alpha_0}{s_{ij}}\sum_{k=0}^\ell \left(\frac{\alpha_0\,S_{\eps}}{2\pi}\right)^k\,\left(\frac{s_{ij}}{\mu^2}\right)^{-k\eps}\,P_{ij}^{(k)}(z)\,\langle\cM^{(0)}(\ldots,P,\ldots)|\cM^{(\ell-k)}(\ldots,P,\ldots)\rangle\,,
\esp\eeq
where $|\cM^{(\ell)}(\ldots,p_i,p_j,\ldots)\rangle$ is the $\ell$-loop amplitude and $\alpha_0$ the bare strong coupling constant and $P=p_i+p_j$ the four-momentum of the parent parton. We work in dimensional regularisation in $D=4-2\eps$ dimensions, and
\beq
c_\Gamma = e^{\gamma_E\eps}\frac{\Gamma(1-\eps)^2\,\Gamma(1+\eps)}{\Gamma(1-2\eps)} {\rm~~and~~} 
S_\eps = (4\pi)^\eps\,e^{-\gamma_E\eps}\,,
\eeq
with $\gamma_E = -\Gamma'(1)$ the Euler-Mascheroni constant. The `$\simeq$' sign denotes `equality up to terms that are power-suppressed in the limit'. The function $P_{ij}^{(\ell)}(z)$ is the $\ell$-loop splitting amplitude, and depends on the colour and spin of the particles involved in the splitting, as well as on the momentum fraction $z$ carried by particle $i$, $p_i = z\,P$. Note that, in contrast to the well-known Altarelli-Parisi splitting functions, the splitting amplitudes considered in this paper only describe the collinear behaviour of a system of two real partons with momenta $p_i$ and $p_j$, i.e., they do not include purely virtual 1PI corrections and multiple real radiation corrections. The (spin-averaged) tree-level splitting amplitudes are given by~\cite{Altarelli:1977zs,Catani:1998nv}
\beq\bsp
P_{gg}^{(0)}(z) &\,= 2\,C_A\,\frac{(1-z+z^2)^2}{z\,(1-z)}\,,\\
P_{gq}^{(0)}(z) &\,= C_F\,\left[\frac{1+(1-z)^2}{z}-\eps\,z\right]\,,\\
P_{q\bar{q}}^{(0)}(z) &\,= \frac{1}{2}\,\left[1-2\frac{z\,(1-z)}{1-\eps}\right]\,,
\esp\eeq
and $P_{qg}^{(0)}(z) = P_{gq}^{(0)}(1-z)$. Here $C_A=N$ and $C_F=V/(2N)$ where $N$ is the number of $SU(N)$ colours and $V=N^2-1$ is the number of adjoint colours.

In the remainder of this section we summarise the main properties of splitting amplitudes.
It is obvious that Bose symmetry implies
\beq\label{eq:symmetry}
P_{gg}^{(\ell)}(1-z) = P_{gg}^{(\ell)}(z) {\rm~~and~~} P_{q\bar{q}}^{(\ell)}(1-z) = P_{q\bar{q}}^{(\ell)}(z)\,.
\eeq

The limits where $z\to0$ or $z\to1$ correspond to the limits where one of the particles involved in the splitting becomes soft. The splitting amplitude for $g\to q\,\bar{q}$ does obviously not have any soft singularity, and so the residue of $P_{q\bar{q}}^{(\ell)}(z)$ at $z=0$ and $z=1$ vanishes. The splitting amplitude $P_{gq}^{(\ell)}(z)$ only develops a soft singularity if the gluon becomes soft, i.e., $z\to 0$, and the residue at the pole is given by the QCD soft current, known up to two loops in perturbative QCD~\cite{treesoft,Catani:2000pi,Badger:2004uk,Duhr:2013msa,Li:2013lsa},
\beq\label{eq:gq_soft}
P_{gq}^{(\ell)}(z) = \frac{C_F\,c_\Gamma^\ell}{2^{\ell-1}}\,z^{-1-\ell\eps}\,r_S^{(\ell)}+\ord(z^0)\,,\textrm{  as }z\to0\,,
\eeq
where $r_S^{(\ell)}$ denotes the $\ell$-loop QCD soft current (in the normalisation of ref.~\cite{Duhr:2013msa}).
Similarly, the gluon splitting amplitude becomes singular if any of the two final-state gluons becomes soft,
\beq\bsp\label{eq:gg_soft}
P_{gg}^{(\ell)}(z) &\, = \frac{C_A\,c_\Gamma^\ell}{2^{\ell-1}}\,z^{-1-\ell\eps}\,r_S^{(\ell)}+\ord(z^0)\,,\textrm{  as }z\to0\,,\\
\phantom{P_{gg}^{(\ell)}(z)} &\, = \frac{C_A\,c_\Gamma^\ell}{2^{\ell-1}}\,(1-z)^{-1-\ell\eps} \,r_S^{(\ell)}+\ord((1-z)^0)\,,\textrm{  as }z\to1\,.
\esp\eeq

The pole structure of $P^{(\ell)}_{ij}(z)$ in dimensional regularisation is completely fixed up to two loops. At one-loop order, we have
\beq
P_{ij}^{(1)}(z) = {\bf I}^{(1)}_{C,ij}(z,\eps)\,P_{ij}^{(0)}(z)+\ord(\eps^0)\,,
\eeq
with
\beq\bsp
{\bf I}^{(1)}_{C,gg}(z,\eps) &\,= -\frac{e^{-\gamma_E\eps}\,C_A}{2\,\eps^2\,\Gamma(1-\eps)}\,[z^{-\eps}+(1-z)^{-\eps}-1]\,,\\
{\bf I}^{(1)}_{C,g\bar{q}}(z,\eps) &\,=-\frac{e^{-\gamma_E\eps}}{2\,\eps^2\,\Gamma(1-\eps)}\,\left[N\,z^{-\eps}-\frac{1}{N}\,((1-z)^{-\eps}-1)\right]\,,\\
{\bf I}^{(1)}_{C,q\bar{q}}(z,\eps) &\,= -\frac{e^{-\gamma_E\eps}}{2\,\Gamma(1-\eps)}\left[\frac{N}{\eps^2}\,(z^{-\eps}+(1-z)^{-\eps}-2)-\frac{1}{N\,\eps^2}+\frac{3\,C_F-2\beta_0}{\eps}\right]\,,
\esp\eeq
where $\beta_{l}$ is the $l$-loop QCD $\beta$ function. At two-loop order we have~\cite{Catani:1998bh,Badger:2004uk}
\beq
P_{ij}^{(2)}(z) = {\bf I}^{(2)}_{C,ij}(z,\eps)\,P_{ij}^{(0)}(z)+{\bf I}^{(1)}_{C,ij}(z,\eps)\,P_{ij}^{(1)}(z)+\ord(\eps^0)\,,
\eeq
with
\begin{align}\label{eq:IS2}
 {\bf I}^{(2)}_{C,ij}(z,\eps)& = -\frac{1}{2}\,\left[ {\bf I}^{(1)}_{C,ij}(z,\eps)\right]^2 -\frac{\beta_0}{\eps}\, {\bf I}^{(1)}_{C,ij}(z,\eps)
 +\frac{e^{-\gamma_E\eps}\,\Gamma(1-2\eps)}{\Gamma(1-\eps)}\,\left(\frac{\beta_0}{\eps}+K\right)\, {\bf I}^{(1)}_{C,ij}(z,2\eps)\nonumber\\
 &\, + \frac{e^{\gamma_E\eps}}{4\,\eps\,\Gamma(1-\eps)}\,\left(H_i^{(2)} + H_j^{(2)} - H_{i+j}^{(2)} +\beta_1-\beta_0\, K\right)\,,
 \end{align}
where the $H_l^{(2)}$ depend on the flavour of the parton,
\beq\bsp
H_q^{(2)} &\,= \frac{ N^2-1}{N}\,N_f\,\left(\frac{1}{8}\,\zeta _2-\frac{25}{216}\right)-\frac{1}{16}\,\zeta _2-\frac{1}{4}\,\zeta _3+N^2\,\left(-\frac{11}{16}\, \zeta _2+\frac{7}{4}\, \zeta _3+\frac{409}{864}\right)\\
&\,+\frac{1}{N^2}\,\left(\frac{3}{4}\, \zeta _2-\frac{3 }{2}\,\zeta _3-\frac{3}{32}\right)-\frac{41}{108}\,,\\
H_g^{(2)} &\,= N\,N_f\,\left(-\frac{1}{12}\,\zeta _2-\frac{89}{108}\right) +\frac{5 }{27}\,N_f^2-\frac{N_f}{4 N}+N^2\,\left(\frac{11 }{24}\,\zeta _2+\frac{1}{2}\,\zeta _3+\frac{5}{12}\right)\,,
\esp\eeq
and
 \begin{eqnarray}
\beta_0 &=& \frac{11C_A - 2N_f}{6}\, \nonumber \\
\beta_1 & = & \frac{17 C_A^2 - 5 C_A N_f - 3 C_F N_f}{6}\,  \\
 K&=&\left(\frac{67}{18}-\zeta_2\right)\,C_A-\frac{5}{9}\,N_f\,,\nonumber
\end{eqnarray}
with $N_f$ the number of flavours. We recall that the above expressions refer to the splitting amplitudes describing the 
collinear behaviour of the {\em unrenormalized} one-loop and two-loop matrix elements, as in ~\cite{Badger:2004uk}. 
Formulae for the factorization of the pole structure can also be derived after renormalization~\cite{Catani:1998bh}.

Splitting amplitudes have been computed in the literature up to two-loops. The tree-level and one-loop splitting amplitudes are known to all orders in the dimensional regulator $\eps$ (in a variety of schemes)~\cite{Altarelli:1977zs,Bern:1998sc}. The two-loop splitting amplitudes have so far only been computed up to finite terms~\cite{Badger:2004uk,Bern:2004cz}. Our goal is to compute the two-loop splitting amplitudes to $\ord(\eps)$ in conventional dimensional regularisation (CDR). This result is needed in order to construct the collinear counterterms for the double-virtual-real contributions to the inclusive N$^3$LO cross section for the production of a heavy colourless state.


\section{Computation of the splitting amplitudes}
\label{sec:splitting_amps}
In this section we describe the computation of the two-loop splitting amplitudes to $\ord(\eps)$ in CDR. The splitting amplitudes can be extracted most conveniently from the two-loop amplitude for a heavy colourless state decaying into three massless partons. In particular, we have
\beq\bsp\label{eq:3partons_factorization}
\langle\cM^{(0)}_{H\to ggg}&|\cM^{(\ell)}_{H\to ggg}\rangle\simeq \frac{8\pi\,\alpha_0}{s_{13}}\sum_{k=0}^\ell \left(\frac{\alpha_0\,S_{\eps}}{2\pi}\right)^k\,\left(\frac{s_{13}}{\mu^2}\right)^{-k\eps}\,P_{gg}^{(k)}(z)\,\langle\cM^{(0)}_{H\to gg}|\cM^{(\ell-k)}_{H\to gg}\rangle\,,\\
\langle\cM^{(0)}_{H\to q\bar{q}g}&|\cM^{(\ell)}_{H\to q\bar{q}g}\rangle\simeq \frac{8\pi\,\alpha_0}{s_{12}}\sum_{k=0}^\ell \left(\frac{\alpha_0\,S_{\eps}}{2\pi}\right)^k\,\left(\frac{s_{12}}{\mu^2}\right)^{-k\eps}\,P_{q\bar{q}}^{(k)}(z)\,\langle\cM^{(0)}_{H\to gg}|\cM^{(\ell-k)}_{H\to gg}\rangle\,,\\
\langle\cM^{(0)}_{\gamma^*\to q\bar{q}g}&|\cM^{(\ell)}_{\gamma^*\to q\bar{q}g}\rangle\simeq \frac{8\pi\,\alpha_0}{s_{13}}\sum_{k=0}^\ell \left(\frac{\alpha_0\,S_{\eps}}{2\pi}\right)^k\,\left(\frac{s_{13}}{\mu^2}\right)^{-k\eps}\,P_{gq}^{(k)}(z)\,\langle\cM^{(0)}_{\gamma^*\to q\bar{q}}|\cM^{(\ell-k)}_{\gamma^*\to q\bar{q}}\rangle\,,
\esp\eeq
with $z=s_{23}/Q^2$, where $Q^2$ is the virtuality off the initial state. The hard matrix elements are given by
\beq\bsp
\langle\cM^{(0)}_{H\to gg}|\cM_{H\to gg}^{(\ell)}\rangle &\,= \left(\frac{S_\eps\,\alpha_0}{2\pi}\right)^\ell\,\frac{\lambda_0^2\,(N^2-1)}{2^{\ell+1}}\,(1-\eps)\,e^{i\pi\ell\eps}\,(Q^2)^{-1-\ell\eps}\,\cF^{(\ell)}_g\,,\\
\langle\cM^{(0)}_{\gamma^*\to q\bar{q}}|\cM_{\gamma^*\to q\bar{q}}^{(\ell)}\rangle &\,=  \left(\frac{S_\eps\,\alpha_0}{2\pi}\right)^\ell\,\frac{e_q^2\,N}{2^{\ell-2}}\,(1-\eps)\,e^{i\pi\ell\eps}\,(Q^2)^{-1-\ell\eps}\,\cF^{(\ell)}_q\,,
\esp\eeq
where $\lambda_0$ and $e_q$ denote the bare coupling of the Higgs boson to the gluons (in the large top mass limit) and the bare electromagnetic coupling of the quarks, and $\cF^{(\ell)}_g$ and $\cF^{(\ell)}_q$ denote the $\ell$-loop gluon and quark form factors, known up to three loops in QCD~\cite{formfactor}. As a consequence, if we know the the matrix elements in the left-hand side in CDR through $\ord(\eps)$ up to two loops, we can immediately extract the corresponding splitting amplitudes through $\ord(\eps)$. The two-loop matrix element for $\gamma^*\to q\,\bar{q}\,g$ in CDR was presented to all orders in $\eps$ in terms of the master integrals in ref.~\cite{Garland:2001tf}. In ref.~\cite{Gehrmann:2011aa} the two-loop $D$-dimensional tensor coefficients as well as the two-loop helicity coefficients (in the 't Hooft Veltman) scheme for $H$ to three partons were presented. In Appendix~\ref{app:HCDR} we show how to construct the matrix elements in CDR from the tensor coefficients of ref.~\cite{Gehrmann:2011aa}.

The two-loop matrix elements are given, to all orders in the dimensional regulator, in terms of the master integrals of ref.~\cite{Gehrmann:2000zt,Gehrmann:2001ck}, where the complete set of master integrals was evaluated in terms of harmonic polylogarithms (HPLs) and two-dimensional harmonic polylogarithms (2dHPLs)~\cite{Remiddi:1999ew,Gehrmann:2000zt} up to transcendental weight four. This is sufficient to compute the corresponding matrix elements up to finite terms (some of the master integrals are known to all orders in $\eps$, see, e.g., ref.~\cite{Gehrmann:1999as}). Since our goal is to compute the splitting amplitudes to $\ord(\eps)$, we need to compute all the masters integrals to one order higher in the $\eps$ expansion, i.e., up to terms of weight five. Moreover, it turns out to be convenient to have the master integrals in a form where all the divergent logarithms in the collinear limit are resummed in $\epsilon$, because the $\ell$-loop splitting amplitude can then immediately be read off from the coefficient of $s_{ij}^{-\ell\eps}$ in the residue of the matrix element at $s_{ij}=0$ (and dividing out the hard matrix element, cf. eq.~\eqref{eq:3partons_factorization}). We have therefore explicitly computed all the master integrals of ref.~\cite{Gehrmann:2000zt,Gehrmann:2001ck} up to transcendental weight five using the method of differential equations~\cite{Gehrmann:1999as,DiffEqs}. 
We choose a canonical basis of master integrals where all the master integrals are of uniform transcendental weight~\cite{Henn:2013pwa}. In this basis, every master integral is a function of three variables $x_{ij}$ (after dividing out an overall scale), defined by
\beq
x_{ij} = \frac{s_{ij}}{Q^2}\,,
\eeq
and constrained by $x_{12}+x_{13}+x_{23}=1$.
Every uniformly transcendental master integral then admits a representation of the form\footnote{Note that a similar representation obviously exists where any other combination of two out of the three variables $x_{ij}$ are singled out.}
\beq
U_i\left(x_{12},x_{13},x_{23};\eps\right) = \sum_{m,n=0}^2x_{13}^{-m\eps}\,x_{23}^{-n\eps}\,\sum_{k=0}^\infty\eps^k\,u_{i,m,n,k}(x_{13},x_{23})\,,
\eeq
with $u_{i,m,n,k}(x_{13},x_{23})$ a linear combination of 2dHPLs with rational coefficients that are analytic in a neighbourhood of $(x_{13},x_{23})=(0,0)$. We have explicitly computed the functions $u_{i,m,n,k}(x_{13},x_{23})$ up to weight $k=5$, which is sufficient to compute the two-loop matrix elements up to order $\eps$. The initial conditions for the differential equations were obtained from the leading term in the expansion of the master integrals around the soft limit~\cite{Duhr:2013msa}. We checked that our results agree with the first few subleading terms in the soft expansion of the master integrals.

Once the master integrals are known up to terms of transcendental weight five, we can immediately compute the residue at $x_{13}=0$ by expanding in this variable (while keeping the logarithmic divergencies resummed in the form $x_{13}^{-n\eps}$). The variable $x_{13}$ only enters the matrix elements through rational functions, as well as through 2dHPLs of the form $G(\vec a;x_{13})$, where $\vec a$ is independent of $x_{13}$, and the last entry in $\vec a$ is non zero. 2dHPLs of this type can easily be expanded into a power series close to $x_{13}=0$ by using the series representation of multiple polylogarithms,
\beq\bsp
G&(\vec 0_{m_1-1},a_1,\ldots,\vec 0_{m_k-1},a_{k};x_{13}) \\
&\,=(-1)^k\,\sum_{n=1}^\infty\frac{1}{n^{m_1}}\,\left(\frac{x_{13}}{a_1}\right)^n\,Z_{m_2,\ldots,m_k}\left(n-1;\frac{a_1}{a_2},\ldots,\frac{a_{k-1}}{a_{k}}\right)\,,
\esp\eeq
with $\vec 0_m = (\underbrace{0,\ldots,0}_{m\textrm{ times}})$ and 
the $Z_{m_2,\ldots,m_k}$ denote $Z$ sums~\cite{Moch:2001zr}. The splitting amplitudes up to two-loops and up to $\ord(\eps)$ can then immediately be read off by comparing the expression to eq.~\eqref{eq:3partons_factorization}. 
At one loop we find
\beq\bsp\label{eq:P1}
P_{ij}^{(1)}(z) &\,= -C_A\,\frac{\Gamma(1+\eps)\Gamma(1-\eps)}{\eps^2}\,P_{ij}^{(0)}(z)\,z^{-\eps}\,(1-z)^{\eps} \\
&\,+2P_{ij}^{(0)}(z)\,\left[C_{i+j}\,A^{(1)}(z,\eps)+(C_{i+j}-C_A)\,B^{(1)}(z,\eps)\right] + R_{ij}^{(1)}(z,\eps)\,,
\esp\eeq
where $C_{i+j}$ is the Casimir of $SU(N)$ in the representation of the parent parton, and $A^{(1)}(z,\eps)$ and $B^{(1)}(z,\eps)$ are functions that are analytic close to $z=0$. They are independent of the identities of the partons involved in the splitting and can be written as a combination of harmonic polylogarithms of uniform weight,
\beq\bsp
A^{(1)}(z,\eps) &\,=-\frac{1}{\epsilon }\,H_1-\epsilon  \left(H_{2,1}+H_{1,2}+H_{1,1,1}+H_3\right) - \epsilon ^3 \left(H_{4,1}+H_{3,2}+H_{3,1,1}+H_{2,3}\right.\\
&\,\left.+H_{2,2,1}+H_{2,1,2}+H_{2,1,1,1}+H_{1,4}+H_{1,3,1}+H_{1,2,2}+H_{1,2,1,1}+H_{1,1,3}\right.\\
&\,\left.+H_{1,1,2,1}+H_{1,1,1,2}+H_{1,1,1,1,1}+H_5\right)+\ord(\eps^4)\,,\\
B^{(1)}(z,\eps) &\,=-\frac{1}{\epsilon ^2}-\left(H_{1,1}+H_2\right) -\epsilon ^2 \left(H_{3,1}+H_{2,2}+H_{2,1,1}+H_{1,3}+H_{1,2,1}+H_{1,1,2}\right.\\
&\,\left.+H_{1,1,1,1}+H_4\right)+\ord(\eps^3)\,,
\esp\eeq
where we used a shorthand for harmonic polylogarithms, $H_{i,\ldots,j} \equiv H_{i,\ldots,j}(z)$. The function $R_{ij}^{(1)}(z,\eps)$ are rational functions of $z$ order by order in the $\eps$ expansion and depend on the identity of the partons involved in the splitting,
\beq\bsp
R_{gg}^{(1)}(z,\eps)&\, =\frac{N}{3}\,[1-2\eps\,z\,(1-z)]\,\left[N-N_f+\frac{\eps}{3} \,  \left(11 N-14 N_f\right)+\frac{\eps^2}{9}  \left(85 N-127 N_f\right)\right.\\
&\,\left.+\frac{ \epsilon^3}{27} \left(575 N-956 N_f\right)
+\ord(\eps^4)\right]\,,\\
R_{gq}^{(1)}(z,\eps)&\, =\frac{N^4-1}{8 N^2}\,\left(\frac{2-z}{1-2 \epsilon }+z\right) + \ord(\eps^4)\,,\\
R_{qq}^{(1)}(z,\eps)&\, =
P_{qq}^{(0)}(z) \,\frac{1}{2\eps(1-2 \epsilon) }\,\left(N\,\frac{13-8 \epsilon }{3-2 \epsilon }-4N_f\frac{1-\epsilon }{3-2 \epsilon }+\frac{3+2 \epsilon }{N }\right)+\ord(\eps^4)\,.
\esp\eeq
We emphasise that the previous expressions are valid for a time-like splitting where the collinear pair is in the final state. The analytic continuation to a space-like splitting can easily be performed by the replacement
\beq\label{eq:PAC}
z^{-k\eps} \to e^{-i\pi k\eps}\,z^{-k\eps}
\eeq
in eq.~\eqref{eq:P1}.
At two loops, we write
\beq
P_{ij}^{(2)}(z) =\sum_{k=0}^2 z^{-k\eps}\,P_{ij}^{(2,k)}(z)\,,
\label{eq:twoloopz}
\eeq
where $P_{ij}^{(2,k)}(z)$ are analytic in a neighbourhood of $z=0$. Their analytical expressions are given in
 Appendix~\ref{app:twoloop}.

We checked that the tree-level and one-loop are correctly reproduced from our two-loop computation. Moreover, we checked that our results have the correct symmetry properties and soft limits, eq.~(\ref{eq:symmetry} - \ref{eq:gg_soft}), as well as the the correct pole structure in dimensional regularisation, eq.~\eqref{eq:IS2}.


\section{Two-loop single-real emission contributions to Higgs production at N$^3$LO}
\label{sec:VVR_xsecs}
As an application of the splitting amplitudes up to two loops computed in the previous section, we present in this section the integration over phase space of the matrix elements for the scattering processes 
\beq\label{eq:subprocesses}
g\,g\to H\,g\,,\qquad q\,\bar{q}\to H\,g\,,\qquad q\,g\to H\,q\,.
\eeq
These processes are important because they contribute to the inclusive Higgs production cross section at N$^3$LO in perturbative QCD. Indeed, the final-state parton may become collinear to any of the two incoming partons, and the matrix elements develop singularities which are described by the splitting amplitudes we have just computed. The knowledge of the spitting amplitudes therefore enables us to write down explicit counterterms that we can subtract from the matrix element in order to render the phase-space integrations finite. Moreover, the counterterms will turn out to be trivial to integrate over the unresolved phase space, giving rise to a pole in dimensional regularisation. This pole, in turn, is at the origin of the appearance of the $\ord(\eps)$ part of the splitting amplitudes computed in the previous section in the finite terms which enter the cross section. 
We start by giving some generic considerations about the phase-space parametrisation we will use in the following to perform the integration, and we discuss the different channels separately in subsequent subsections.

We denote the momenta of the two incoming partons by $q_1$ and $q_2$, and the momentum of the final-state parton is denoted by $q_3$. Furthermore, we define
\beq\label{eq:PSparam}
s = 2q_1\cdot q_2\,,\qquad t = 2q_1\cdot q_3\,,\qquad u = 2q_2\cdot q_3\,.
\eeq
Note that $s,t,u>0$ and $s=m_H^2+t+u$. The phase space integrals we want to compute are
\beq
\frac{\cN_X}{2s}\,\int d\Phi_2\,2\,\textrm{Re}\langle\cM^{(0)}_X|\cM^{(2)}_X\rangle\,,
\eeq
where $X$ denotes the channel under consideration, and $\cN_X$ denotes the averaging factor over the initial-state colours and spins,
\beq
\cN_{g\,g\to H\,g} = \frac{1}{4\,V^2\,(1-\eps)^2}\,,\qquad
\cN_{q\,g\to H\,q} = \frac{1}{4\,N^2\,(1-\eps)}\,,\qquad
\cN_{q\,\bar{q}\to H\,g} = \frac{1}{4\,N\,V}\,.
\eeq
The $D$-dimensional phase-space measure is given by
\beq
d\Phi_2 = (2\pi)^D\,\delta^{(D)}(q_H+q_3-q_1-q_2)\,\frac{d^Dq_H}{(2\pi)^{D-1}}\,\delta_+(q_H^2-m_H^2)\,\frac{d^Dq_3}{(2\pi)^{D-1}}\,\delta_+(q_3^2)\,.
\eeq
 We parametrise the phase space by
 \beq\label{eq:PS_parametrization}
 m_H^2= (1-\bar{z})\,,\qquad t = s\,\bar{z}\,\lambda\,,\qquad u = s\,\bar{z}\,(1-\lambda)\,.
 \eeq
In this parametrization the phase-space measure becomes
 \beq
 d\Phi_2 = \frac{S_\eps\,e^{\gamma_E\eps}\,s^{-\eps}\,\bar{z}^{1-2\eps}}{8\pi\Gamma(1-\eps)}\,d\lambda\,[\lambda(1-\lambda)]^{-\eps}\,\Theta(\lambda)\,\Theta(1-\lambda)\,,
 \eeq
 where $\Theta$ denotes the step function.
In the following we discuss the integration over phase space of the three different processes in eq.~\eqref{eq:subprocesses}, and we separately discuss the processes with different initial states.

\subsection{The $q\bar{q}$ initial state}
The two-loop matrix element for $q\,\bar{q}\to H\,g$ can be written as, with $s=\mu^2=1$
\beq\bsp
\langle\cM_{q\bar{q}\to Hg}^{(0)}|\cM_{q\bar{q}\to Hg}^{(2)}\rangle&\,=8\pi^2\,\lambda_0^2\,\left(\frac{\alpha_0}{2\pi}\right)^{3}\,\left(\frac{S_\eps\,c_\Gamma}{2}\right)^2\,C_F\,C_A\,e^{2i\pi\eps}\,(1-\bar{z})^{1-2\eps}\\
&\,\times 
M_{q\bar{q}g}^{(2)}(1+x_t+x_u,-x_t,-x_u;\eps)\,,
\esp\eeq
with 
\beq
x_t = \frac{t}{m_H^2} = s\,\frac{\bar{z}}{1-\bar{z}}\,\lambda {\rm~~and~~}x_u = \frac{u}{m_H^2} = s\,\frac{\bar{z}}{1-\bar{z}}\,(1-\lambda )\,,
\eeq
and $M_{q\bar{q}g}^{(2)}$ is a Laurent series in $\eps$ whose coefficients are rational functions of the $x_{ij}$ multiplied by 2dHPLs. The expression of $M_{q\bar{q}g}^{(2)}$ in terms of the tensor coefficients of ref.~\cite{Gehrmann:2011aa} can be found in Appendix~\ref{app:HCDR}.
Note that $M_{q\bar{q}g}^{(2)}$ develops an imaginary part, because the master integrals have branch points at $x_t=0$ and $x_u=0$,
\beq
U_i(1+x_t+x_u,-x_t,-x_u;\eps) =\sum_{m,n=0}^2\,e^{-(m+n)i\pi\eps}\,x_t^{-m\eps}\,x_u^{-n\eps}\,\sum_{k=0}^\infty\eps^k\,u_{i,m,n,k}(-x_{t},-x_{u})\,.
\eeq
The functions $u_{i,m,n,k}(x_{t},x_{u})$ are real, because they are analytic in a neighbourhood of $(x_t,x_u)=(0,0)$. In the previous section, we argued that the amplitude is only singular if the quark pair becomes collinear. This implies that $\langle\cM_{q\bar{q}\to Hg}^{(0)}|\cM_{q\bar{q}\to Hg}^{(2)}\rangle$ is finite everywhere inside the phase space, and so we can simply expand in $\eps$ and perform the integral over $\lambda$ order by order in $\epsilon$. The integrand contains 2dHPLs of the form $G(\vec a(x_u);-x_t)$, where $x_t$ and $x_u$ are rational functions of $\lambda$ and $\bar{z}$ (cf. eq.~\eqref{eq:PS_parametrization}). Interpreting the 2dHPLs as special instances of multiple polylogarithms~\cite{Gonchpoly}, and using the techniques of ref.~\cite{symbols}, we can easily convert all the 2dHPLs of the form $G(\vec a(x_u);-x_t)$ into multiple polylogarithms of the form $G(\vec b(\bar{z});\lambda)$ and multiple polylogarithms in $\bar{z}$. As an example, the following relation holds
\beq\bsp
G\left(\frac{(\lambda-1) \bar{z}}{\bar{z}-1},0,1;\frac{\lambda \bar{z}}{\bar{z}-1}\right)&\, =-G\left(1,0,\frac{\bar{z}-1}{\bar{z}};\lambda \right)-G\left(1,\frac{2 \bar{z}-1}{\bar{z}},1;\lambda \right)\\
&\,-G\left(1,\frac{2 \bar{z}-1}{\bar{z}},\frac{\bar{z}-1}{\bar{z}};\lambda \right)\,.
\esp\eeq
Similar identities can easily be derived in all other cases, and the phase-space integration over $\lambda$ can easily be performed using the the definition of the multiple polylogarithms,
\beq\label{eq:MPL_def}
\int_0^1\frac{d\lambda}{\lambda-b_1(\bar{z})}\,G(\vec b(\bar{z});\lambda) = G(b_1(\bar{z}),\vec b(\bar{z});1)\,.
\eeq
After a final massaging, we find that all the phase space integrals can be expressed in terms of HPLs in $1-\bar{z}$, in agreement with the expectations.

\subsection{The $qg$ initial state}
The two-loop matrix element for $q\,g\to H\,q$ can be written as, with $s=\mu^2=1$,
\beq\bsp
\langle\cM_{qg\to Hq}^{(0)}|\cM_{qg\to Hq}^{(2)}\rangle&\,=-8\pi^2\,\lambda_0^2\,\left(\frac{\alpha_0}{2\pi}\right)^{3}\,\left(\frac{S_\eps\,c_\Gamma}{2}\right)^2\,C_F\,C_A\,e^{2i\pi\eps}\,(1-\bar{z})^{1-2\eps}\\
&\,\times M_{q\bar{q}g}^{(2)}(-x_t,1+x_t+x_u,-x_u;\eps)\,.
\esp\eeq
The function $M_{q\bar{q}g}^{(2)}(-x_t,1+x_t+x_u,-x_u;\eps)$ is identical to the corresponding quantity in the $q\,\bar{q}\to H\,g$ case, up to analytic continuation.
Note that there is an additional minus sign, coming from the crossing of a single fermion to the initial state. The analytic continuation of the function $M_{q\bar{q}g}^{(2)}(-x_t,1+x_t+x_u,-x_u;\eps)$ can be performed in a similar way as in the $q\,\bar{q}\to H\,g$ case.

The integrand becomes singular when the initial and final-state quarks are collinear, and the singularity is controlled by the splitting amplitudes computed in Section~\ref{sec:splitting_amps}. More precisely, after inserting the parametrisation~\eqref{eq:PSparam}, the integrand has a simple pole as $\lambda\to0$, and from eq.~\eqref{eq:3partons_factorization}, we have
\beq
\langle\cM^{(0)}_{qg \to Hq}|\cM^{(2)}_{qg \to Hq}\rangle = \sum_{k=0}^2\lambda^{-1-k\eps}\,{\bf C}^{(2,k)}_{q\bar{q}}(\bar{z};\eps) + \ord(\lambda^0)\,,
\eeq
with
\beq
{\bf C}^{(2,k)}_{q\bar{q}}(\bar{z};\eps) = 16\pi^2\,\left(\frac{\alpha_0}{2\pi}\right)^{k+1}\,S_{\eps}^k\,\bar{z}^{-1-k\eps}\,e^{-ki\pi\eps}\,P_{q\bar{q}}^{(k)}\left(\frac{\bar{z}}{\bar{z}-1}\right)\,\langle\cM^{(0)}_{gg\to H}|\cM^{(2-k)}_{gg\to H}\rangle\,.
\eeq
Note that the splitting amplitude is evaluated at $\lambda=0$, so that we only subtract the residue of the matrix element at $\lambda=0$. 
The phase-space integral then becomes 
\beq\bsp\label{eq:qg_integral}
\int_0^1&d\lambda\,[\lambda(1-\lambda)]^{-\eps}\,\langle\cM^{(0)}_{qg \to Hq}|\cM^{(2)}_{qg \to Hq}\rangle
=-\frac{1}{\eps}\sum_{k=0}^2\frac{{\bf C}^{(2,k)}_{q\bar{q}}(\bar{z};\eps)}{(k+2)}\\
&+\int_0^1d\lambda\,\left\{[\lambda(1-\lambda)]^{-\eps}\,\langle\cM^{(0)}_{qg \to Hq}|\cM^{(2)}_{qg \to Hq}\rangle -\sum_{k=0}^2\lambda^{-1-(k+1)\eps}\,{\bf C}^{(2,k)}_{q\bar{q}}(\bar{z};\eps)\right\}\,.
\esp\eeq
The integral in the second line is convergent, and so we can expand in $\eps$ under the integration sign and integrate order by order in $\epsilon$. At this point, however, we need to face a technical difficulty: performing the integrations using eq.~\eqref{eq:MPL_def} requires the integrations to be performed term by term. In general, these integrals will be divergent, because only the sum in the second line of eq.~\eqref{eq:qg_integral} is finite. We therefore introduce a cut-off $\delta$ for the lower integration limit\footnote{The integrand does not develop poles for $\lambda\to1$, and so no cut-off is required for the upper integration limit.} and perform all the integrations term by term. As a result, all the integrals are convergent, but they explicitly depend on the cut-off. We observe that all divergences cancel when we expand the result around $\delta=0$, in agreement with the expectation that the integral in the second line of eq.~\eqref{eq:qg_integral} is convergent.

\subsection{The $gg$ initial state}
The two-loop matrix element for $g\,g\to H\,g$ can be written as
\beq\bsp
\langle\cM_{gg\to Hg}^{(0)}|\cM_{gg\to Hg}^{(2)}\rangle&\,=8\pi^2\,s^{1-2\eps}\,|\lambda_0|^2\,\left(\frac{\alpha_0}{2\pi}\right)^{3}\,\left(\frac{S_\eps\,c_\Gamma}{2}\right)^2\,e^{2i\pi\eps}\,(1-\bar{z})^{1-2\epsilon}\,V\,C_A\\
&\,\times M_{ggg}^{(2)}\left(1+x_{t}+x_u,-x_{t},-x_{u};\eps\right)\,.
\esp\eeq
The expression of $M_{ggg}^{(2)}$ in terms of the tensor coefficients of ref.~\cite{Gehrmann:2011aa} is given in Appendix~\ref{app:HCDR}.
Its analytic continuation can be performed in the same way as in the previous cases. The phase-space integral diverges in the collinear limits where either $t$ or $u$ vanish. Moreover, the gluon amplitude also develops a singularity when the final state gluon becomes soft, corresponding to a simultaneous vanishing of $t$ and $u$. In terms of our phase space parametrisation~\eqref{eq:PS_parametrization} this corresponds to $\bar{z}\to0$, and in this limit the matrix element behaves like
\beq
\langle\cM_{gg\to Hg}^{(0)}|\cM_{gg\to Hg}^{(2)}\rangle = 
\sum_{k=0}^2\bar{z}^{-2-2k\eps}\,[\lambda\,(1-\lambda)]^{-1-k\eps}\,{\bf S}_g^{(2,k)}(\eps) + \ord(\bar{z}^0)\,,
\eeq
with
\beq\label{eq:Sg_deg}
{\bf S}_g^{(2,k)}(\eps) = -32\pi^2\,\left(\frac{\alpha_0}{2\pi}\right)^{k+1}\,\left(\frac{S_\eps}{2}\right)^k\,C_A\,e^{-k\pi i\eps}\,r_S^{(k)}(\eps)\,\langle\cM^{(0)}_{gg\to H}|\cM^{(2-k)}_{gg\to H}\rangle_{\big|\bar{z}=0}\,.
\eeq
We define the \emph{soft-regularized} amplitude, defined as the amplitude with the soft singularity at $\bar{z}=0$ subtracted,
\beq
\langle\cM_{gg\to Hg}^{(0)}|\cM_{gg\to Hg}^{(2)}\rangle^{\textrm{reg}}\equiv \langle\cM_{gg\to Hg}^{(0)}|\cM_{gg\to Hg}^{(2)}\rangle - \sum_{k=0}^2\bar{z}^{-2-2k\eps}\,[\lambda\,(1-\lambda)]^{-1-k\eps}\,{\bf S}_g^{(2,k)}(\eps)\,.
\eeq
The soft-regularised amplitude is free of soft singularities, but still contains non-overlapping collinear divergences. The residues at the collinear poles at $\lambda=0$ and $\lambda=1$ can be obtained in a way similar to the $q\,g\to H\,q$ case considered in the previous section. By Bose symmetry, the residues at the two poles are identical. In particular, the residue at $\lambda=0$ is given by
\beq
\langle\cM^{(0)}_{gg \to Hg}|\cM^{(2)}_{gg \to Hg}\rangle^{\textrm{reg}} = \sum_{k=0}^2\lambda^{-1-k\eps}\,{\bf C}^{(2,k),\textrm{reg}}_{g\bar{g}}(\bar{z};\eps) + \ord(\lambda^0)\,,
\eeq
with
\beq
{\bf C}^{(2,k),\textrm{reg}}_{gg}(\bar{z};\eps)= {\bf C}^{(2,k)}_{gg}(\bar{z};\eps)- \bar{z}^{-1-2k\eps}\,{\bf S}_g^{(2,k)}(\eps)\,,
\eeq
and
\beq
{\bf C}^{(2,k)}_{gg}(\bar{z};\eps) = -16\pi^2\,\left(\frac{\alpha_0}{2\pi}\right)^{k+1}\,S_{\eps}^k\,\bar{z}^{-1-k\eps}\,e^{-ki\pi\eps}\,P_{gg}^{(k)}\left(\frac{\bar{z}}{\bar{z}-1}\right)\,\langle\cM^{(0)}_{gg\to H}|\cM^{(2-k)}_{gg\to H}\rangle\,.
\eeq
The phase-space integral can now be written as
\beq\bsp\label{eq:gg_reg}
\bar{z}^{1-2\eps}\int_0^1&d\lambda\,[\lambda(1-\lambda)]^{-\eps}\,\langle\cM^{(0)}_{gg \to Hg}|\cM^{(2)}_{gg \to Hg}\rangle\\
&
=\sum_{k=0}^2\bar{z}^{-1-2(k+1)\eps}\,\frac{\Gamma(-(k+1)\eps)^2}{\Gamma(-2(k+1)\eps)}\,{\bf S}^{(2,k)}_{g}(\eps)\\
&\,\qquad
+ \bar{z}^{1-2\eps}\int_0^1d\lambda\,[\lambda(1-\lambda)]^{-\eps}\,\langle\cM^{(0)}_{gg \to Hg}|\cM^{(2)}_{gg \to Hg}\rangle^{\textrm{reg}}\\
&=\sum_{k=0}^2\left\{\bar{z}^{-1-2(k+1)\eps}\,\frac{\Gamma(-(k+1)\eps)^2}{\Gamma(-2(k+1)\eps)}\,{\bf S}^{(2,k)}_{g}(\eps)-\frac{2\,\bar{z}^{1-2\eps}}{\eps}\,\frac{{{\bf C}}^{(2,k),\textrm{reg}}_{gg}(\bar{z};\eps)}{(k+2)}\right\}\\
&\,\qquad+\bar{z}^{1-2\eps}\int_0^1d\lambda\,\Big\{[\lambda(1-\lambda)]^{-\eps}\,\langle\cM^{(0)}_{g\,g \to H\, g}|\cM^{(2)}_{g\,g \to H\, g}\rangle^{\textrm{reg}} \\
&\,\qquad-\sum_{k=0}^2\left[\lambda^{-1-(k+1)\eps}+(1-\lambda)^{-1-(k+1)\eps}\right]\,{{\bf C}}^{(2,k),\textrm{reg}}_{gg}(\bar{z};\eps)\Big\}\,,
\esp\eeq
The remaining integral is convergent and can be performed order by order in $\eps$ using eq.~\eqref{eq:MPL_def}. Individual terms in the integrand can develop poles as $\lambda$ approaches either 0 or 1. We therefore introduce a cut-off for each integration limit, and we observe that all the singularities cancel in the sum over all terms when the cut-offs are removed, confirming the expectation that the remaining integral in eq.~\eqref{eq:gg_reg} is convergent.


\section{Conclusion}
\label{sec:conclusion}

The single collinear factorization of scattering amplitudes in massless QCD is determined by so-called splitting amplitudes, 
which can be expanded perturbatively in the number of virtual loops~\cite{Kosower:1999xi}. We derived the two-loop corrections 
to all splitting amplitudes, finding agreement with previously known results~\cite{Bern:2004cz,Badger:2004uk} up to
 finite terms in the 
dimensional regulator, and deriving the subleading terms for the first time. When integrated over the unresolved single 
collinear phase space in actual N$^3$LO calculations, these subleading terms produce finite contributions. 
We used our result to analytically compute the 
two-loop single-real contribution to the coefficient function of inclusive Higgs boson production in gluon fusion, improving 
upon the previously available soft approximation~\cite{Duhr:2013msa,Li:2013lsa} of this function. 

When combined with either expanded or full results for the other channels (one-loop double-real and triple-real), 
our expression will allow to derive further terms in the threshold expansion~\cite{Anastasiou:2014vaa} or 
to complete an exact calculation of the full N$^3$LO coefficient function for Higgs production in gluon fusion.


\section*{Acknowledgements}
We would like to thank Falko Dulat and Bernhard Mistlberger as well as Lance Dixon and Hua-Xing Zhu for 
useful discussions and comparisons with closely related results prior to publication. 
This research was supported in part by the Swiss National Science Foundation (SNF) under contract  200020-149517
and by 
the Research Executive Agency (REA) of the European Union under the Grant Agreements PITN--GA---2010-264564 ({\it LHCPhenoNet}), PITN--GA--2012--316704 ({\it HiggsTools}), and the ERC Advanced Grant {\it MC@NNLO} (340983).


\appendix

\section{Splitting amplitudes at two loops}
\label{app:twoloop}
The two-loop splitting amplitudes can be expanded according to the powers $z^{-k\epsilon}$, with coefficients 
that are analytic around $z=0$, as defined in (\ref{eq:twoloopz}). 
We write these coefficients as:
\beq\bsp
P_{ij}^{(2,2)}(z) &\,= P_{ij}^{(0)}(z)\,U_{ij}^{(2,2)}(z,\eps) + V_{ij}^{(2,2)}(z,\eps) \\
&\,+ \left(\frac{ N }{4 \epsilon ^3}\,\beta _0\,-\frac{N}{2 \epsilon ^2}\,K\right)\,P_{ij}^{(0)}(z)\,(1-z)^{2 \epsilon }\,,\\
P_{ij}^{(2,1)}(z) &\,= P_{ij}^{(0)}(z)\,U_{ij}^{(2,1)}(z,\eps) + V_{ij}^{(2,1)}(z,\eps)\\
&\,-\frac{N }{\epsilon ^2}\,\Gamma (1-\epsilon ) \Gamma (1+\epsilon )\,R_{ij}^{(1)}(z,\eps) (1-z)^{\epsilon }\,,\\
P_{ij}^{(2,0)}(z) &\,= P_{ij}^{(0)}(z)\,U_{ij}^{(2,0)}(z,\eps) + V_{ij}^{(2,0)}(z,\eps)-\frac{ N}{\epsilon ^2}\,\beta _0\,P_{ij}^{(0)}(z)\,H_1\,.
\esp\eeq 
Note that these expressions are again valid for a time-like splitting, and the space-like case can again be obtained by the replacement~\eqref{eq:PAC}.
The functions $U_{ij}^{(2,k)}(z)$ are of uniform weight, and are given by
\begin{align}
U_{gg}^{(2,2)}(z,\eps) &= N^2\,\Bigg[\frac{1}{2\eps^4} - \frac{1}{\eps^3}\,H_1+\frac{1}{\eps^2}\,\big(2\,H_{1,1}+\zeta_2\big) + \frac{1}{\eps}\big(-4\,H_{1,1,1}+\frac{1}{2}\,\zeta _3-2\,\zeta_2\,H_1\big)\nonumber \\
&+ 3\,\zeta_2\,H_{1,1}+2\,H_{3,1}+3\,H_{2,2}+H_{2,1,1}+2\,H_{1,3}+H_{1,2,1}+H_{1,1,2}+\frac{29 }{4}\,\zeta_4\nonumber\\
&+8\,H_{1,1,1,1}-\zeta_2\,H_2+2\,\zeta_3\,H_1+4\,H_4 + \eps\,\big(-\zeta_2\,H_{1,2}-5\,\zeta_2\,H_{1,1,1}\\
&+10\,H_{3,2}-3\,H_{3,1,1}+10\,H_{2,3}-4\,H_{2,2,1}-3\,H_{2,1,1,1}+8\,H_{1,4}-2\,H_{1,3,1}\nonumber\\
&-H_{1,2,2}-4\,H_{1,2,1,1}+2\,H_{1,1,3}-2\,H_{1,1,2,1}-3\,H_{1,1,1,2}-16\,H_{1,1,1,1,1}\nonumber\\
&+4\,\zeta_2\,\zeta_3-\frac{37}{2}\,\zeta_5-4\,\zeta_2\,H_3-2\,\zeta_3\,H_2-\frac{35}{4}\,\zeta_4\,H_1+24\,H_5-12\,\zeta_3\,H_{1,1}\big)\nonumber\\
&+\ord(\eps^2)\Bigg]\nonumber\,,
\end{align}
\begin{align}
U_{gg}^{(2,1)}(z,\eps) &\,= N^2\,\Bigg[\frac{2}{\eps^3}\,H_1-\frac{4}{\eps^2}\,H_{1,1}+\frac{1}{\eps}\,\big(2\,H_{2,1}+2\,H_{1,2}+8\,H_{1,1,1}+2\,\zeta_2\,H_1+2\,H_3\big)\nonumber\\
&-4\,\zeta_2\,H_{1,1}-4\,H_{3,1}-2\,H_{2,2}-6\,H_{2,1,1}-2\,H_{1,3}-6\,H_{1,2,1}-4\,H_{1,1,2}\\
&-16\,H_{1,1,1,1}-12\,\zeta_3\,H_1
+\eps\,\big(2\,\zeta_2\,H_{2,1}+2\,\zeta_2\,H_{1,2}+8\,\zeta_2\,H_{1,1,1}+24\,\zeta_3\,H_{1,1}\nonumber\\
&+2\,H_{4,1}+2\,H_{3,2}+8\,H_{3,1,1}+2\,H_{2,3}+6\,H_{2,2,1}+4\,H_{2,1,2}+14\,H_{2,1,1,1}\nonumber\\
&+2\,H_{1,4}+6\,H_{1,3,1}+4\,H_{1,2,2}+14\,H_{1,2,1,1}+4\,H_{1,1,3}+12\,H_{1,1,2,1}+8\,H_{1,1,1,2}\nonumber\\
&+32\,H_{1,1,1,1,1}+2\,\zeta_2\,H_3-\frac{29}{2}\,\zeta_4\,H_1+2\,H_5\big)+\ord(\eps^2)\Bigg]\,,\nonumber\\
%
U_{gg}^{(2,0)}(z,\eps) &\,= N^2\,\Bigg[\frac{4}{\eps^2}\,H_{1,1}+\frac{2}{\eps}\,\zeta_2\,H_1+\,\zeta_2\,H_{1,1}+6\,H_{3,1}+H_{2,2}+11\,H_{2,1,1}+2\,H_{1,3}\nonumber\\
&+11\,H_{1,2,1}+7\,H_{1,1,2}+16\,H_{1,1,1,1}+\zeta_2\,H_2+11\,\zeta_3\,H_1-4\,H_4\\
&+\eps\,\big(-3\,\zeta_2\,H_{1,2}-3\,\zeta_2\,H_{1,1,1}-2\,\zeta_3\,H_{1,1}+12\,H_{4,1}+8\,H_{3,2}+7\,H_{3,1,1}+8\,H_{2,3}\nonumber\\
&+8\,H_{2,2,1}+8\,H_{2,1,2}+3\,H_{2,1,1,1}+6\,H_{1,4}+6\,H_{1,3,1}+11\,H_{1,2,2}+4\,H_{1,2,1,1}\nonumber\\
&+6\,H_{1,1,3}+2\,H_{1,1,2,1}+11\,H_{1,1,1,2}-2\,\zeta_3\,H_2+\frac{11}{4}\,\zeta_4\,H_1+8\,H_5\big)+\ord(\eps^2)\Bigg]\,,\nonumber\\
U_{gq}^{(2,2)}(z,\eps) &= N^2\,\Bigg[\frac{1}{2\eps^4} - \frac{1}{\eps^3}\,H_1+\frac{1}{\eps^2}\,\big(2\,H_{1,1}+\zeta_2\big) +\frac{1}{\eps}\,\bigg(\frac{1}{2}\zeta_3-H_{2,1}-\frac{3}{2}\,H_{1,2}-\frac{9}{2}\,H_{1,1,1}\nonumber\\
&-\frac{3}{2}\,\zeta_2\,H_1-2\,H_3\bigg)
+3\,\zeta_2\,H_{1,1}+2\,H_{3,1}-2\,H_{2,2}+\frac{5}{2}\,H_{2,1,1}-3\,H_{1,3}\\
&+3\,H_{1,2,1}+H_{1,1,2}+\frac{19}{2}\,H_{1,1,1,1}+\frac{29\,\zeta_4}{4}+\zeta_2\,H_2+3\,\zeta_3\,H_1-8\,H_4\nonumber\\
&+\eps\,\big(-\frac{9}{2}\,\zeta_2\,H_{2,1}-\frac{9}{2}\,\zeta_2\,H_{1,2}-7\,\zeta_2\,H_{1,1,1}-\frac{21}{2}\,\zeta_3\,H_{1,1}+8\,H_{4,1}-7\,H_{3,2}\nonumber\\
&-\frac{7}{2}\,H_{3,1,1}-7\,H_{2,3}+\frac{5}{2}\,H_{2,2,1}+\frac{1}{2}\,H_{2,1,2}-\frac{11}{2}\,H_{2,1,1,1}-12\,H_{1,4}+3\,H_{1,3,1}\nonumber\\
&-\frac{9}{2}\,H_{1,2,2}-\frac{15}{2}\,H_{1,2,1,1}+H_{1,1,3}-\frac{5}{2}\,H_{1,1,2,1}-3\,H_{1,1,1,2}-\frac{39}{2}\,H_{1,1,1,1,1}\nonumber\\
&-\frac{37}{2}\,\zeta_5+4\,\zeta_2\,\zeta_3-4\,\zeta_2\,H_3+\frac{7}{2}\,\zeta_3\,H_2+\frac{17}{8}\,\zeta_4\,H_1-32\,H_5\big)\Bigg]\nonumber\\
&+\frac{1}{\eps}\,\bigg(-2\,H_{2,1}-\frac{3}{2}\,H_{1,2}-\frac{3}{2}\,H_{1,1,1}+\frac{1}{2}\,\zeta_2\,H_1-2\,H_3\bigg)
-7\,H_{3,1}-10\,H_{2,2}\nonumber\\
&-\frac{3}{2}\,H_{2,1,1}-7\,H_{1,3}-3\,H_{1,2,1}-6\,H_{1,1,2}+\frac{3}{2}\,H_{1,1,1,1}+\zeta_2\,H_2+2\,\zeta_3\,H_1-12\,H_4\nonumber\\
&+\eps\,\bigg(-\frac{13}{2}\,\zeta_2\,H_{2,1}-\frac{11}{2}\,\zeta_2\,H_{1,2}-3\,\zeta_2\,H_{1,1,1}-\frac{1}{2}\,\zeta_3\,H_{1,1}-29\,H_{4,1}-52\,H_{3,2}\nonumber\\
&-\frac{19}{2}\,H_{3,1,1}-43\,H_{2,3}-\frac{7}{2}\,H_{2,2,1}-\frac{47}{2}\,H_{2,1,2}-\frac{13}{2}\,H_{2,1,1,1}-28\,H_{1,4}-14\,H_{1,3,1}\nonumber\\
&-\frac{55}{2}\,H_{1,2,2}-\frac{9}{2}\,H_{1,2,1,1}-25\,H_{1,1,3}+\frac{7}{2}\,H_{1,1,2,1}-7\,H_{1,1,1,2}-\frac{15}{2}\,H_{1,1,1,1,1}\nonumber\\
&-7\,\zeta_2\,H_3+\frac{7}{2}\,\zeta_3\,H_2+\frac{109}{8}\,\zeta_4\,H_1-56\,H_5\bigg)\nonumber
\end{align}
\begin{align}
&+\frac{1}{N^2}\,\Bigg[-\frac{1}{\eps}\big(H_{2,1}+H_{1,1,1})-7\,H_{3,1}-5\,H_{2,2}-3\,H_{2,1,1}-2\,H_{1,3}-5\,H_{1,2,1}\nonumber\\
&-6\,H_{1,1,2}-\zeta_2\,H_2+\zeta_3\,H_1
+\eps\,\big(-2\,\zeta_2\,H_{2,1}-2\,\zeta_2\,H_{1,2}-\zeta_2\,H_{1,1,1}-2\,\zeta_3\,H_{1,1}\nonumber\\
&-37\,H_{4,1}-35\,H_{3,2}-9\,H_{3,1,1}-26\,H_{2,3}-10\,H_{2,2,1}-24\,H_{2,1,2}-4\,H_{2,1,1,1}\nonumber\\
&-8\,H_{1,4}-19\,H_{1,3,1}-24\,H_{1,2,2}-H_{1,2,1,1}-24\,H_{1,1,3}+4\,H_{1,1,2,1}-7\,H_{1,1,1,2}\nonumber\\
&-4\,H_{1,1,1,1,1}-7\,\zeta_2\,H_3-2\,\zeta_3\,H_2+\frac{11}{4}\,\zeta_4\,H_1\big)\Bigg]+\ord(\eps^2)\,,\nonumber\\
%
%
U_{gq}^{(2,1)}(z,\eps) &= N^2\,\Bigg[\frac{1}{\eps^3}\,H_1-\frac{1}{\eps^2}\,\big(-3\,H_{1,1}-H_2\big) +\frac{1}{\eps}\,\big(3\,H_{2,1}+2\,H_{1,2}+7\,H_{1,1,1}+\zeta_2\,H_1+H_3\big)\nonumber\\
&-3\,\zeta_2\,H_{1,1}-3\,H_{3,1}-2\,H_{2,2}-7\,H_{2,1,1}-2\,H_{1,3}-6\,H_{1,2,1}-4\,H_{1,1,2}\\
&-15\,H_{1,1,1,1}-\zeta_2\,H_2-6\,\zeta_3\,H_1-H_4
+\eps\,\big(3\,\zeta_2\,H_{2,1}+2\,\zeta_2\,H_{1,2}+7\,\zeta_2\,H_{1,1,1}\nonumber\\
&+18\,\zeta_3\,H_{1,1}+3\,H_{4,1}+2\,H_{3,2}+7\,H_{3,1,1}+2\,H_{2,3}+6\,H_{2,2,1}+4\,H_{2,1,2}\nonumber\\
&+15\,H_{2,1,1,1}+2\,H_{1,4}+6\,H_{1,3,1}+4\,H_{1,2,2}+14\,H_{1,2,1,1}+4\,H_{1,1,3}+12\,H_{1,1,2,1}\nonumber\\
&+8\,H_{1,1,1,2}+31\,H_{1,1,1,1,1}+\zeta_2\,H_3+6\,\zeta_3\,H_2-\frac{29}{4}\,\zeta_4\,H_1+H_5\big)\Bigg]\nonumber\\
&-\frac{1}{\eps^3}\,H_1+\frac{1}{\eps^2}\,\big(\,H_{1,1}-H_2\big)+\frac{1}{\eps}\,\big(\,H_{2,1}-H_{1,1,1}-\zeta_2\,H_1-H_3\big)
+\zeta_2\,H_{1,1}+H_{3,1}\nonumber\\
%
&-H_{2,1,1}+H_{1,1,1,1}-\zeta_2\,H_2+6\,\zeta_3\,H_1-H_4
+\eps\,\big(\,\zeta_2\,H_{2,1}-\zeta_2\,H_{1,1,1}-6\,\zeta_3\,H_{1,1}\nonumber\\
&+H_{4,1}-H_{3,1,1}+H_{2,1,1,1}-H_{1,1,1,1,1}-\zeta_2\,H_3+6\,\zeta_3\,H_2+\frac{29}{4}\,\zeta_4\,H_1-H_5\big)+\ord(\eps^2)\,,\nonumber\\
U_{gq}^{(2,0)}(z,\eps) &= N^2\,\Bigg[\frac{1}{\eps^2}\,H_{1,1}+\frac{1}{\eps}\,\big(-H_{2,1}+\frac{1}{2}\,H_{1,2}-\frac{5}{2}\,H_{1,1,1}+\frac{1}{2}\,\zeta_2\,H_1+2\,H_3\big)
-2\,H_{3,1}\nonumber\\
&-H_{2,2}+\frac{5}{2}\,H_{2,1,1}-2\,H_{1,3}+H_{1,2,1}+H_{1,1,2}+\frac{11}{2}\,H_{1,1,1,1}+8\,\zeta_3\,H_1-4\,H_4\nonumber\\
&
+\eps\,\big(-\frac{1}{2}\,\zeta_2\,H_{2,1}-\frac{3}{2}\,\zeta_2\,H_{1,2}-\frac{21}{2}\,\zeta_3\,H_{1,1}+8\,H_{4,1}+13\,H_{3,2}+\frac{7}{2}\,H_{3,1,1}\\
&+7\,H_{2,3}+\frac{1}{2}\,H_{2,2,1}+\frac{5}{2}\,H_{2,1,2}-\frac{11}{2}\,H_{2,1,1,1}+11\,H_{1,4}+2\,H_{1,3,1}+\frac{19}{2}\,H_{1,2,2}\nonumber\\
&-\frac{5}{2}\,H_{1,2,1,1}+2\,H_{1,1,3}-\frac{11}{2}\,H_{1,1,2,1}-H_{1,1,1,2}-\frac{23}{2}\,H_{1,1,1,1,1}-2\,\zeta_2\,H_3\nonumber\\
&-\frac{21}{2}\,\zeta_3\,H_2+\frac{75}{8}\,\zeta_4\,H_1+16\,H_5\big)\Bigg]\nonumber\\
&-\frac{2}{\eps^2}\,H_{1,1}+\frac{1}{\eps}\,\big(2\,H_{2,1}+\frac{3}{2}\,H_{1,2}+\frac{3}{2}\,H_{1,1,1}-\frac{3}{2}\,\zeta_2\,H_1+2\,H_3\big)
-\zeta_2\,H_{1,1}-H_{3,1}\nonumber\\
&+4\,H_{2,2}-\frac{3}{2}\,H_{2,1,1}-2\,H_{1,2,1}+3\,H_{1,1,2}-\frac{7}{2}\,H_{1,1,1,1}-6\,\zeta_3\,H_1
+\eps\,\big(-\frac{1}{2}\,\zeta_2\,H_{2,1}\nonumber\\
&+\frac{7}{2}\,\zeta_2\,H_{1,2}-\frac{5}{2}\,\zeta_3\,H_{1,1}+11\,H_{4,1}+21\,H_{3,2}+\frac{19}{2}\,H_{3,1,1}+13\,H_{2,3}+\frac{5}{2}\,H_{2,2,1}\nonumber\\
&+\frac{25}{2}\,H_{2,1,2}+\frac{19}{2}\,H_{2,1,1,1}+2\,H_{1,4}+10\,H_{1,3,1}+\frac{25}{2}\,H_{1,2,2}+\frac{15}{2}\,H_{1,2,1,1}\nonumber\\
&+11\,H_{1,1,3}+\frac{7}{2}\,H_{1,1,2,1}+6\,H_{1,1,1,2}+\frac{15}{2}\,H_{1,1,1,1,1}+\zeta_2\,H_3-\frac{9}{2}\,\zeta_3\,H_2\nonumber\\
&-\frac{45}{8}\,\zeta_4\,H_1+8\,H_5\big)\nonumber
\end{align}
\begin{align}
&+\frac{1}{N^2}\,\Bigg[\frac{1}{\eps^2}\,H_{1,1}+\frac{1}{\eps}\,\big(3\,H_{2,1}+H_{1,2}+4\,H_{1,1,1}\big)
+7\,H_{3,1}+6\,H_{2,2}+7\,H_{2,1,1}\nonumber\\
&+4\,H_{1,3}+8\,H_{1,2,1}+9\,H_{1,1,2}+7\,H_{1,1,1,1}+\zeta_2\,H_2-3\,\zeta_3\,H_1
+\eps\,\big(2\,\zeta_2\,H_{1,2}\nonumber\\
&-3\,\zeta_2\,H_{1,1,1}+6\,\zeta_3\,H_{1,1}+15\,H_{4,1}+16\,H_{3,2}+13\,H_{3,1,1}+14\,H_{2,3}\nonumber\\
&+10\,H_{2,2,1}+18\,H_{2,1,2}+18\,H_{2,1,1,1}-3\,H_{1,4}+14\,H_{1,3,1}+14\,H_{1,2,2}+14\,H_{1,2,1,1}\nonumber\\
&+15\,H_{1,1,3}+11\,H_{1,1,2,1}+18\,H_{1,1,1,2}+19\,H_{1,1,1,1,1}+3\,\zeta_2\,H_3+4\,\zeta_3\,H_2\nonumber\\
&-\frac{49}{4}\,\zeta_4\,H_1\big)\Bigg]+\ord(\eps^2)\,,\nonumber\\
%
%
U_{qq}^{(2,2)}(z,\eps) &=N^2\,\Bigg[\frac{1}{2\eps^4}-\frac{1}{\eps^3}\,H_1+\frac{1}{\eps^2}\,\big(2\,H_{1,1}+\zeta_2\big)
+\frac{1}{\eps}\bigg(\frac{1}{2}\,H_{2,1}+H_{1,2}-\frac{7}{2}\,H_{1,1,1}-\zeta_3\nonumber\\
&-2\,\zeta_2\,H_1+H_3\bigg)
+\frac{5}{2}\,\zeta_2\,H_{1,1}+H_{3,1}+\frac{11}{2}\,H_{2,2}+\frac{1}{2}\,H_{2,1,1}+4\,H_{1,3}\\
&-H_{1,2,1}+\frac{3}{2}\,H_{1,1,2}+\frac{13}{2}\,H_{1,1,1,1}+\frac{35\,\zeta_4}{8}-\frac{3}{2}\,\zeta_2\,H_2+5\,\zeta_3\,H_1+8\,H_4\nonumber\\
&+\eps\,\bigg(\frac{3}{2}\,\zeta_2\,H_{2,1}+2\,\zeta_2\,H_{1,2}-\frac{5}{2}\,\zeta_2\,H_{1,1,1}-\frac{35}{2}\,\zeta_3\,H_{1,1}-4\,H_{4,1}\nonumber\\
&+18\,H_{3,2}+H_{3,1,1}+18\,H_{2,3}-\frac{15}{2}\,H_{2,2,1}+\frac{1}{2}\,H_{2,1,2}-\frac{5}{2}\,H_{2,1,1,1}+16\,H_{1,4}\nonumber\\
&
-4\,H_{1,3,1}+2\,H_{1,2,2}+H_{1,2,1,1}+6\,H_{1,1,3}-\frac{3}{2}\,H_{1,1,2,1}-\frac{7}{2}\,H_{1,1,1,2}-\frac{25}{2}\,H_{1,1,1,1,1}\nonumber\\
&-\frac{57}{2}\,\zeta_5-\frac{1}{2}\,\zeta_2\,\zeta_3-2\,\zeta_2\,H_3-\frac{3}{2}\,\zeta_3\,H_2-3\,\zeta_4\,H_1+40\,H_5\bigg)\Bigg]\nonumber\\
&+\frac{1}{\eps}\,\bigg(\frac{1}{2}\,H_{2,1}+H_{1,2}+\frac{1}{2}\,H_{1,1,1}-\frac{3\,\zeta_3}{2}+H_3\bigg)
-\frac{1}{2}\,\zeta_2\,H_{1,1}-H_{3,1}+\frac{5}{2}\,H_{2,2}\nonumber\\
&-\frac{1}{2}\,H_{2,1,1}+2\,H_{1,3}-2\,H_{1,2,1}+\frac{1}{2}\,H_{1,1,2}-\frac{3}{2}\,H_{1,1,1,1}-\frac{23\,\zeta_4}{8}-\frac{1}{2}\,\zeta_2\,H_2+3\,\zeta_3\,H_1\nonumber\\
&+4\,H_4
+\eps\,\bigg(\frac{3}{2}\,\zeta_2\,H_{2,1}+3\,\zeta_2\,H_{1,2}+\frac{5}{2}\,\zeta_2\,H_{1,1,1}-\frac{11}{2}\,\zeta_3\,H_{1,1}-4\,H_{4,1}+8\,H_{3,2}\nonumber\\
&+4\,H_{3,1,1}+8\,H_{2,3}-\frac{7}{2}\,H_{2,2,1}+\frac{1}{2}\,H_{2,1,2}+\frac{1}{2}\,H_{2,1,1,1}+8\,H_{1,4}-2\,H_{1,3,1}\nonumber\\
&+3\,H_{1,2,2}+5\,H_{1,2,1,1}+4\,H_{1,1,3}+\frac{1}{2}\,H_{1,1,2,1}-\frac{1}{2}\,H_{1,1,1,2}+\frac{7}{2}\,H_{1,1,1,1,1}-10\,\zeta_5\nonumber\\
&-\frac{9}{2}\,\zeta_2\,\zeta_3+2\,\zeta_2\,H_3+\frac{1}{2}\,\zeta_3\,H_2+\frac{23}{4}\,\zeta_4\,H_1+16\,H_5\bigg)+\ord(\eps^2)\,,\nonumber\\
U_{qq}^{(2,1)}(z,\eps) &=N^2\,\Bigg[-\frac{1}{\eps^4}+\frac{3}{\eps^3}\,H_1-\frac{1}{\eps^2}\,\big(5\,H_{1,1}+\zeta_2\big)+\frac{1}{\eps}\,\big(2\,H_{2,1}+2\,H_{1,2}+9\,H_{1,1,1}+6\,\zeta_3\nonumber\\
&+3\,\zeta_2\,H_1+2\,H_3\big)
-5\,\zeta_2\,H_{1,1}-4\,H_{3,1}-2\,H_{2,2}-6\,H_{2,1,1}-2\,H_{1,3}\\
&-6\,H_{1,2,1}-4\,H_{1,1,2}-17\,H_{1,1,1,1}+\frac{29\,\zeta_4}{4}-18\,\zeta_3\,H_1
+\eps\,\big(2\,\zeta_2\,H_{2,1}+2\,\zeta_2\,H_{1,2}\nonumber\\
&+9\,\zeta_2\,H_{1,1,1}+30\,\zeta_3\,H_{1,1}+2\,H_{4,1}+2\,H_{3,2}+8\,H_{3,1,1}+2\,H_{2,3}+6\,H_{2,2,1}\nonumber\\
&+4\,H_{2,1,2}+14\,H_{2,1,1,1}+2\,H_{1,4}+6\,H_{1,3,1}+4\,H_{1,2,2}+14\,H_{1,2,1,1}+4\,H_{1,1,3}\nonumber\\
&+12\,H_{1,1,2,1}+8\,H_{1,1,1,2}+33\,H_{1,1,1,1,1}+42\,\zeta_5+6\,\zeta_2\,\zeta_3+2\,\zeta_2\,H_3-\frac{87}{4}\,\zeta_4\,H_1+2\,H_5\big)\nonumber
\end{align}
\begin{align}
&-\frac{1}{\eps^4}+\frac{1}{\eps^3}\,H_1-\frac{1}{\eps^2}\,\big(H_{1,1}+\zeta_2\big)+\frac{1}{\eps}\,\big(H_{1,1,1}+6\,\zeta_3+\zeta_2\,H_1\big)
-\zeta_2\,H_{1,1}-H_{1,1,1,1}\nonumber\\
&+\frac{29}{4}\,\zeta_4-6\,\zeta_3\,H_1
+\eps\big(\zeta_2\,H_{1,1,1}+6\,\zeta_3\,H_{1,1}+H_{1,1,1,1,1}+42\,\zeta_5+6\,\zeta_2\,\zeta_3\nonumber\\
&-\frac{29}{4}\,\zeta_4\,H_1\big)+\ord(\eps^2)\,,\nonumber\\
%
%
U_{qq}^{(2,0)}(z,\eps) &=N^2\,\Bigg[\frac{1}{2\eps^4}-\frac{2}{\eps^3}\,H_1+\frac{1}{\eps^2}\,\bigg(4\,H_{1,1}-\frac{1}{2}\,\zeta_2\bigg)+\frac{1}{\eps}\,\bigg(-\frac{5}{2}\,H_{2,1}-3\,H_{1,2}-\frac{5}{2}\,H_{1,1,1}\nonumber\\
&-2\,\zeta_3+2\,\zeta_2\,H_1-3\,H_3\bigg)
+\frac{5}{2}\,\zeta_2\,H_{1,1}+9\,H_{3,1}-\frac{1}{2}\,H_{2,2}+\frac{25}{2}\,H_{2,1,1}\\
&+3\,H_{1,3}+14\,H_{1,2,1}+\frac{11}{2}\,H_{1,1,2}+\frac{35}{2}\,H_{1,1,1,1}+\frac{43\,\zeta_4}{8}+\frac{5}{2}\,\zeta_2\,H_2+11\,\zeta_3\,H_1\nonumber\\
&-3\,H_4
+\eps\,\big(\frac{9}{2}\,\zeta_2\,H_{2,1}-\zeta_2\,H_{1,2}+\frac{3}{2}\,\zeta_2\,H_{1,1,1}-\frac{7}{2}\,\zeta_3\,H_{1,1}-3\,H_{3,2}-3\,H_{3,1,1}\nonumber\\
&-4\,H_{2,3}-\frac{3}{2}\,H_{2,2,1}-\frac{1}{2}\,H_{2,1,2}-\frac{5}{2}\,H_{2,1,1,1}-7\,H_{1,4}-6\,H_{1,3,1}-6\,H_{1,2,1,1}\nonumber\\
&-6\,H_{1,1,3}-\frac{15}{2}\,H_{1,1,2,1}+\frac{5}{2}\,H_{1,1,1,2}-\frac{11}{2}\,H_{1,1,1,1,1}+34\,\zeta_5+\frac{1}{2}\,\zeta_2\,\zeta_3+2\,\zeta_2\,H_3\nonumber\\
&-\frac{7}{2}\,\zeta_3\,H_2+\frac{11}{4}\,\zeta_4\,H_1-5\,H_5\big)\nonumber\\
&+\frac{1}{\eps^4}-\frac{2}{\eps^3}\,H_1+\frac{1}{2\eps^2}\,\zeta_2+\frac{1}{\eps}\,\bigg(-\frac{5}{2}\,H_{2,1}-3\,H_{1,2}-\frac{5}{2}\,H_{1,1,1}-5\,\zeta_3-3\,H_3\bigg)\nonumber\\
&+\frac{3}{2}\,\zeta_2\,H_{1,1}+3\,H_{3,1}-\frac{3}{2}\,H_{2,2}+\frac{3}{2}\,H_{2,1,1}+H_{1,3}+3\,H_{1,2,1}-\frac{3}{2}\,H_{1,1,2}+\frac{3}{2}\,H_{1,1,1,1}\nonumber\\
&-\frac{1}{8}\,\zeta_4+\frac{3}{2}\,\zeta_2\,H_2+H_4
+\eps\,\bigg(\frac{9}{2}\,\zeta_2\,H_{2,1}+2\,\zeta_2\,H_{1,2}+\frac{9}{2}\,\zeta_2\,H_{1,1,1}-\frac{3}{2}\,\zeta_3\,H_{1,1}\nonumber\\
&-12\,H_{4,1}-11\,H_{3,2}-10\,H_{3,1,1}-12\,H_{2,3}-\frac{19}{2}\,H_{2,2,1}-\frac{17}{2}\,H_{2,1,2}-\frac{11}{2}\,H_{2,1,1,1}\nonumber\\
&-13\,H_{1,4}-12\,H_{1,3,1}-11\,H_{1,2,2}-10\,H_{1,2,1,1}-12\,H_{1,1,3}-\frac{19}{2}\,H_{1,1,2,1}\nonumber\\
&-\frac{17}{2}\,H_{1,1,1,2}-\frac{11}{2}\,H_{1,1,1,1,1}+31\,\zeta_5+\frac{9}{2}\,\zeta_2\,\zeta_3+2\,\zeta_2\,H_3-\frac{3}{2}\,\zeta_3\,H_2-13\,H_5\bigg)\nonumber\\
&+\frac{1}{N^2}\,\Bigg[\frac{1}{2\eps^4}-\frac{3}{\eps}\,\zeta_3\,-\frac{11}{2}\,\zeta_4+\eps\,\big(4\,\zeta_2\,\zeta_3-3\,\zeta_5\big)\Bigg]+\ord(\eps^2)\,.\nonumber
\end{align}
\enlargethispage{1cm}
The functions $V_{ij}^{(2,k)}(z)$ are not of maximal weight, and are given by
\begin{align}
V_{gg}^{(2,2)}(z,\eps) &=N^3\,\Bigg[\frac{1}{\eps}\bigg(\frac{12z^3-33z^2+32z-22}{6(1-z)z}\,H_2-\frac{1}{3}(1-z)\,H_1\\
&-\frac{44z^4-76z^3+99z^2-56z+22}{6(1-z)z}\,\zeta_2-\frac{404z^4-799z^3+1185z^2-772z+386}{54(1-z)z}\bigg)\nonumber\\
&+\frac{(z+1)\left(11z^2-10z+11\right)}{3z}\,H_{2,1}-\frac{44z^3-100z^2+105z-60}{6(1-z)}\,H_{1,2}\nonumber\\
&+\frac{22z^3-34z^2+35z-11}{6z}\,H_{1,1,1}+\frac{5-6z}{6}\,H_{1,1}\nonumber\\
&+\frac{72z^3-201z^2+196z-134}{18(1-z)z}\,H_2+\frac{88z^4-164z^3+231z^2-144z+66}{6(1-z)z}\,\zeta_2\,\,H_1\nonumber
\end{align}
\begin{align}
\phantom{V_{gg}^{(2,2)}(z,\eps)}
&+\frac{404z^4-823z^3+1224z^2-787z+386}{27(1-z)z}\,H_1-\frac{22z^4-56z^3+99z^2-76z+44}{3(1-z)z}\,H_3\nonumber\\
&-\frac{268z^4-464z^3+597z^2-334z+134}{18(1-z)z}\,\zeta_2+\frac{66z^4-144z^3+231z^2-164z+88}{3(1-z)z}\,\zeta_3\nonumber\\
&-\frac{2428z^4-4757z^3+7041z^2-4568z+2284}{162(1-z)z}\nonumber\\
&+\eps\bigg(-\frac{14576z^4-28387z^3+42027z^2-27280z+13640}{486(1-z)z}\nonumber\\
&+\frac{2428z^4-4913z^3+7308z^2-4733z+2338}{81(1-z)z}\,H_1\nonumber\\
&+\frac{405z^3-1221z^2+1292z-880}{54(1-z)z}\,H_2\nonumber\\
&-\frac{1616z^4-3322z^3+4941z^2-3163z+1544}{54(1-z)z}\,H_{1,1}\nonumber\\
&-\frac{134z^4-334z^3+600z^2-467z+268}{9(1-z)z}\,H_3+\frac{134z^3+16z^2+13z+134}{18z}\,H_{2,1}\nonumber\\
&-\frac{268z^3-572z^2+546z-309}{18(1-z)}\,H_{1,2}+\frac{134z^3-146z^2+157z-67}{18z}\,H_{1,1,1}\nonumber\\
&-4\,\frac{22z^4-56z^3+99z^2-76z+44}{3(1-z)z}\,H_4+\frac{22z^4-56z^3+99z^2-76z+44}{3(1-z)z}\,H_{3,1}\nonumber\\
&-\frac{110z^4-328z^3+447z^2-328z+110}{6(1-z)z}\,H_{2,2}\nonumber\\
&+\frac{22z^4+52z^3-108z^2+122z-55}{6(1-z)z}\,H_{2,1,1}\nonumber\\
&-\frac{44z^4-112z^3+108z^2-62z-11}{3(1-z)z}\,H_{1,3}-\frac{88z^3-100z^2+107z-11}{6z}\,H_{1,2,1}\nonumber\\
&-\frac{22z^4-68z^3+102z^2-78z+33}{6(1-z)z}\,H_{1,1,2}-\frac{22z^3-34z^2+35z-11}{2z}\,H_{1,1,1,1}\nonumber\\
&-\frac{1616z^4-2755z^3+3453z^2-1766z+664}{54(1-z)z}\,\zeta_2\nonumber\\
&+\frac{22z^4-8z^3-33z^2+52z-44}{6(1-z)z}\,\zeta_2\,H_2\nonumber\\
&+\frac{536z^4-1024z^3+1446z^2-891z+402}{18(1-z)z}\,\zeta_2\,H_1\nonumber\\
&-\frac{154z^4-284z^3+426z^2-274z+143}{6(1-z)z}\,\zeta_2\,H_{1,1}\nonumber\\
&+\frac{804z^4-1734z^3+2835z^2-2039z+1072}{18(1-z)z}\,\zeta_3\nonumber\\
&-\frac{264z^4-588z^3+867z^2-598z+275}{6(1-z)z}\,\zeta_3\,H_1\nonumber\\
&-\frac{110z^4+128z^3-627z^2+708z-528}{8(1-z)z}\,\zeta_4\bigg)\Bigg]\nonumber
\end{align}
\begin{align}
\phantom{V_{gg}^{(2,2)}(z,\eps)}
&+N\,N_f\,\Bigg[\frac{1}{\eps}\bigg(-\frac{3z^3-6z^2+5z-4}{6(1-z)z}\,H_2+\frac{1}{3}(1-z)\,H_1\nonumber\\
&+\frac{8z^4-13z^3+18z^2-11z+4}{6(1-z)z}\,\zeta_2+\frac{56z^4-103z^3+141z^2-76z+38}{54(1-z)z}\bigg)\nonumber\\
&-\frac{(z+1)\left(2z^2-z+2\right)}{3z}\,H_{2,1}+\frac{8z^3-19z^2+24z-15}{6(1-z)}\,H_{1,2}\nonumber\\
&-\frac{4z^3-7z^2+8z-2}{6z}\,H_{1,1,1}+\frac{6z-5}{6}\,H_{1,1}-\frac{9z^3-15z^2+11z-10}{9(1-z)z}\,H_2\nonumber\\
&-\frac{16z^4-29z^3+42z^2-27z+12}{6(1-z)z}\,\zeta_2\,H_1-2\frac{28z^4-68z^3+99z^2-50z+19}{27(1-z)z}\,H_1\nonumber\\
&+\frac{4z^4-11z^3+18z^2-13z+8}{3(1-z)z}\,H_3+\frac{20z^4-31z^3+42z^2-26z+10}{9(1-z)z}\,\zeta_2\nonumber\\
&-\frac{12z^4-27z^3+42z^2-29z+16}{3(1-z)z}\,\zeta_3+\frac{164z^4-265z^3+330z^2-130z+65}{81(1-z)z}\nonumber\\
&+\eps\,\bigg(\frac{1952z^4-2761z^3+3183z^2-844z+422}{486(1-z)z}\nonumber\\
&-\frac{328z^4-785z^3+1098z^2-497z+184}{81(1-z)z}\,H_1-\frac{81z^3-177z^2+224z-184}{54(1-z)z}\,H_2\nonumber\\
&+4\frac{28z^4-74z^3+108z^2-53z+19}{27(1-z)z}\,H_{1,1}+\frac{20z^4-52z^3+87z^2-65z+40}{9(1-z)z}\,H_3\nonumber\\
&-\frac{20z^3+22z^2+19z+20}{18z}\,H_{2,1}+\frac{40z^3-62z^2+51z-39}{18(1-z)}\,H_{1,2}\nonumber\\
&-\frac{(z+1)\left(10z^2+z-5\right)}{9z}\,H_{1,1,1}+4\frac{4z^4-11z^3+18z^2-13z+8}{3(1-z)z}\,H_4\nonumber\\
&-\frac{4z^4-11z^3+18z^2-13z+8}{3(1-z)z}\,H_{3,1}+\frac{\left(4z^2-7z+4\right)\left(5z^2-8z+5\right)}{6(1-z)z}\,H_{2,2}\nonumber\\
&-\frac{4z^4+16z^3-27z^2+23z-10}{6(1-z)z}\,H_{2,1,1}+\frac{8z^4-22z^3+27z^2-17z-2}{3(1-z)z}\,H_{1,3}\nonumber\\
&+\frac{16z^3-19z^2+26z-2}{6z}\,H_{1,2,1}+\frac{4z^4-14z^3+21z^2-15z+6}{6(1-z)z}\,H_{1,1,2}\nonumber\\
&+\frac{4z^3-7z^2+8z-2}{2z}\,H_{1,1,1,1}+\frac{224z^4-295z^3+303z^2-32z-32}{54(1-z)z}\,\zeta_2\nonumber\\
&-\frac{80z^4-166z^3+249z^2-153z+60}{18(1-z)z}\,\zeta_2\,H_1-\frac{4z^4+z^3-6z^2+7z-8}{6(1-z)z}\,\zeta_2\,H_2\nonumber\\
&+\frac{28z^4-50z^3+75z^2-49z+26}{6(1-z)z}\,\zeta_2\,H_{1,1}\nonumber\\
&-\frac{120z^4-258z^3+441z^2-323z+160}{18(1-z)z}\,\zeta_3\nonumber\\
&+\frac{48z^4-111z^3+165z^2-112z+50}{6(1-z)z}\,\zeta_3\,H_1\nonumber\\
&+\frac{20z^4+47z^3-114z^2+105z-96}{8(1-z)z}\,\zeta_4\bigg)\Bigg]+\ord(\eps^2)\,,\nonumber
\end{align}
\begin{align}
V_{gg}^{(2,1)}(z,\eps) &=2\,N^2\,(N-N_f)\,\eps\,\zeta_3+\ord(\eps^2)\,,\\
V_{gg}^{(2,1)}(z,\eps) &= N^3\,\Bigg[\frac{1}{\eps}\bigg(-\frac{12z^3-33z^2+32z-22}{6(1-z)z}\,\,H_2\\
&-\frac{134z^4-271z^3+402z^2-265z+134}{9(1-z)z}\,\,H_1+\frac{22z^3-32z^2+33z-12}{6(1-z)}\,\zeta_2\nonumber\\
&+\frac{6z^3-9z^2-13z+22}{18(1-z)}\bigg)
-\frac{44z^4-76z^3+139z^2-96z+66}{2(1-z)z}\,H_{1,2}\nonumber\\
&-\frac{77z^4-166z^3+264z^2-186z+99}{3(1-z)z}\,H_{2,1}\nonumber\\
&-\frac{154z^4-296z^3+459z^2-306z+165}{6(1-z)z}\,H_{1,1,1}-\frac{5-6z}{6}\,H_{1,1}\nonumber\\
&+\frac{12z^3+7z^2-8z+22}{2(1-z)z}\,\zeta_2\,H_1-\frac{96z^3-237z^2+208z-134}{18(1-z)z}\,H_2\nonumber\\
&-\frac{66z^4-144z^3+231z^2-164z+88}{3(1-z)z}\,H_3\nonumber\\
&-\frac{2\left(386z^4-784z^3+1167z^2-769z+395\right)}{27(1-z)z}\,H_1+\frac{134z^3-184z^2+195z-78}{18(1-z)}\,\zeta_2\nonumber\\
&-\frac{22z^3-32z^2+33z-12}{3(1-z)}\,\zeta_3-\frac{84z^3-201z^2+715z-646}{54(1-z)}\nonumber\\
&+\eps\,\bigg(-\frac{1188z^3-2461z^2+4884z-3715}{54(1-z)}-\frac{18z^3-31z^2-35z+24}{18(1-z)z}\,H_{1,1}\nonumber\\
&+\frac{22z^3-34z^2+35z-11}{2z}\,H_{1,1,1,1}-\frac{22z^3-32z^2+33z-12}{1-z}\,H_{3,1}\nonumber\\
&-\frac{525z^3-1401z^2+1208z-808}{54(1-z)z}\,H_2+\frac{22z^4-80z^3+125z^2-100z+44}{2(1-z)z}\,H_{2,2}\nonumber\\
&-\frac{22z^4-20z^3+10z^2+10z-11}{2(1-z)z}\,H_{2,1,1}-\frac{22z^4-20z^3+30z^2-10z+11}{3(1-z)z}\,H_{1,3}\nonumber\\
&-\frac{22z^4-8z^3-33z^2+52z-44}{3(1-z)z}\,H_4+\frac{66z^4-180z^3+240z^2-170z+55}{6(1-z)z}\,H_{1,1,2}\nonumber\\
&-\frac{132z^4-252z^3+333z^2-202z+77}{6(1-z)z}\,H_{1,2,1}\nonumber\\
&-\frac{268z^4-480z^3+890z^2-611z+402}{6(1-z)z}\,H_{1,2}\nonumber\\
&-\frac{402z^4-876z^3+1407z^2-1000z+536}{9(1-z)z}\,H_3\nonumber\\
&-\frac{938z^4-2038z^3+3249z^2-2283z+1206}{18(1-z)z}\,H_{2,1}\nonumber\\
&-\frac{938z^4-1864z^3+2901z^2-1908z+1005}{18(1-z)z}\,H_{1,1,1}\nonumber\\
&-2\,\frac{2230z^4-4538z^3+6870z^2-4589z+2383}{81(1-z)z}\,H_1\nonumber\\
&-\frac{22z^3-56z^2+59z-36}{2(1-z)}\,\zeta_2\,H_2+\frac{76z^3+52z^2-61z+134}{6(1-z)z}\,\zeta_2\,H_1\nonumber
\end{align}
\begin{align}
\phantom{V_{gg}^{(2,2)}(z,\eps)}
&-\frac{66z^4-228z^3+252z^2-178z+11}{6(1-z)z}\,\zeta_2\,H_{1,1}\nonumber\\
&+\frac{264z^4-540z^3+775z^2-510z+231}{2(1-z)z}\,\zeta_3\,H_1\nonumber\\
&+\frac{952z^3-1367z^2+1605z-714}{54(1-z)}\,\zeta_2-\frac{268z^3-386z^2+417z-165}{18(1-z)}\,\zeta_3\nonumber\\
&+53\frac{22z^3-32z^2+33z-12}{24(1-z)}\,\zeta_4\bigg)\Bigg]\nonumber\\
&+N^2\,N_f\,\Bigg[\frac{1}{\eps}\,\bigg(\frac{3z^3-6z^2+5z-4}{6(1-z)z}\,H_2+\frac{20z^4-43z^3+60z^2-37z+20}{9(1-z)z}\,H_1\nonumber\\
&-\frac{4z^3-5z^2+6z-3}{6(1-z)}\,\zeta_2-\frac{6z^3-9z^2-17z+26}{18(1-z)}\bigg)\nonumber\\
&+\frac{\left(2z^2-3z+3\right)\left(7z^2-5z+6\right)}{3(1-z)z}\,H_{2,1}+\frac{8z^4-13z^3+22z^2-15z+12}{2(1-z)z}\,H_{1,2}\nonumber\\
&\,+\frac{28z^4-53z^3+81z^2-54z+30}{6(1-z)z}\,H_{1,1,1}+\frac{5-6z}{6}\,H_{1,1}-\frac{3z^3-2z^2+z+4}{2(1-z)z}\,\zeta_2\,H_1\nonumber\\
&+\frac{21z^3-33z^2+17z-10}{9(1-z)z}\,H_2+\frac{12z^4-27z^3+42z^2-29z+16}{3(1-z)z}\,H_3\nonumber\\
&+\frac{76z^4-185z^3+246z^2-137z+94}{27(1-z)z}\,H_1-\frac{10z^3-5z^2+12z-12}{9(1-z)}\,\zeta_2\nonumber\\
&+\frac{4z^3-5z^2+6z-3}{3(1-z)}\,\zeta_3+\frac{90z^3-222z^2+869z-803}{54(1-z)}\nonumber\\
&+\eps\,\bigg(\frac{2872z^3-5998z^2+12421z-9635}{108(1-z)}-\frac{4z^3-7z^2+8z-2}{2z}\,H_{1,1,1,1}\nonumber\\
&+\frac{4z^3-5z^2+6z-3}{1-z}\,H_{3,1}+2\frac{9z^3-16z^2-5z+6}{9(1-z)z}\,H_{1,1}+\frac{273z^3-465z^2+176z-112}{54(1-z)z}\,H_2\nonumber\\
&-\frac{4z^4-17z^3+26z^2-19z+8}{2(1-z)z}\,H_{2,2}+\frac{4z^4-2z^3+z^2+z-2}{2(1-z)z}\,H_{2,1,1}\nonumber\\
&+\frac{4z^4-2z^3+3z^2-z+2}{3(1-z)z}\,H_{1,3}+\frac{4z^4+z^3-6z^2+7z-8}{3(1-z)z}\,H_4\nonumber\\
&-\frac{12z^4-36z^3+51z^2-35z+10}{6(1-z)z}\,H_{1,1,2}+\frac{24z^4-45z^3+63z^2-40z+14}{6(1-z)z}\,H_{1,2,1}\nonumber\\
&+\frac{2\left(30z^4-69z^3+105z^2-71z+40\right)}{9(1-z)z}\,H_3+\frac{40z^4-78z^3+137z^2-89z+60}{6(1-z)z}\,H_{1,2}\nonumber\\
&+\frac{70z^4-161z^3+249z^2-153z+75}{9(1-z)z}\,H_{1,1,1}+\frac{140z^4-334z^3+513z^2-339z+180}{18(1-z)z}\,H_{2,1}\nonumber\\
&+\frac{152z^4-559z^3+870z^2-517z+512}{81(1-z)z}\,H_1+\frac{\left(4z^3-11z^2+14z-9\right)\zeta_2}{2(1-z)}\,H_2\nonumber\\
&-\frac{22z^3-17z^2+5z+20}{6(1-z)z}\,\zeta_2\,H_1+\frac{12z^4-48z^3+63z^2-43z+2}{6(1-z)z}\,\zeta_2\,H_{1,1}\nonumber\\
&-\frac{48z^4-99z^3+145z^2-96z+42}{2(1-z)z}\,\zeta_3\,H_1-\frac{256z^3-263z^2+597z-462}{54(1-z)}\,\zeta_2\nonumber\\
&+\frac{40z^3-38z^2+111z-93}{18(1-z)}\,\zeta_3-53\,\frac{4z^3-5z^2+6z-3}{24(1-z)}\,\zeta_4\bigg)\Bigg]\nonumber
\end{align}
\begin{align}
\phantom{V_{gg}^{(2,2)}(z,\eps)}
&+N\,N_f^2\,\Bigg[\frac{2}{9\eps}+\frac{2}{27}\left(6z^2-6z+19\right)+\eps\,\bigg(\frac{2}{27}\left(38z^2-38z+75\right)\bigg)\Bigg]\nonumber\\
&+N_f\,\Bigg[-\frac{1}{2}+\eps\,\bigg(\zeta_2-2\zeta_3+\frac{1}{12}\left(-12z^2+12z-37\right)\bigg)\Bigg]+\ord(\eps^2)\,,\nonumber\\
V_{gq}^{(2,2)}(z,\eps) &= N^3\,\Bigg[\frac{1}{\eps}\bigg(\frac{13\left(z^2-2z+2\right)}{24z}\,H_{1,1}+\frac{13z^2-20z+8}{24z}\,H_2+\frac{z-2}{8z}\,H_1\\
&-\frac{31z^2-56z+44}{24z}\,\zeta_2-\frac{202z^2-395z+386}{108z}\bigg)
+\frac{4z^2-11z+26}{12z}\,H_{1,2}\nonumber\\
&-\frac{11z^2-16z+4}{12z}\,H_{2,1}-\frac{16z^2-32z+29}{8z}\,H_{1,1,1}+\frac{32z^2-151z+178}{72z}\,H_{1,1}\nonumber\\
&+\frac{31z^2-59z+53}{12z}\,\zeta_2\,H_1+\frac{16z^2-23z-10}{36z}\,H_2+\frac{17z^2-28z+16}{12z}\,H_3\nonumber\\
&+\frac{101z^2-193z+184}{27z}\,H_1-\frac{52z^2-167z+134}{36z}\,\zeta_2+\frac{79z^2-164z+176}{12z}\,\zeta_3\nonumber\\
&-\frac{152z^2-589z+571}{81z}
+\eps\,\bigg(-\frac{89z^2-172z+169}{24z}\,\zeta_2\,H_{1,1}-\frac{4z^2+46z-115}{36z}\,H_{1,2}\nonumber\\
&+\frac{10z^2-8z-43}{24z}\,H_{1,2,1}-\frac{17z^2-28z+16}{12z}\,H_{3,1}-\frac{19z^2-56z+38}{24z}\,H_{2,2}\nonumber\\
&+\frac{22z^2-14z-19}{24z}\,H_{2,1,1}+\frac{34z^2-74z+113}{12z}\,H_{1,3}-\frac{35z^2-76z+79}{24z}\,H_{1,1,2}\nonumber\\
&-\frac{44z^2-184z+193}{24z}\,H_{1,1,1}-\frac{68z^2-127z-20}{72z}\,H_{2,1}+\frac{118z^2-236z+209}{24z}\,H_{1,1,1,1}\nonumber\\
&-\frac{352z^2-548z+503}{54z}\,H_{1,1}+\frac{35z^2-52z+16}{24z}\,\zeta_2\,H_2+\frac{104z^2-355z+304}{36z}\,\zeta_2\,H_1\nonumber\\
&-\frac{316z^2-644z+641}{24z}\,\zeta_3\,H_1+\frac{17z^2-28z+16}{3z}\,H_4+\frac{26z^2-52z-5}{27z}\,H_2\nonumber\\
&+\frac{28z^2-11z-40}{36z}\,H_3+\frac{608z^2-2311z+2194}{162z}\,H_1+3\frac{89z^2-236z+352}{32z}\,\zeta_4\nonumber\\
&-\frac{151z^2-464z+332}{54z}\,\zeta_2+\frac{472z^2-2057z+2144}{72z}\,\zeta_3-\frac{1823z^2-7054z+6820}{486z}\bigg)\bigg]\nonumber\\
&+N^2\,N_f\,\Bigg[\frac{1}{\eps}\,\bigg(-\frac{z^2-2z+2}{6z}\,H_{1,1}-\frac{z^2-2z+2}{6z}\,H_2+\frac{z^2-2z+2}{6z}\,\zeta_2\nonumber\\
&+\frac{28z^2-47z+38}{108z}\bigg)
-\frac{z^2-5z+5}{9z}\,H_{1,1}+\frac{z^2-2z+2}{6z}\,H_{2,1}-\frac{z^2-2z+2}{3z}\,H_{1,2}\nonumber\\
&+\frac{z^2-2z+2}{2z}\,H_{1,1,1}-\frac{z^2-2z+2}{3z}\,\zeta_2\,H_1-\frac{2\left(z^2-2z+2\right)}{3z}\,H_3\nonumber\\
&-\frac{4z^2-17z+20}{36z}\,H_2-\frac{56z^2-103z+94}{108z}\,H_1-\frac{4\left(z^2-2z+2\right)}{3z}\,\zeta_3\nonumber\\
&+\frac{4z^2-17z+20}{36z}\,\zeta_2+\frac{80z^2-229z+130}{324z}
+\eps\,\bigg(\frac{z^2-2z+2}{3z}\,\zeta_2\,H_{1,1}\nonumber\\
&+\frac{z^2-5z+5}{3z}\,H_{1,1,1}-\frac{z^2-2z+2}{6z}\,H_{2,1,1}-\frac{z^2-2z+2}{3z}\,H_{2,2}\nonumber\\
&+\frac{z^2-2z+2}{3z}\,H_{1,2,1}+\frac{z^2-2z+2}{3z}\,H_{1,1,2}+\frac{2\left(z^2-2z+2\right)}{3z}\,H_{3,1}\nonumber
\end{align}
\begin{align}
\phantom{V_{gg}^{(2,2)}(z,\eps)}
&-\frac{7\left(z^2-2z+2\right)}{6z}\,H_{1,1,1,1}-\frac{4\left(z^2-2z+2\right)}{3z}\,H_{1,3}+\frac{4z^2-17z+20}{36z}\,H_{2,1}\nonumber\\
&-\frac{8z^2-43z+40}{36z}\,H_{1,2}+\frac{77z^2-49z+40}{108z}\,H_{1,1}-\frac{z^2-2z+2}{3z}\,\zeta_2\,H_2\nonumber\\
&+\frac{8\left(z^2-2z+2\right)}{3z}\,\zeta_3\,H_1-\frac{8z^2-37z+40}{36z}\,\zeta_2\,H_1-\frac{8\left(z^2-2z+2\right)}{3z}\,H_4\nonumber\\
&-\frac{4z^2-17z+20}{9z}\,H_3-\frac{35z^2-151z+148}{108z}\,H_2-\frac{160z^2-557z+458}{324z}\,H_1\nonumber\\
&-\frac{3\left(z^2-2z+2\right)}{z}\,\zeta_4+\frac{17z^2-7z-32}{108z}\,\zeta_2-\frac{32z^2-175z+160}{36z}\,\zeta_3\nonumber\\
&+\frac{484z^2-1187z+422}{972z}\bigg)\Bigg]\nonumber\\
&+N\,\Bigg[\frac{1}{\eps}\,\bigg(\frac{z^2-8z+5}{12z}\,H_{1,1}+\frac{z^2-11z+14}{12z}\,H_2+\frac{1}{4z}\,H_1+\frac{14z^2-31z+22}{12z}\,\zeta_2\nonumber\\
&+\frac{485z^2-952z+772}{216z}\bigg)
+\frac{7z^2-14z+11}{8z}\,H_{1,1,1}+\frac{11z^2-16z+4}{12z}\,H_{2,1}\nonumber\\
&+\frac{31z^2-80z+38}{24z}\,H_{1,2}+\frac{67z^2-254z+155}{72z}\,H_{1,1}-\frac{53z^2-100z+88}{24z}\,\zeta_2\,H_1\nonumber\\
&+\frac{8z^2-25z+28}{6z}\,H_3+\frac{67z^2-350z+308}{72z}\,H_2-\frac{485z^2-961z+736}{108z}\,H_1\nonumber\\
&-\frac{85z^2-152z+176}{12z}\,\zeta_3+\frac{113z^2-424z+268}{72z}\,\zeta_2+\frac{1337z^2-3814z+2284}{324z}\nonumber\\
&
+\eps\,\bigg(\frac{140z^2-358z+313}{24z}\,\zeta_2\,H_{1,1}+\frac{7\left(5z^2-16z+7\right)}{12z}\,H_{1,3}\nonumber\\
&+\frac{11z^2-100z+109}{24z}\,H_{2,1,1}-\frac{13z^2-20z+38}{12z}\,H_{3,1}-\frac{31z^2-14z+11}{24z}\,H_{1,1,1,1}\nonumber\\
&+\frac{43z^2-146z+143}{48z}\,H_{1,1,1}+\frac{44z^2-202z+151}{24z}\,H_{1,1,2}+\frac{50z^2-163z+127}{12z}\,H_{2,2}\nonumber\\
&-\frac{70z^2-104z+65}{24z}\,H_{1,2,1}+\frac{181z^2-146z+50}{144z}\,H_{2,1}+\frac{251z^2-781z+670}{72z}\,H_{1,2}\nonumber\\
&+\frac{4679z^2-12079z+9424}{432z}\,H_{1,1}+\frac{17z^2-85z+82}{12z}\,\zeta_2\,H_2\nonumber\\
&-\frac{118z^2-503z+329}{72z}\,\zeta_2\,H_1+\frac{352z^2-716z+713}{24z}\,\zeta_3\,H_1+\frac{2\left(8z^2-25z+28\right)}{3z}\,H_4\nonumber\\
&+\frac{277z^2-1310z+1232}{72z}\,H_3+\frac{799z^2-5324z+3536}{432z}\,H_2-\frac{1337z^2-3850z+2194}{162z}\,H_1\nonumber\\
&-\frac{188z^2-245z+528}{16z}\,\zeta_4-\frac{917z^2-3466z+4288}{144z}\,\zeta_3+\frac{2207z^2-7330z+2656}{432z}\,\zeta_2\nonumber\\
&+\frac{3463z^2-8630z+3410}{243z}\bigg)\nonumber\\
&+N_f\,\Bigg[\frac{1}{\eps}\,\bigg(\frac{z^2-2z+2}{6z}\,H_{1,1}+\frac{z^2-2z+2}{6z}\,H_2-\frac{z^2-2z+2}{6z}\,\zeta_2-\frac{28z^2-47z+38}{108z}\bigg)\nonumber\\
&
+\frac{z^2-5z+5}{9z}\,H_{1,1}-\frac{z^2-2z+2}{6z}\,H_{2,1}+\frac{z^2-2z+2}{3z}\,H_{1,2}-\frac{z^2-2z+2}{2z}\,H_{1,1,1}\nonumber
\end{align}
\begin{align}
\phantom{V_{gg}^{(2,2)}(z,\eps)}
&+\frac{z^2-2z+2}{3z}\,\zeta_2\,H_1+\frac{2\left(z^2-2z+2\right)}{3z}\,H_3+\frac{4z^2-17z+20}{36z}\,H_2\nonumber\\
&+\frac{56z^2-103z+94}{108z}\,H_1+\frac{4\left(z^2-2z+2\right)}{3z}\,\zeta_3-\frac{4z^2-17z+20}{36z}\,\zeta_2-\frac{80z^2-229z+130}{324z}\nonumber\\
&
+\eps\,\bigg(-\frac{z^2-2z+2}{3z}\,\zeta_2\,H_{1,1}-\frac{z^2-5z+5}{3z}\,H_{1,1,1}+\frac{z^2-2z+2}{6z}\,H_{2,1,1}\nonumber\\
&+\frac{z^2-2z+2}{3z}\,H_{2,2}-\frac{z^2-2z+2}{3z}\,H_{1,2,1}-\frac{z^2-2z+2}{3z}\,H_{1,1,2}-\frac{2\left(z^2-2z+2\right)}{3z}\,H_{3,1}\nonumber\\
&+\frac{7\left(z^2-2z+2\right)}{6z}\,H_{1,1,1,1}+\frac{4\left(z^2-2z+2\right)}{3z}\,H_{1,3}-\frac{4z^2-17z+20}{36z}\,H_{2,1}\nonumber\\
&+\frac{8z^2-43z+40}{36z}\,H_{1,2}-\frac{77z^2-49z+40}{108z}\,H_{1,1}+\frac{z^2-2z+2}{3z}\,\zeta_2\,H_2\nonumber\\
&-\frac{8\left(z^2-2z+2\right)}{3z}\,\zeta_3\,H_1+\frac{8z^2-37z+40}{36z}\,\zeta_2\,H_1+\frac{8\left(z^2-2z+2\right)}{3z}\,H_4\nonumber\\
&+\frac{4z^2-17z+20}{9z}\,H_3+\frac{35z^2-151z+148}{108z}\,H_2+\frac{160z^2-557z+458}{324z}\,H_1\nonumber\\
&+\frac{3\left(z^2-2z+2\right)}{z}\,\zeta_4-\frac{17z^2-7z-32}{108z}\,\zeta_2+\frac{32z^2-175z+160}{36z}\,\zeta_3\nonumber\\
&-\frac{484z^2-1187z+422}{972z}\bigg)\Bigg]\nonumber\\
&+\frac{1}{N}\,\Bigg[\frac{1}{\eps}\,\bigg(-\frac{3\left(z^2-2z+2\right)}{8z}\,H_{1,1}-\frac{3z^2-10z+12}{8z}\,H_2-\frac{z-2}{8z}\,H_1\nonumber\\
&+\frac{2-3z}{8}\,\zeta_2\bigg)
+\frac{3\left(2z^2-4z+5\right)}{8z}\,H_{1,1,1}-\frac{3z^2-5z+3}{2z}\,H_{1,2}\nonumber\\
&-\frac{4z^2-15z+18}{8z}\,H_{1,1}-\frac{2z-3}{2z}\,H_{2,1}+\frac{3z^2-4z+3}{4z}\,\zeta_2\,H_1-\frac{2z^2-9z+12}{4z}\,H_2\nonumber\\
&-\frac{9z^2-22z+24}{4z}\,H_3-\frac{z-2}{4z}\,H_1+\frac{2-3z}{4}\,\zeta_3+\frac{3-2z}{4}\,\zeta_2\nonumber\\
&
+\eps\,\bigg(-\frac{15z^2-26z+27}{8z}\,\zeta_2\,H_{1,1}+\frac{3z^2-2z+3}{8z}\,H_{1,1,2}\nonumber\\
&-\frac{3\left(4z^2-8z+11\right)}{8z}\,H_{1,1,1,1}-\frac{3\left(7z^2-18z+14\right)}{8z}\,H_{2,2}+\frac{8z^2-34z+45}{8z}\,H_{1,1,1}\nonumber\\
&-\frac{8z^2-18z+11}{4z}\,H_{1,2}+\frac{9z^2-22z+24}{4z}\,H_{3,1}-\frac{18z^2-32z+21}{4z}\,H_{1,3}\nonumber\\
&+\frac{18z^2-28z+15}{8z}\,H_{1,2,1}-\frac{3(z-3)(z-2)}{4z}\,H_{1,1}-\frac{13z-24}{8z}\,H_{2,1}\nonumber\\
&+\frac{20z-21}{8z}\,H_{2,1,1}+\frac{4z^2-11z+10}{4z}\,\zeta_2\,H_1+\frac{12z^2-16z+21}{8z}\,\zeta_3\,H_1\nonumber\\
&-\frac{9z^2-22z+24}{z}\,H_4-\frac{12z^2-41z+48}{4z}\,H_3-\frac{3(z-4)(z-2)}{8z}\,\zeta_2\,H_2-\frac{z-2}{2z}\,H_1\nonumber\\
&-\frac{3(z-4)(z-2)}{4z}\,H_2+\frac{15-8z}{8}\,\zeta_3-\frac{3(z-2)}{4}\,\zeta_2-\frac{87}{32}(3z-2)\,\zeta_4\bigg)\Bigg]\nonumber
\end{align}
\begin{align}
\phantom{V_{gg}^{(2,2)}(z,\eps)}
&+\frac{1}{N^3}\,\Bigg[\frac{1}{\eps}\,\bigg(\frac{(1-z)(z-3)}{4z}\,H_{1,1}+\frac{2-z}{4}\,H_2-\frac{1}{4z}\,H_1+\frac{z}{2}\,\zeta_2-\frac{3}{8}(z-2)\bigg)\nonumber\\
&
-\frac{z^2-14z+18}{8z}\,H_{1,2}-\frac{7z^2-30z+19}{8z}\,H_{1,1}+\frac{3(1-z)^2}{8z}\,H_{1,1,1}+\frac{2z-3}{2z}\,H_{2,1}\nonumber\\
&-\frac{9z^2-14z+12}{8z}\,\zeta_2\,H_1+\frac{3z^2-6z-2}{4z}\,H_1-\frac{7z^2-26z+8}{8z}\,H_2+\frac{2-z}{2}\,H_3\nonumber\\
&+\frac{3z+4}{8}\,\zeta_2+\frac{5z+2}{4}\,\zeta_3-\frac{9}{4}(z-2)
+\eps\,\bigg(-\frac{2z^2-36z+21}{8z}\,\zeta_2\,H_{1,1}\nonumber\\
&+\frac{z^2+6z-6}{4z}\,H_{3,1}+\frac{2z^2-4z+21}{8z}\,H_{1,2,1}-\frac{5z^2-30z+33}{4z}\,H_{1,3}\nonumber\\
&-\frac{5z^2-14z+58}{16z}\,H_{2,1}-\frac{6z^2-44z+27}{8z}\,H_{1,1,2}-\frac{11z^2-18z+9}{8z}\,H_{2,1,1}\nonumber\\
&-\frac{57z^2-225z+128}{16z}\,H_{1,1}+\frac{3(1-z)(z-5)}{4z}\,H_{2,2}-\frac{(z+3)^2}{16z}\,H_{1,1,1}\nonumber\\
&-\frac{(z-2)(11z-39)}{8z}\,H_{1,2}+\frac{(1-z)(17z-33)}{8z}\,H_{1,1,1,1}-\frac{3\left(6z^2-15z+17\right)}{8z}\,\zeta_2\,H_1\nonumber\\
&-\frac{24z^2-40z+45}{8z}\,\zeta_3\,H_1+\frac{9z^2-18z-2}{2z}\,H_1-\frac{13z^2-66z+32}{8z}\,H_3\nonumber\\
&-\frac{33z^2-156z+32}{16z}\,H_2+\frac{(1-z)(5z-9)}{2z}\,\zeta_2\,H_2-2(z-2)\,H_4-\frac{5}{16}(5z-22)\,\zeta_2\nonumber\\
&+\left(\frac{1}{16}(13z+42)\right)\,\zeta_3+\left(\frac{1}{16}(185z+22)\right)\,\zeta_4-\frac{21}{2}(z-2)\bigg)\Bigg]+\ord(\eps^2)\,,\nonumber\\
V_{gq}^{(2,1)}(z,\eps) & = 3\eps\,\frac{N^2-1}{2N}\,\zeta_3+\ord(\eps^2)\,,\\
V_{gq}^{(2,0)}(z,\eps) & = N^3\,\Bigg[\frac{1}{\eps}\,\bigg(\frac{31\left(z^2-2z+2\right)}{24z}\,H_{1,1}+\frac{31z^2-68z+80}{24z}\,H_2+\frac{3z-4}{8}\,\zeta_2\\
&-\frac{134z^2-241z+250}{72z}\,H_1+\frac{z+10}{12}\bigg)
-\frac{10z^2-23z+29}{6z}\,H_{2,1}\nonumber\\
&-\frac{35z^2-73z+88}{12z}\,H_{1,2}-\frac{40z^2-80z+89}{24z}\,H_{1,1,1}+\frac{104z^2-367z+358}{72z}\,H_{1,1}\nonumber\\
&-\frac{35z^2-76z+88}{12z}\,H_3+\frac{52z^2-227z+260}{36z}\,H_2-\frac{88z^2-359z+368}{54z}\,H_1\nonumber\\
&-\frac{z-3}{4z}\,\zeta_2\,H_1+\frac{4-3z}{4}\,\zeta_3+\frac{9z-14}{12}\,\zeta_2-\frac{7}{144}(16z-119)\nonumber\\
&
+\eps\,\bigg(\frac{3z^2-4z-3}{8z}\,\zeta_2\,H_{1,1}+\frac{35z^2-88z+124}{12z}\,H_{3,1}+\frac{44z^2-94z+115}{12z}\,H_{1,3}\nonumber\\
&+\frac{58z^2-146z+197}{24z}\,H_{2,1,1}+\frac{58z^2-116z+143}{24z}\,H_{1,1,1,1}\nonumber\\
&+\frac{70z^2-152z+221}{24z}\,H_{1,2,1}-\frac{100z^2-422z+473}{36z}\,H_{1,2}\nonumber\\
&+\frac{115z^2-284z+374}{24z}\,H_{2,2}+\frac{115z^2-236z+257}{24z}\,H_{1,1,2}\nonumber\\
&-\frac{140z^2-625z+736}{72z}\,H_{2,1}-\frac{140z^2-502z+493}{72z}\,H_{1,1,1}\nonumber\\
&+\frac{275z^2-1042z+1006}{108z}\,H_{1,1}+\frac{80z^2-156z+121}{8z}\,\zeta_3\,H_1+\frac{11z^2-28z+40}{3z}\,H_4\nonumber
\end{align}
\begin{align}
\phantom{V_{gg}^{(2,2)}(z,\eps)}
&-\frac{25z^2-107z+125}{9z}\,H_3+\frac{275z^2-1282z+1492}{108z}\,H_2\nonumber\\
&-\frac{527z^2-2149z+2194}{162z}\,H_1+\frac{3(z-2)(z+2)}{8z}\,\zeta_2\,H_2-\frac{5z-9}{6z}\,\zeta_2\,H_1\nonumber\\
&+\frac{17-36z}{24}\,\zeta_3+\frac{33}{32}(3z-4)\,\zeta_4+\frac{24z-53}{18}\,\zeta_2+\frac{22583-4966z}{864}\bigg)\Bigg]\nonumber\\
&+N^2\,N_f\,\Bigg[\frac{1}{\eps}\,\bigg(-\frac{z^2-2z+2}{6z}\,H_{1,1}-\frac{z^2-2z+2}{6z}\,H_2+\frac{5\left(z^2-2z+2\right)}{18z}\,H_1\nonumber\\
&-\frac{1+z}{12}\bigg)
-\frac{z^2-5z+5}{9z}\,H_{1,1}+\frac{z^2-2z+2}{6z}\,H_{2,1}+\frac{z^2-2z+2}{6z}\,H_{1,1,1}\nonumber\\
&+\frac{2\left(z^2-2z+2\right)}{3z}\,H_{1,2}-\frac{1}{12}\,\zeta_2+\frac{2\left(z^2-2z+2\right)}{3z}\,H_3-\frac{4z^2-23z+20}{36z}\,H_2\nonumber\\
&+\frac{26z^2-103z+94}{108z}\,H_1-\frac{z+19}{18}
+\eps\,\bigg(\frac{z^2-5z+5}{9z}\,H_{1,1,1}\nonumber\\
&-\frac{z^2-2z+2}{6z}\,H_{2,1,1}-\frac{z^2-2z+2}{6z}\,H_{1,1,1,1}-\frac{2\left(z^2-2z+2\right)}{3z}\,H_{3,1}\nonumber\\
&-\frac{2\left(z^2-2z+2\right)}{3z}\,H_{2,2}-\frac{2\left(z^2-2z+2\right)}{3z}\,H_{1,3}-\frac{2\left(z^2-2z+2\right)}{3z}\,H_{1,2,1}\nonumber\\
&-\frac{2\left(z^2-2z+2\right)}{3z}\,H_{1,1,2}+\frac{4z^2-23z+20}{36z}\,H_{2,1}+\frac{16z^2-83z+80}{36z}\,H_{1,2}\nonumber\\
&-\frac{17z^2-49z+40}{108z}\,H_{1,1}-\frac{1}{12}\,\zeta_3+\frac{1}{12}\,\zeta_2\,H_1-\frac{z^2-2z+2}{z}\,\zeta_3\,H_1\nonumber\\
&+\frac{4\left(z^2-5z+5\right)}{9z}\,H_3-\frac{2\left(z^2-2z+2\right)}{3z}\,H_4-\frac{17z^2-55z+40}{108z}\,H_2\nonumber\\
&+\frac{160z^2-557z+458}{324z}\,H_1+\left(\frac{1}{36}(13-3z)\right)\,\zeta_2+\frac{1}{216}(190z-1331)\bigg)\Bigg]\nonumber\\
&+N\,\Bigg[\frac{1}{\eps}\,\bigg(-\frac{z^2-8z+5}{12z}\,H_{1,1}-\frac{z^2-11z+14}{12z}\,H_2+\frac{134z^2-259z+259}{36z}\,H_1\nonumber\\
&+\frac{3-z}{4}\,\zeta_2+\frac{20-11z}{24}\bigg)
-\frac{67z^2-272z+155}{72z}\,H_{1,1}+\frac{79z^2-176z+182}{12z}\,H_{2,1}\nonumber\\
&+\frac{143z^2-316z+370}{24z}\,H_{1,2}+\frac{155z^2-310z+319}{24z}\,H_{1,1,1}-\frac{67z^2-386z+272}{72z}\,H_2\nonumber\\
&+\frac{73z^2-176z+176}{12z}\,H_3+\frac{379z^2-1472z+1382}{108z}\,H_1-\frac{(z-6)^2}{8z}\,\zeta_2\,H_1\nonumber\\
&+\frac{52-21z}{24}\,\zeta_2+\frac{z-3}{2}\,\zeta_3+\frac{1}{144}(635-344z)
+\eps\,\bigg(\frac{34z^2-42z+15}{8z}\,\zeta_2\,H_{1,1}\nonumber\\
&+\frac{8z^2-10z+25}{12z}\,H_{1,3}-\frac{28z^2-101z+65}{12z}\,H_{2,2}+\frac{31z^2-14z+11}{24z}\,H_{1,1,1,1}\nonumber\\
&+\frac{31z^2+16z-43}{24z}\,H_{2,1,1}+\frac{47z^2-64z+34}{12z}\,H_{3,1}-\frac{50z^2-166z+85}{24z}\,H_{1,1,2}\nonumber\\
&+\frac{100z^2-212z+203}{24z}\,H_{1,2,1}+\frac{535z^2-2321z+2216}{72z}\,H_{1,2}\nonumber
\end{align}
\begin{align}
\phantom{V_{gg}^{(2,2)}(z,\eps)}
&+\frac{959z^2-3814z+3859}{144z}\,H_{1,1,1}+\frac{1175z^2-5050z+4858}{144z}\,H_{2,1}\nonumber\\
&-\frac{2203z^2-7775z+4256}{432z}\,H_{1,1}-\frac{36z^2-227z+327}{24z}\,\zeta_2\,H_1\nonumber\\
&-\frac{194z^2-368z+365}{8z}\,\zeta_3\,H_1+\frac{z^2+25z-28}{6z}\,H_4+\frac{643z^2-2468z+2072}{72z}\,H_3\nonumber\\
&+\frac{1135z^2-3929z+3164}{162z}\,H_1-\frac{2203z^2-8924z+4256}{432z}\,H_2\nonumber\\
&+\frac{3(2z-1)(3z-2)}{4z}\,\zeta_2\,H_2+\frac{1}{144}(1126-309z)\,\zeta_2-\frac{5}{16}(19z-31)\,\zeta_4\nonumber\\
&+\frac{1}{48}(39z-346)\,\zeta_3+\frac{1}{864}(16733-9026z)\bigg)\Bigg]\nonumber\\
&+N_f\,\Bigg[\frac{1}{\eps}\,\bigg(-\frac{z^2-2z+2}{6z}\,H_{1,1}-\frac{z^2-2z+2}{6z}\,H_2-\frac{5\left(z^2-2z+2\right)}{9z}\,H_1+\frac{z-1}{12}\bigg)\nonumber\\
&
-\frac{z^2-5z+5}{9z}\,H_{1,1}-\frac{5\left(z^2-2z+2\right)}{6z}\,H_{2,1}-\frac{5\left(z^2-2z+2\right)}{6z}\,H_{1,1,1}\nonumber\\
&-\frac{4\left(z^2-2z+2\right)}{3z}\,H_{1,2}+\frac{1}{12}\,\zeta_2-\frac{4\left(z^2-2z+2\right)}{3z}\,H_3-\frac{4z^2-17z+20}{36z}\,H_2\nonumber\\
&-\frac{52z^2-215z+206}{108z}\,H_1+\frac{1}{36}(8z+7)
+\eps\,\bigg(-\frac{5\left(z^2-5z+5\right)}{9z}\,H_{1,1,1}\nonumber\\
&-\frac{2\left(z^2-2z+2\right)}{3z}\,H_{3,1}-\frac{2\left(z^2-2z+2\right)}{3z}\,H_{2,2}-\frac{2\left(z^2-2z+2\right)}{3z}\,H_{1,3}\nonumber\\
&-\frac{2\left(z^2-2z+2\right)}{3z}\,H_{1,2,1}-\frac{2\left(z^2-2z+2\right)}{3z}\,H_{1,1,2}-\frac{7\left(z^2-2z+2\right)}{6z}\,H_{2,1,1}\nonumber\\
&-\frac{7\left(z^2-2z+2\right)}{6z}\,H_{1,1,1,1}-\frac{20z^2-103z+100}{36z}\,H_{2,1}-\frac{32z^2-163z+160}{36z}\,H_{1,2}\nonumber\\
&-\frac{35z^2-175z+184}{108z}\,H_{1,1}+\frac{1}{12}\,\zeta_3-\frac{1}{12}\,\zeta_2\,H_1+\frac{2\left(z^2-2z+2\right)}{z}\,\zeta_3\,H_1\nonumber\\
&-\frac{8\left(z^2-5z+5\right)}{9z}\,H_3-\frac{2\left(z^2-2z+2\right)}{3z}\,H_4-\frac{35z^2-169z+184}{108z}\,H_2\nonumber\\
&-\frac{320z^2-1213z+1114}{324z}\,H_1+\left(\frac{1}{36}(3z-13)\right)\,\zeta_2+\frac{151-z}{54}\bigg)\Bigg]\nonumber\\
&+\frac{1}{N}\,\Bigg[\frac{1}{\eps}\,\bigg(-\frac{35\left(z^2-2z+2\right)}{24z}\,H_{1,1}-\frac{35z^2-58z+52}{24z}\,H_2\nonumber\\
&-\frac{134z^2-295z+286}{72z}\,H_1+\frac{1}{8}(3z-2)\,\zeta_2-\frac{11}{12}\bigg)
-\frac{35z^2-76z+106}{12z}\,H_{1,2}\nonumber\\
&-\frac{53z^2-118z+124}{12z}\,H_{2,1}-\frac{100z^2-383z+374}{72z}\,H_{1,1}\nonumber\\
&-\frac{106z^2-212z+221}{24z}\,H_{1,1,1}-\frac{35z^2-82z+88}{12z}\,H_3-\frac{50z^2-169z+142}{36z}\,H_2\nonumber\\
&-\frac{230z^2-889z+862}{108z}\,H_1-\frac{2z-15}{4z}\,\zeta_2\,H_1+\frac{1}{4}(2-3z)\,\zeta_3+\frac{z-2}{4}\,\zeta_2\nonumber\\
&+\frac{1}{144}(132z-793)
+\eps\,\bigg(-\frac{21z^2-46z+3}{8z}\,\zeta_2\,H_{1,1}-\frac{70z^2-128z+149}{12z}\,H_{1,3}\nonumber
\end{align}
\begin{align}
\phantom{V_{gg}^{(2,2)}(z,\eps)}
&-\frac{77z^2-142z+199}{24z}\,H_{1,1,2}-\frac{77z^2-118z+190}{24z}\,H_{2,2}-\frac{97z^2-158z+140}{12z}\,H_{3,1}\nonumber\\
&-\frac{100z^2-572z+725}{36z}\,H_{1,2}-\frac{140z^2-280z+253}{24z}\,H_{1,1,1,1}\nonumber\\
&-\frac{140z^2-220z+199}{24z}\,H_{2,1,1}-\frac{194z^2-364z+379}{24z}\,H_{1,2,1}\nonumber\\
&-\frac{298z^2-1157z+1130}{108z}\,H_{1,1}-\frac{344z^2-1477z+1612}{72z}\,H_{2,1}\nonumber\\
&-\frac{344z^2-1396z+1477}{72z}\,H_{1,1,1}-\frac{3\left(7z^2-10z-4\right)}{8z}\,\zeta_2\,H_2\nonumber\\
&+\frac{152z^2-296z+289}{8z}\,\zeta_3\,H_1-\frac{35z^2-55z+52}{6z}\,H_4-\frac{50z^2-259z+286}{18z}\,H_3\nonumber\\
&-\frac{149z^2-511z+430}{54z}\,H_2-\frac{689z^2-2671z+2590}{162z}\,H_1-\frac{4z-9}{z}\,\zeta_2\,H_1\nonumber\\
&+\frac{1}{8}(23-4z)\,\zeta_3+\frac{z-2}{2}\,\zeta_2+\left(\frac{3}{32}(73z-62)\right)\,\zeta_4+\frac{1}{864}(4758z-20335)\bigg)\Bigg]\nonumber\\
&
+\frac{N_f}{N^2}\,\Bigg[\frac{1}{\eps}\,\bigg(\frac{z^2-2z+2}{3z}\,H_{1,1}+\frac{5\left(z^2-2z+2\right)}{18z}\,H_1+\frac{z^2-2z+2}{3z}\,H_2+\frac{1}{6}\bigg)\nonumber\\
&
+\frac{2\left(z^2-5z+5\right)}{9z}\,H_{1,1}+\frac{2\left(z^2-2z+2\right)}{3z}\,H_{2,1}+\frac{2\left(z^2-2z+2\right)}{3z}\,H_{1,2}\nonumber\\
&+\frac{2\left(z^2-2z+2\right)}{3z}\,H_{1,1,1}+\frac{2\left(z^2-5z+5\right)}{9z}\,H_2+\frac{2\left(z^2-2z+2\right)}{3z}\,H_3\nonumber\\
&+\frac{13z^2-56z+56}{54z}\,H_1+\frac{1}{36}(31-6z)
+\eps\,\bigg(\frac{4\left(z^2-5z+5\right)}{9z}\,H_{2,1}\nonumber\\
&+\frac{4\left(z^2-5z+5\right)}{9z}\,H_{1,2}+\frac{4\left(z^2-5z+5\right)}{9z}\,H_{1,1,1}+\frac{4\left(z^2-2z+2\right)}{3z}\,H_{3,1}\nonumber\\
&+\frac{4\left(z^2-2z+2\right)}{3z}\,H_{2,2}+\frac{4\left(z^2-2z+2\right)}{3z}\,H_{2,1,1}+\frac{4\left(z^2-2z+2\right)}{3z}\,H_{1,3}\nonumber\\
&+\frac{4\left(z^2-2z+2\right)}{3z}\,H_{1,2,1}+\frac{4\left(z^2-2z+2\right)}{3z}\,H_{1,1,2}+\frac{4\left(z^2-2z+2\right)}{3z}\,H_{1,1,1,1}\nonumber\\
&+\frac{13z^2-56z+56}{27z}\,H_{1,1}-\frac{z^2-2z+2}{z}\,\zeta_3\,H_1+\frac{4\left(z^2-5z+5\right)}{9z}\,H_3\nonumber\\
&+\frac{4\left(z^2-2z+2\right)}{3z}\,H_4+\frac{4\left(10z^2-41z+41\right)}{81z}\,H_1\nonumber\\
&+\frac{13z^2-56z+56}{27z}\,H_2+\frac{1}{216}(727-186z)\bigg)\Bigg]\nonumber\\
&+\frac{1}{N^3}\,\Bigg[\frac{1}{\eps}\,\bigg(-\frac{(1-z)(z-3)}{4z}\,H_{1,1}+\frac{z-2}{4}\,H_2+\frac{1-z}{4z}\,H_1-\frac{z}{2}\,\zeta_2+\frac{3(z-2)}{8}\bigg)\nonumber\\
&
-\frac{z^2-6z-6}{8z}\,H_{1,2}+\frac{7z^2-32z+19}{8z}\,H_{1,1}+\frac{2-z}{2}\,H_{2,1}-\frac{3(1-z)^2}{8z}\,H_{1,1,1}\nonumber\\
&+\frac{z^2-5z+8}{4z}\,H_1+\frac{7z^2-30z+4}{8z}\,H_2+\frac{z-6}{8}\,\zeta_2\,H_1+\frac{6-z}{4}\,H_3\nonumber
\end{align}
\begin{align}
\phantom{V_{gg}^{(2,2)}(z,\eps)}
&-\frac{1}{8}(z+4)\,\zeta_2+z\,\zeta_3+\frac{3}{16}(12z-25)
+\eps\,\bigg(-\frac{16z^2+9}{8z}\,\zeta_2\,H_{1,1}\nonumber\\
&+\frac{z^2+2z+9}{16z}\,H_{1,1,1}+\frac{4z^2-24z+9}{8z}\,H_{1,1,2}+\frac{5z^2-2z-6}{4z}\,H_{3,1}\nonumber\\
&+\frac{6z^2-8z+3}{4z}\,H_{1,3}+\frac{8z^2-15}{8z}\,H_{1,2,1}-\frac{15z^2-37z-20}{8z}\,H_{1,2}\nonumber\\
&+\frac{17z^2-30z+15}{8z}\,H_{2,1,1}-\frac{23z^2-94z+18}{16z}\,H_{2,1}+\frac{85z^2-305z+176}{16z}\,H_{1,1}\nonumber\\
&+\frac{3(z-3)(z+1)}{4z}\,H_{2,2}-\frac{(1-z)(17z-33)}{8z}\,H_{1,1,1,1}-\frac{3\left(3z^2-2z+2\right)}{4z}\,\zeta_2\,H_2\nonumber\\
&-\frac{38z^2-84z+45}{8z}\,\zeta_3\,H_1+\frac{z^2-11z+20}{2z}\,H_1-\frac{27z^2-64z-8}{8z}\,H_3\nonumber\\
&-\frac{(1-z)(12z-25)}{8z}\,\zeta_2\,H_1+(2(z-2))\,H_4+\frac{(5z-16)(17z-4)}{16z}\,H_2\nonumber\\
&+\frac{1}{4}(1-16z)\,\zeta_4+\frac{1}{16}(5z-62)\,\zeta_2+\frac{1}{16}(19z+58)\,\zeta_3+\frac{19}{32}(18z-37)\bigg)\Bigg]+\ord(\eps^2)\,,\nonumber\\
V_{qq}^{(2,2)}(z,\eps) & = N^2\,\Bigg[\frac{1}{\eps}\,\bigg(\frac{1-z}{8}\,H_1+\frac{1}{8}(4z-3)\,H_2+\frac{1}{24}\left(-88z^2+76z-35\right)\,\zeta_2\nonumber\\
&-\frac{101}{54}\left(2z^2-2z+1\right)\bigg)
+\frac{1}{8}\left(-6z^2-2z+3\right)\,H_{2,1}+\frac{1}{8}(3z-2)\,H_{1,1}\nonumber\\
&+\frac{1}{4}(1-z)(3z-2)\,H_{1,1,1}-\frac{1}{8}(2z-1)(6z-5)\,H_{1,2}\nonumber\\
&+\frac{1}{24}\left(176z^2-164z+79\right)\,\zeta_2\,H_1+\frac{1}{2}\left(-3z^2+5z-3\right)\,H_3\nonumber\\
&+\frac{1}{216}\left(1616z^2-1697z+862\right)\,H_1+\frac{1}{8}(11z-5)\,H_2\nonumber\\
&+\frac{1}{72}\left(-800z^2+701z-205\right)\,\zeta_2+\frac{1}{8}\left(62z^2-70z+37\right)\,\zeta_3\nonumber\\
&+\frac{1}{162}\left(-1820z^2+1820z-607\right)
+\eps\,\bigg(\frac{1}{24}\left(-334z^2+310z-155\right)\,\zeta_2\,H_{1,1}\nonumber\\
&+\frac{1}{216}\left(-3232z^2+3475z-1751\right)\,H_{1,1}+\frac{1}{8}\left(-18z^2-5z+5\right)\,H_{2,1}\nonumber\\
&+\frac{1}{8}\left(-18z^2+19z-4\right)\,H_{1,1,1}+\frac{1}{2}\left(-6z^2+10z-3\right)\,H_{1,3}\nonumber\\
&+\frac{1}{8}\left(-6z^2+14z-7\right)\,H_{1,1,2}+\frac{1}{2}\left(3z^2-5z+3\right)\,H_{3,1}-\frac{3}{4}\left(5z^2-11z+5\right)\,H_{2,2}\nonumber\\
&+\frac{1}{8}\left(6z^2+26z-15\right)\,H_{2,1,1}+\frac{1}{8}\left(24z^2-28z+9\right)\,H_{1,2,1}\nonumber\\
&-\frac{3}{4}(1-z)(3z-2)\,H_{1,1,1,1}-\frac{1}{8}(4z-1)(9z-8)\,H_{1,2}\nonumber\\
&+\frac{1}{8}\left(-124z^2+144z-71\right)\,\zeta_3\,H_1+\frac{3}{4}\left(z^2+z-1\right)\,\zeta_2\,H_2\nonumber\\
&+\frac{1}{72}\left(1600z^2-1537z+500\right)\,\zeta_2\,H_1-2\left(3z^2-5z+3\right)\,H_4\nonumber\\
&+\frac{7}{648}\left(2080z^2-2161z+740\right)\,H_1+\frac{1}{2}(1-z)(9z-5)\,H_3+\frac{1}{8}(25z-12)\,H_2\nonumber\\
&+\frac{1}{216}\left(-5632z^2+4957z-1238\right)\,\zeta_2+\frac{1}{48}\left(-629z^2+107z+77\right)\,\zeta_4\nonumber
\end{align}
\begin{align}
\phantom{V_{gg}^{(2,2)}(z,\eps)}
&+\frac{1}{24}\left(562z^2-619z+239\right)\,\zeta_3-\frac{2}{243}\left(3187z^2-3187z+911\right)\bigg)\Bigg]\nonumber\\
&+N\,N_f\,\Bigg[\frac{1}{\eps}\,\bigg(\frac{1}{3}\left(2z^2-2z+1\right)\,\zeta_2+\frac{7}{27}\left(2z^2-2z+1\right)\bigg)\nonumber\\
&
-\frac{2}{3}\left(2z^2-2z+1\right)\,\zeta_2\,H_1-\frac{14}{27}\left(2z^2-2z+1\right)\,H_1+\frac{1}{2}\left(-2z^2+2z-1\right)\,\zeta_3\nonumber\\
&+\frac{1}{9}\left(16z^2-16z+5\right)\,\zeta_2+\frac{1}{81}\left(124z^2-124z+41\right)
+\eps\,\bigg(\frac{4}{3}\left(2z^2-2z+1\right)\,\zeta_2\,H_{1,1}\nonumber\\
&+\frac{28}{27}\left(2z^2-2z+1\right)\,H_{1,1}+\left(2z^2-2z+1\right)\,\zeta_3\,H_1-\frac{2}{9}\left(16z^2-16z+5\right)\,\zeta_2\,H_1\nonumber\\
&-\frac{2}{81}\left(124z^2-124z+41\right)\,H_1+\frac{1}{6}\left(-16z^2+16z-5\right)\,\zeta_3+\frac{19}{12}\left(2z^2-2z+1\right)\,\zeta_4\nonumber\\
&+\frac{4}{27}\left(26z^2-26z+7\right)\,\zeta_2+\frac{4}{243}\left(215z^2-215z+61\right)\bigg)\Bigg]\nonumber\\
&+\frac{1}{\eps}\,\bigg(\frac{1}{8}(3-2z)\,H_2+\frac{1-z}{8}\,H_1+\frac{1}{8}(2z-3)\,\zeta_2\bigg)
+\frac{1}{8}\left(6z^2-2z-3\right)\,H_{2,1}\nonumber\\
&+\frac{1}{8}\left(12z^2-14z+1\right)\,H_{1,2}+\frac{1}{8}(3z-2)\,H_{1,1}-\frac{1}{4}(1-z)(3z-1)\,H_{1,1,1}\nonumber\\
&+\frac{1}{2}\left(3z^2-4z+3\right)\,H_3+\frac{1}{8}(3-2z)\,\zeta_2\,H_1+\frac{1}{8}(2-3z)\,H_1+\frac{9(1-z)}{8}\,H_2\nonumber\\
&+\frac{1}{8}\left(-18z^2+22z-15\right)\,\zeta_3+\frac{1}{8}(9z-7)\,\zeta_2
+\eps\,\bigg(\frac{1}{8}\left(-6z^2+10z-5\right)\,\zeta_2\,H_{1,1}\nonumber\\
&+\frac{1}{8}\left(-24z^2+26z-3\right)\,H_{1,2,1}+\frac{1}{8}\left(-6z^2-10z+9\right)\,H_{2,1,1}\nonumber\\
&+\frac{1}{2}\left(-3z^2+4z-3\right)\,H_{3,1}+\frac{1}{8}\left(6z^2-10z+5\right)\,H_{1,1,2}\nonumber\\
&+\frac{1}{8}\left(22z^2-41z+16\right)\,H_{1,1,1}+\frac{1}{8}\left(22z^2-5z-9\right)\,H_{2,1}\nonumber\\
&+\frac{1}{8}\left(44z^2-59z+18\right)\,H_{1,2}+z(3z-4)\,H_{1,3}+\frac{3}{4}(1-z)(3z-1)\,H_{1,1,1,1}\nonumber\\
&-\frac{3}{4}(1-z)(5z-3)\,H_{2,2}+\frac{1}{8}(9z-5)\,H_{1,1}+\frac{1}{8}\left(36z^2-46z+21\right)\,\zeta_3\,H_1\nonumber\\
&+2\left(3z^2-4z+3\right)\,H_4+\frac{1}{2}\left(11z^2-16z+9\right)\,H_3+\frac{1}{8}(10-13z)\,\zeta_2\,H_1\nonumber\\
&+\frac{3}{4}(1-z^2)\,\zeta_2\,H_2+\frac{1}{8}(16-23z)\,H_2+\frac{1}{8}(4-7z)\,H_1+\frac{1}{8}\left(-66z^2+87z-35\right)\,\zeta_3\nonumber\\
&-\frac{3}{16}\left(23z^2-52z+55\right)\,\zeta_4+\frac{1}{8}(23z-14)\,\zeta_2\bigg)+\ord(\eps^2)\,,\nonumber\\
V_{qq}^{(2,1)}(z,\eps) & = N^2\,\Bigg[\frac{13}{2}\left(2z^2-2z+1\right)\,\zeta_3 + \eps\,\bigg(-\frac{13}{2}\left(2z^2-2z+1\right)\,\zeta_3\,H_1\\
&+\frac{39}{4}\left(2z^2-2z+1\right)\,\zeta_4+\frac{1}{3}\left(119z^2-119z+40\right)\,\zeta_3\bigg)\Bigg]\nonumber\\
&+N\,N_f\,\Bigg[-2\left(2z^2-2z+1\right)\,\zeta_3 +\eps\,\bigg(2\left(2z^2-2z+1\right)\,\zeta_3\,H_1\nonumber
\end{align}
\begin{align}
\phantom{V_{gg}^{(2,2)}(z,\eps)}
&-3\left(2z^2-2z+1\right)\,\zeta_4-\frac{2}{3}\left(16z^2-16z+5\right)\,\zeta_3\bigg)\Bigg]\nonumber\\
&+\frac{9}{2}\left(2z^2-2z+1\right)\,\zeta_3 + \eps\,\bigg(-\frac{9}{2}\left(2z^2-2z+1\right)\,\zeta_3\,H_1+\frac{27}{4}\left(2z^2-2z+1\right)\,\zeta_4\nonumber\\
&+3\left(11z^2-11z+4\right)\,\zeta_3\bigg)+\ord(\eps^2)\,,\nonumber\\
V_{qq}^{(2,0)}(z,\eps) & = N^2\,\Bigg[\frac{37}{24\eps^3}\left(2z^2-2z+1\right)+\frac{1}{\eps^2}\,\bigg(\frac{1}{48}\left(754z^2-754z+303\right)\\
&-\frac{13}{6}\left(2z^2-2z+1\right)\,H_1\bigg)
+\frac{1}{\eps}\,\bigg(\frac{1}{24}\left(-496z^2+499z-199\right)\,H_1+\frac{1}{8}(3-4z)\,H_2\nonumber\\
&+\frac{1}{24}\left(-52z^2+64z-35\right)\,\zeta_2+\frac{1}{864}\left(50926z^2-50926z+18677\right)\bigg)\nonumber\\
&
+\frac{1}{8}\left(-146z^2+154z-79\right)\,H_{2,1}+\frac{1}{8}\left(-140z^2+128z-69\right)\,H_{1,2}\nonumber\\
&+\frac{1}{4}\left(-73z^2+71z-36\right)\,H_{1,1,1}+\frac{1}{8}(2-3z)\,H_{1,1}\nonumber\\
&+\frac{1}{72}\left(-3856z^2+3883z-1202\right)\,H_1+\frac{1}{2}\left(-35z^2+37z-19\right)\,H_3\nonumber\\
&+\frac{1}{8}(12z-1)\,\zeta_2\,H_1+\frac{1}{8}(7-15z)\,H_2+\frac{1}{72}\left(-782z^2+899z-358\right)\,\zeta_2\nonumber\\
&+\frac{1}{24}\left(-550z^2+526z-257\right)\,\zeta_3+\frac{1016318z^2-1016318z+355381}{5184}\nonumber\\
&
+\eps\,\bigg(\frac{1}{8}\left(-18z^2+50z-19\right)\,\zeta_2\,H_{1,1}+\frac{1}{24}\left(-1330z^2+1327z-452\right)\,H_{1,1,1}\nonumber\\
&+\frac{1}{24}\left(-1330z^2+1405z-491\right)\,H_{2,1}+\frac{1}{24}\left(-1276z^2+1201z-416\right)\,H_{1,2}\nonumber\\
&+\frac{1}{8}\left(-36z^2+32z-13\right)\,H_{1,2,1}+\frac{1}{8}\left(-18z^2-6z+7\right)\,H_{2,1,1}\nonumber\\
&+\frac{1}{4}\left(-6z^2-2z+1\right)\,H_{1,3}+\frac{1}{8}\left(18z^2-34z+15\right)\,H_{1,1,2}+\frac{1}{8}(5-9z)\,H_{1,1}\nonumber\\
&+\frac{3}{4}(1-z)(3z-2)\,H_{1,1,1,1}+\frac{1}{4}(3z-7)(3z-2)\,H_{2,2}-\frac{3}{2}z(3z-1)\,H_{3,1}\nonumber\\
&+\frac{1}{4}\left(-9z^2+15z-5\right)\,\zeta_2\,H_2+\frac{1}{8}\left(528z^2-540z+263\right)\,\zeta_3\,H_1\nonumber\\
&+\frac{1}{216}\left(-25808z^2+25997z-7228\right)\,H_1+\frac{1}{12}\left(-638z^2+671z-235\right)\,H_3\nonumber\\
&-\frac{3}{2}\left(z^2+z-1\right)\,H_4+\frac{1}{8}(35z-12)\,\zeta_2\,H_1+\frac{1}{8}(18-37z)\,H_2\nonumber\\
&+\frac{1}{432}\left(-17402z^2+19076z-7111\right)\,\zeta_2+\frac{1}{72}\left(-9758z^2+9551z-3991\right)\,\zeta_3\nonumber\\
&+\frac{1}{16}\left(77z^2+29z-41\right)\,\zeta_4+\frac{6380090z^2-6380090z+2173727}{10368}\bigg)\Bigg]\nonumber\\
&N\,N_f\,\Bigg[-\frac{5}{12\eps^3}\left(2z^2-2z+1\right)+\frac{1}{\eps^2}\,\bigg(\frac{2}{3}\left(2z^2-2z+1\right)\,H_1\nonumber\\
&+\frac{1}{18}\left(-101z^2+101z-43\right)\bigg) +\frac{1}{\eps}\,\bigg(\frac{1}{3}\left(14z^2-14z+5\right)\,H_1\nonumber
\end{align}
\begin{align}
\phantom{V_{gg}^{(2,2)}(z,\eps)}
&+\frac{1}{3}\left(2z^2-2z+1\right)\,\zeta_2+\frac{1}{72}\left(-1622z^2+1622z-609\right)\bigg)
+2\left(2z^2-2z+1\right)\,H_{2,1}\nonumber\\
&+2\left(2z^2-2z+1\right)\,H_{1,2}+2\left(2z^2-2z+1\right)\,H_{1,1,1}+2\left(2z^2-2z+1\right)\,H_3\nonumber\\
&+\frac{14}{9}\left(7z^2-7z+2\right)\,H_1+\frac{5}{6}\left(2z^2-2z+1\right)\,\zeta_3+\frac{1}{18}\left(50z^2-50z+19\right)\,\zeta_2\nonumber\\
&+\frac{-97742z^2+97742z-34273}{1296}
+\eps\,\bigg(\frac{2}{3}\left(16z^2-16z+5\right)\,H_{2,1}\nonumber\\
&+\frac{2}{3}\left(16z^2-16z+5\right)\,H_{1,2}+\frac{2}{3}\left(16z^2-16z+5\right)\,H_{1,1,1}-6\left(2z^2-2z+1\right)\,\zeta_3\,H_1\nonumber\\
&+\frac{2}{3}\left(16z^2-16z+5\right)\,H_3+\frac{2}{27}\left(311z^2-311z+82\right)\,H_1-\frac{13}{4}\left(2z^2-2z+1\right)\,\zeta_4\nonumber\\
&+\frac{1}{18}\left(464z^2-464z+217\right)\,\zeta_3+\frac{1}{108}\left(1046z^2-1046z+373\right)\,\zeta_2\nonumber\\
&+\frac{-1816750z^2+1816750z-615149}{7776}\bigg)\Bigg]\nonumber\\
&+N_f\,\Bigg[\frac{2}{9\eps^2}\left(2z^2-2z+1\right)+\frac{4}{27\eps}\left(13z^2-13z+5\right)+\frac{2}{27}\left(80z^2-80z+27\right)\nonumber\\
&+\frac{32\eps}{243}\left(121z^2-121z+38\right)\Bigg]\nonumber\\
&+\frac{55}{24\eps^3}\left(2z^2-2z+1\right)+\frac{1}{\eps^2}\,\bigg(\frac{1}{72}\left(1504z^2-1504z+587\right)-\frac{3}{2}\left(2z^2-2z+1\right)\,H_1\bigg)\nonumber\\
&+\frac{1}{\eps}\,\bigg(\frac{1}{8}\left(-88z^2+89z-33\right)\,H_1+\frac{1}{8}(2z-3)\,H_2+\frac{1}{24}\left(-28z^2+22z-5\right)\,\zeta_2\nonumber\\
&+\frac{47}{216}\left(334z^2-334z+119\right)\bigg)
+\frac{1}{8}\left(-36z^2+42z-11\right)\,H_{1,2}\nonumber\\
&+\frac{1}{8}\left(-30z^2+26z-9\right)\,H_{2,1}+\frac{1}{4}\left(-15z^2+16z-7\right)\,H_{1,1,1}+\frac{1}{8}(2-3z)\,H_{1,1}\nonumber\\
&+\frac{1}{2}\left(-9z^2+8z-3\right)\,H_3-\frac{3}{8}\left(72z^2-73z+22\right)\,H_1-\frac{1}{8}(6z+7)\,\zeta_2\,H_1\nonumber\\
&+\frac{1}{8}(5z-7)\,H_2+\frac{1}{24}\left(-538z^2+550z-287\right)\,\zeta_3+\frac{1}{72}\left(-248z^2+185z-1\right)\,\zeta_2\nonumber\\
&+\frac{292994z^2-292994z+99403}{1296}
+\eps\,\bigg(\frac{1}{8}\left(18z^2-34z+5\right)\,\zeta_2\,H_{1,1}\nonumber\\
&+\frac{1}{8}\left(-132z^2+167z-54\right)\,H_{1,2}+\frac{1}{8}\left(-110z^2+95z-27\right)\,H_{2,1}\nonumber\\
&+\frac{1}{8}\left(-110z^2+129z-48\right)\,H_{1,1,1}+\frac{1}{8}\left(-18z^2+26z-9\right)\,H_{1,1,2}\nonumber\\
&+\frac{1}{4}\left(-9z^2+18z-10\right)\,H_{2,2}+\frac{1}{4}\left(6z^2-2z+1\right)\,H_{1,3}+\frac{1}{8}\left(18z^2-6z-5\right)\,H_{2,1,1}\nonumber\\
&+\frac{1}{8}\left(36z^2-34z+11\right)\,H_{1,2,1}+\frac{1}{8}(5-9z)\,H_{1,1}+\frac{3}{2}z(3z-2)\,H_{3,1}\nonumber\\
&-\frac{3}{4}(1-z)(3z-1)\,H_{1,1,1,1}+\frac{1}{4}\left(9z^2-12z+1\right)\,\zeta_2\,H_2\nonumber\\
&+\frac{1}{8}\left(-472z^2+479z-132\right)\,H_1+\frac{1}{4}\left(-66z^2+57z-17\right)\,H_3-\frac{1}{8}(25z+2)\,\zeta_2\,H_1\nonumber\\
&+\frac{1}{8}(6z+11)\,\zeta_3\,H_1-\frac{3}{2}(1-z^2)\,H_4+\frac{1}{8}(11z-10)\,H_2\nonumber
\end{align}
\begin{align}
\phantom{V_{gg}^{(2,2)}(z,\eps)}
&+\frac{1}{72}\left(-10466z^2+10619z-4615\right)\,\zeta_3+\frac{1}{216}\left(-640z^2+181z+430\right)\,\zeta_2\nonumber\\
&+\frac{1}{16}\left(-405z^2+352z-123\right)\,\zeta_4+\frac{5179370z^2-5179370z+1710703}{7776}\bigg)\nonumber\\
&+\frac{N_f}{N}\,\Bigg[-\frac{5}{12\eps^3}\left(2z^2-2z+1\right)+\frac{1}{18\eps^2}\left(-67z^2+67z-26\right)\nonumber\\
&
+\frac{1}{\eps}\,\bigg(\frac{1}{6}\left(-2z^2+2z-1\right)\,\zeta_2+\frac{1}{216}\left(-2786z^2+2786z-991\right)\bigg)\nonumber\\
&
+\frac{1}{9}\left(-17z^2+17z-7\right)\,\zeta_2-\frac{5}{3}\left(2z^2-2z+1\right)\,\zeta_3-\frac{7\left(6830z^2-6830z+2221\right)}{1296}\nonumber\\
&
+\eps\,\bigg(\frac{1}{108}\left(-910z^2+910z-353\right)\,\zeta_2+\frac{1}{9}\left(-116z^2+116z-43\right)\,\zeta_3\nonumber\\
&-3\left(2z^2-2z+1\right)\,\zeta_4+\frac{-728594z^2+728594z-220867}{7776}\bigg)\Bigg]\nonumber\\
&+\frac{1}{N^2}\,\Bigg[\frac{3}{4\eps^3}\left(2z^2-2z+1\right)+\frac{1}{16\eps^2}\left(106z^2-106z+41\right)\nonumber\\
&
+\frac{1}{\eps}\,\bigg(\frac{3}{4}\left(2z^2-2z+1\right)\,\zeta_2+\frac{1}{32}\left(654z^2-654z+221\right)\bigg)
-\frac{15}{4}\left(2z^2-2z+1\right)\,\zeta_3\nonumber\\
&+\frac{1}{8}\left(70z^2-70z+29\right)\,\zeta_2+\frac{1}{64}\left(3610z^2-3610z+1151\right)\nonumber\\
&
+\eps\,\bigg(\frac{1}{2}\left(-107z^2+107z-46\right)\,\zeta_3-\frac{39}{8}\left(2z^2-2z+1\right)\,\zeta_4\nonumber\\
&+\frac{7}{16}\left(82z^2-82z+31\right)\,\zeta_2+\frac{1}{128}\left(18702z^2-18702z+5741\right)\bigg)\Bigg]+\ord(\eps^2)\,.\nonumber
\end{align}

\section{The CDR matrix elements for $H$ to three partons}
\label{app:HCDR}
\subsection{The matrix element for $H\to ggg$}
In this section we show how to obtain the two-loop matrix element in CDR from the $D$-dimensional tensor coefficients of ref.~\cite{Gehrmann:2011aa}.
The amplitude for $H\to ggg$ can be written as
\beq
|\cM_{H\to ggg}\rangle = S_{\mu\nu\rho}\,\eps^\mu_1\,\eps^\nu_2\,\eps^\rho_3\,,
\eeq
where the gluon tensor is given by,
\beq
S_{\mu\nu\rho} = \lambda_0\,\sqrt{4\pi \alpha_0}\,f^{a_1a_2a_3}\,\sum_{\ell=0}^\infty\left(\frac{\alpha_0}{2\pi}\right)^\ell\,\left(\frac{S_\eps\,c_\Gamma}{2}\right)^\ell\,e^{i\pi\ell\epsilon}\,(m_H^2)^{-\ell\eps}\,S^{(\ell)}_{\mu\nu\rho}\,,
\eeq
where $\lambda_0$ and $\alpha_0$ denote the bare coupling constants, and
\beq
S_\eps = (4\pi)^\eps\,e^{-\gamma_E\eps} {\rm~~and~~} c_\Gamma=e^{\gamma_E\eps}\,\frac{\Gamma(1-\eps)^2\,\Gamma(1+\eps)}{\Gamma(1-2\eps)}\,,
\eeq
where $\gamma_E=-\Gamma'(1)$ denotes the Euler-Mascheroni constant.
After tensor decomposition, we get
\beq
 S_{\mu\nu\rho}^{(\ell)}\,\eps^\mu_1\,\eps^\nu_2\,\eps^\rho_3 = \frac{1}{m_H^2}\,\left[T_{232}\,A_{232}^{(\ell)} + T_{211}\,A_{211}^{(\ell)}+ T_{311}\,A_{311}^{(\ell)} + T_{312}\,A_{312}^{(\ell)}\right]\,,
\eeq
where $T_{ijk}$ are the tensors given in eq.~(3.7) of ref.~\cite{Gehrmann:2011aa}, and the $A_{ijk}^{(\ell)}$ are scalars\footnote{We slightly changed the normalisation of the $A_{ijk}^{(\ell)}$ compared to ref.~\cite{Gehrmann:2011aa}, in order to have the colour tensor and the overall coupling constant explicit.}.

Our goal is to compute the interference $\langle\cM^{(0)}_{H\to ggg}|\cM^{(\ell)}_{H\to ggg}\rangle$ in CDR, summed over colours and spins. The colour sum is trivial, and gives
\beq
f^{a_1a_2a_3}\,f^{a_1a_2a_3} = V\,C_A\,.
\eeq
The polarisation sum read,
\beq
\sum_{\textrm{pol.}}\eps^*_{i\mu}\,\eps_{i\nu} = -\eta_{\mu\nu} + \textrm{gauge dependent terms.}
\eeq
The gauge dependent terms are proportional to the gluon momentum. The tensors $T_{ijk}\equiv T_{ijk}^{\mu\nu\rho}\,\eps^\mu_1\,\eps^\nu_2\,\eps^\rho_3$ of ref.~\cite{Gehrmann:2011aa} are transverse, i.e., they vanish whenever they are contracted with an external gluon momentum,
\beq
T_{ijk}^{\mu\nu\rho}\,p_{1\mu}=T_{ijk}^{\mu\nu\rho}\,p_{2\nu}=T_{ijk}^{\mu\nu\rho}\,p_{3\rho} = 0\,.
\eeq
As a consequence, the gauge dependent terms will always drop out, and we can use the `naive' polarisation sum
\beq
\sum_{\textrm{pol.}}\eps^*_{i\mu}\,\eps_{i\nu} \to -\eta_{\mu\nu}\,.
\eeq
We get
\beq\bsp\label{eq:MEggg}
\langle\cM^{(0)}_{H\to ggg}|\cM^{(\ell)}_{H\to ggg}\rangle &\,= 8\pi^2\,|\lambda_0|^2\,\left(\frac{\alpha_0}{2\pi}\right)^{\ell+1}\,\left(\frac{S_\eps\,c_\Gamma}{2}\right)^\ell\,e^{i\pi\ell\eps}\,(m_H^2)^{1-\ell\epsilon}\,V\,C_A\\
&\,\times M_{ggg}^{(\ell)}\left(x_{12},x_{13},x_{23};\eps\right)\,,
\esp\eeq
with
\beq\bsp
M&_{ggg}^{(\ell)}\left(x_{12},x_{13},x_{23};\eps\right)\\
&=A_{211}^{(\ell)}\,\frac{x_{12}}{2 x_{23}}\Big[(D-2) (x_{12}^2+x_{12}\,x_{13}+x_{12}\,x_{23} + x_{13}\, x_{23})+x_{13}^2+x_{23}^2\Big]\\
&\,-A_{311}^{(\ell)} \,\frac{x_{13}}{2 x_{23}}\, \Big[(D-2) (x_{13}^2+x_{12}\,x_{13}+x_{12}\,x_{23}+ x_{13}\, x_{23})+x_{12}^2+x_{23}^2\Big]\\
&\,+A_{232}^{(\ell)}\,\frac{ x_{23}}{2 x_{13}}\, \Big[(D-2) (x_{23}^2+x_{12}\,x_{13}+x_{12}\,x_{23}+ x_{13}\, x_{23})+x_{12}^2+x_{13}^2\Big]\\
&\,-\frac{1}{2} A_{312}^{(\ell)} \Big[(D-2) (x_{12}^2+x_{23}^2+x_{13}^2)+(3 D-8) (x_{12}\,x_{13}+x_{12}\,x_{23}+ x_{13} \,x_{23})\Big]\,,
\esp\eeq
where $x_{ij}=s_{ij}/m_H^2$ and $x_{12}+x_{13}+x_{23}=1$, and where we have used the fact that
\beq\bsp
A_{211}^{(0)} = \frac{2}{x_{13}}\,,\qquad &A_{311}^{(0)} = -\frac{2}{x_{12}}\,,\\
A_{232}^{(0)} = \frac{2}{x_{12}}\,,\qquad &A_{312}^{(0)} = -\frac{2}{x_{12}}-\frac{2}{x_{13}}-\frac{2}{x_{23}}\,.
\esp\eeq

\subsection{The matrix element for $H\to q\bar{q}g$}
The amplitude for $H\to q\bar{q}g$ can be written as
\beq\bsp
|\cM_{H\to q\bar{q}g}\rangle &\,= \lambda_0\sqrt{4\pi\alpha_0}\,T_{ij}^a\,\sum_{\ell=0}^\infty \left(\frac{\alpha_0}{2\pi}\right)^\ell\,\left(\frac{S_\eps\,c_\Gamma}{2}\right)^\ell\,e^{i\pi\ell\eps}\,(m_H^2)^{-\ell\eps}\, T^{(\ell)}_\rho\,\eps_3^{\rho}\\
&\,=\lambda_0\sqrt{4\pi\alpha_0}\,T_{ij}^a\,\left[T_1\,A_1+T_2\,A_2\right]\,,
\esp\eeq
where as in the gluon case we have factored out explicitly the overall coupling and colour structure. We have
\beq
T_i = \bar{u}_1\gamma_\mu v_2\,\left[p_3^\mu\,\eps_3\cdot p_i - \eps_3^\mu\,p_3\cdot p_i\right]\,.
\eeq
The colour sum is trivial and the quake spin sum give
\beq\bsp
\sum_{\textrm{spins}}[\bar{u}_1\gamma_\mu v_2]^\dagger\,\bar{u}_1\gamma_\nu v_2 &\,= 
\textrm{Tr}\left[\slashed{p_1}\gamma_{\nu}\slashed{p_2}\gamma_\mu\right]
=4\,\left(p_{1\mu}p_{2\nu}+p_{1\nu}p_{2\mu}-\frac{1}{2}\,s_{12}\,\eta_{\mu\nu}\right)\,.
\esp\eeq
The tensors $T_i\equiv T_{i\rho}\,\eps_3^\rho$ are transverse, $T_{i\rho}\,p_3^\rho=0$, and so we can again use the `naive' polarisation sum for the gluons. We get,
\beq\bsp
\langle\cM^{(0)}_{H\to q\bar{q}g}|\cM^{(\ell)}_{H\to q\bar{q}g}\rangle &\,= 8\pi^2\,|\lambda_0|^2\,\left(\frac{\alpha_0}{2\pi}\right)^{\ell+1}\,\left(\frac{S_\eps\,c_\Gamma}{2}\right)^\ell\,C_F\,C_A\,e^{i\pi\ell\eps}\,(m_H^2)^{1-\ell\eps}\\
&\,\times M_{q\bar{q}g}^{(\ell)}(x_{12},x_{13},x_{23};\eps)\,,
\esp\eeq
where we used the fact that $A_i^{(0)}=1/x_{12}$ and where we defined
\beq\bsp
M_{q\bar{q}g}^{(\ell)}(x_{12},x_{13},x_{23};\eps) &\,= \frac{1}{2} A^{(\ell)}_1\, x_{13} \left[(D-2) x_{13}+(D-4) x_{23}\right]\\
&\,+\frac{1}{2} A^{(\ell)}_2 x_{23} \left[(D-4) x_{13}+(D-2) x_{23}\right]\,.
\esp\eeq

\section{Two-loop single-real contributions to Higgs production at N$^3$LO}
\label{app:coeff}
In this appendix we present the result for the contributions of the two-loop amplitude for $H$ + 3 partons to the inclusive gluon-fusion cross section at N$^3$LO computed in Section~\ref{sec:VVR_xsecs}. The results are expressed in terms of harmonic polylogarithms up to weight five, and we use the shorthand $H_{i,\ldots,j} \equiv H_{i,\ldots,j}(z)$, where $z=m_H^2/s$. The results for the three different initial states are given in the subsequent sections. In all cases we have factored out the leading-order inclusive cross section,
\beq
\sigma_0 = \frac{\pi \lambda_0^2}{8Vv^2}\,,
\eeq
where $v\simeq 246$ GeV is the vacuum expectation value of the Higgs field.

\subsection{The $gg$ initial state}
The contribution of the two-loop amplitude for $gg\to Hg$ to the inclusive Higgs cross section at N$^3$LO can be written as
\beq
\hat{\sigma}_{gg\to Hg}^{(3)} = 2\,\left(\frac{\alpha_0}{2\pi}\right)^3\,s^{-1-3\eps}\,\sigma_0\,\left(\hat{\sigma}_{gg\to Hg}^{(3),sing} + \hat{\sigma}_{gg\to Hg}^{(3),reg}\right)\,.
\eeq
The first term represents the soft-virtual term, which is entirely determined by the QCD soft current,
\beq
\hat{\sigma}_{gg\to Hg}^{(3),sing} = 
\sum_{k=0}^2\bar{z}^{-1-2(k+1)\eps}\,\frac{\Gamma(-(k+1)\eps)^2}{\Gamma(-2(k+1)\eps)}\,{\bf S}^{(2,k)}_{g}(\eps)\,,
\eeq
where ${\bf S}^{(2,k)}_{g}(\eps)$ is defined in eq.~\eqref{eq:Sg_deg}. The second term is regular as $z\to 1$, and can be written as
\beq
\hat{\sigma}_{gg\to Hg}^{(3),reg} = \sum_{k=-5}^0\eps^k\left[N^3\,A_{gg}^{(k)} + N^2\,N_f\,B_{gg}^{(k)} + N\,N_f^2\,C_{gg}^{(k)}+N_f\,D_{gg}^{(k)}\right]\,,
\eeq
with
\begin{align}
A_{gg}^{(-5)} &= \frac{19}{6}z\left(z^2-z+2\right)\,,\\
B_{gg}^{(-5)} &= 0\,,\\
C_{gg}^{(-5)} &= 0\,,\\
D_{gg}^{(-5)} &=0\,,\\
\nonumber\\
A_{gg}^{(-4)} &= \frac{7\left(z^2-z+1\right)^2}{3(1-z)}\,H_0+9z\left(z^2-z+2\right)\,H_1\\
&+\frac{1}{72}\left(371z^3-965z^2+1039z-297\right)\,,\nonumber\\
B_{gg}^{(-4)} &= \frac{7}{18}z\left(z^2-z+2\right)\,,\\
C_{gg}^{(-4)} &= 0\,,\\
D_{gg}^{(-4)} &=0\,,
\end{align}
\begin{align}
A_{gg}^{(-3)} & = \frac{8\left(z^2-z+1\right)^2}{3(1-z)}\,H_{1,0}-\frac{19\left(z^2-z+1\right)^2}{3(1-z)}\,H_{0,0}+30z\left(z^2-z+2\right)\,H_{1,1}\\
&-\frac{\left(z^2-z+1\right)^2}{9(1-z)}\,H_0+\frac{8\left(z^2-z+1\right)^2}{3(1-z)}\,H_2+\frac{1}{4}\left(69z^3-179z^2+193z-55\right)\,H_1\nonumber\\
&\,-\frac{143}{4}z\left(z^2-z+2\right)\,\zeta_2+\frac{1}{432}\left(5291z^3-16889z^2+15361z-6723\right)\,,\nonumber\\
B_{gg}^{(-3)} & = \frac{4\left(z^2-z+1\right)^2}{9(1-z)}\,H_0+z\left(z^2-z+2\right)\,H_1\\
&+\frac{1}{216}\left(323z^3-473z^2+745z-99\right)\,,\nonumber\\
C_{gg}^{(-3)} &= 0\,,\\
D_{gg}^{(-3)} &=0\,,\\
\nonumber\\
A_{gg}^{(-2)} &= -\frac{8\left(z^2-z+1\right)^2}{3(1-z)}\,H_{2,1}-\frac{8\left(z^2-z+1\right)^2}{3(1-z)}\,H_{1,2}-\frac{16\left(z^2-z+1\right)^2}{3(1-z)}\,H_{1,1,0}\\
&-\frac{40\left(z^2-z+1\right)^2}{3(1-z)}\,H_{2,0}-\frac{40\left(z^2-z+1\right)^2}{3(1-z)}\,H_{1,0,0}+\frac{17\left(z^2-z+1\right)^2}{1-z}\,H_{0,0,0}\nonumber\\
&+116z\left(z^2-z+2\right)\,H_{1,1,1}+\frac{1}{6}\left(389z^3-1027z^2+1097z-319\right)\,H_{1,1}\nonumber\\
&-\frac{10z^4-30z^3+30z^2-10z-1}{1-z}\,H_{0,0}-\frac{95z^4-388z^3+582z^2-388z+95}{18(1-z)}\,H_{1,0}\nonumber\\
&-\frac{287\left(z^2-z+1\right)^2}{6(1-z)}\,\zeta_2\,H_0-\frac{22\left(z^2-z+1\right)^2}{3(1-z)}\,H_3\nonumber\\
&+\frac{1}{72}\left(2915z^3-8969z^2+8361z-3467\right)\,H_1-\frac{97z^4-284z^3+291z^2-104z-2}{9(1-z)}\,H_2\nonumber\\
&+\frac{545z^4-1090z^3+1635z^2-1090z-16}{6(1-z)}\,\zeta_2\,H_1\nonumber\\
&-\frac{274z^4-653z^3+1035z^2-656z+274}{54(1-z)}\,H_0\nonumber\\
&+\frac{57601z^3-149875z^2+147119z-58485}{1296}\nonumber\\
&+\frac{1}{144}\left(-6853z^3+19051z^2-17645z+5379\right)\,\zeta_2\nonumber\\
&+\frac{121z^4-242z^3+363z^2-242z-10}{3(1-z)}\,\zeta_3\,,\nonumber\\
B_{gg}^{(-2)} &=
\frac{4\left(z^2-z+1\right)^2}{9(1-z)}\,H_{1,0}-\frac{4\left(z^2-z+1\right)^2}{3(1-z)}\,H_{0,0}+\frac{10}{3}z\left(z^2-z+2\right)\,H_{1,1}\\
&+\frac{4\left(z^2-z+1\right)^2}{9(1-z)}\,H_2+\frac{1}{36}\left(151z^3-233z^2+357z-55\right)\,H_1\nonumber\\
&+\frac{64z^4-125z^3+189z^2-128z+64}{54(1-z)}\,H_0+\frac{1}{648}\left(1435z^3-4207z^2+4541z-1509\right)\nonumber\\
&-\frac{239}{36}z\left(z^2-z+2\right)\,\zeta_2\nonumber\,,
\end{align}
\begin{align}C_{gg}^{(-2)} &=\frac{z^2}{27}\,,\\
D_{gg}^{(-2)} &=\frac{1}{2}z\left(z^2-z+2\right)\,,\\
\nonumber\\%
A_{gg}^{(-1)} &=
\frac{324799z^3-729868z^2+781709z-239112}{1944}+\frac{14\left(z^2-z+1\right)^2}{3(1-z)}\,H_{3,1}\\
&-\frac{34\left(z^2-z+1\right)^2}{3(1-z)}\,H_{1,3}+\frac{18\left(z^2-z+1\right)^2}{1-z}\,H_{2,0,0}+\frac{20\left(z^2-z+1\right)^2}{1-z}\,H_4\nonumber\\
&-\frac{32\left(z^2-z+1\right)^2}{1-z}\,H_{1,1,0,0}-\frac{98\left(z^2-z+1\right)^2}{3(1-z)}\,H_{2,1,0}-\frac{98\left(z^2-z+1\right)^2}{3(1-z)}\,H_{1,2,0}\nonumber\\
&-\frac{100\left(z^2-z+1\right)^2}{3(1-z)}\,H_{2,2}+\frac{34\left(z^2-z+1\right)^2}{1-z}\,H_{3,0}+\frac{40\left(z^2-z+1\right)^2}{1-z}\,H_{1,0,0,0}\nonumber\\
&-\frac{124\left(z^2-z+1\right)^2}{3(1-z)}\,H_{2,1,1}-\frac{124\left(z^2-z+1\right)^2}{3(1-z)}\,H_{1,2,1}-\frac{45\left(z^2-z+1\right)^2}{1-z}\,H_{0,0,0,0}\nonumber\\
&-\frac{152\left(z^2-z+1\right)^2}{3(1-z)}\,H_{1,1,2}-\frac{208\left(z^2-z+1\right)^2}{3(1-z)}\,H_{1,1,1,0}+504z\left(z^2-z+2\right)\,H_{1,1,1,1}\nonumber\\
&+\frac{1}{864}\left(-101659z^3+339529z^2-262625z+133107\right)\,\zeta_2\nonumber\\
&-\frac{z\left(77z^3-190z^2+195z-60\right)}{18(1-z)}\,H_3+\frac{1}{3}\left(799z^3-2185z^2+2291z-693\right)\,H_{1,1,1}\nonumber\\
&+\frac{1}{36}\left(5559z^3-16881z^2+15857z-6479\right)\,H_{1,1}\nonumber\\
&+\frac{1}{216}\left(27345z^3-74861z^2+71343z-26931\right)\,H_1\nonumber\\
&-\frac{152z^4-610z^3+915z^2-600z+141}{18(1-z)}\,H_{1,0,0}\nonumber\\
&+\frac{179z^4-570z^3+560z^2-180z-30}{6(1-z)}\,H_{0,0,0}\nonumber\\
&-\frac{323z^4-1372z^3+2058z^2-1372z+323}{9(1-z)}\,H_{1,1,0}\nonumber\\
&-\frac{407z^4-1143z^3+956z^2-220z-286}{18(1-z)}\,H_{0,0}\nonumber\\
&-\frac{499z^4-2030z^3+3045z^2-2020z+488}{18(1-z)}\,H_{1,2}\nonumber\\
&-\frac{526z^4-1520z^3+1545z^2-540z-24}{18(1-z)}\,H_{2,0}\nonumber\\
&+\frac{541z^4+1513z^3-2448z^2-614z+1630}{162(1-z)}\,H_0\nonumber\\
&-\frac{607z^4-1784z^3+1821z^2-644z-20}{9(1-z)}\,H_{2,1}\nonumber\\
&-\frac{1597z^4-5681z^3+8427z^2-5684z+1597}{54(1-z)}\,H_{1,0}\nonumber
\end{align}
\begin{align}
\phantom{A_gg^{(-1)}}
&-\frac{2335z^4-6035z^3+6978z^2-3278z+256}{54(1-z)}\,H_2\nonumber\\
&+\frac{3443z^4-12394z^3+17931z^2-11554z+2498}{36(1-z)}\,\zeta_3\nonumber\\
&-\frac{16783z^4-33566z^3+50349z^2-33566z-2976}{96(1-z)}\,\zeta_4-\frac{11\left(z^2-z+1\right)^2}{1-z}\,\zeta_3\,H_0\nonumber\\
&-\frac{184\left(z^2-z+1\right)^2}{3(1-z)}\,\zeta_2\,H_2-\frac{250\left(z^2-z+1\right)^2}{3(1-z)}\,\zeta_2\,H_{1,0}+\frac{265\left(z^2-z+1\right)^2}{2(1-z)}\,\zeta_2\,H_{0,0}\nonumber\\
&+\frac{7\left(121z^4-242z^3+363z^2-242z-8\right)}{3(1-z)}\,\zeta_2\,H_{1,1}\nonumber\\
&+\frac{2\left(233z^4-466z^3+699z^2-466z-19\right)}{3(1-z)}\,\zeta_3\,H_1\nonumber\\
&-\frac{331z^4-1010z^3+960z^2-270z-87}{6(1-z)}\,\zeta_2\,H_0\nonumber\\
&+\frac{9059z^4-36400z^3+54600z^2-36400z+9389}{72(1-z)}\,\zeta_2\,H_1\,,\nonumber\\
B_{gg}^{(-1)} &= -\frac{8\left(z^2-z+1\right)^2}{9(1-z)}\,H_{2,1}-\frac{16\left(z^2-z+1\right)^2}{9(1-z)}\,H_{1,1,0}+\frac{44}{3}z\left(z^2-z+2\right)\,H_{1,1,1}\nonumber\\
&+\frac{1}{18}\left(281z^3-471z^2+683z-121\right)\,H_{1,1}-\frac{2z^4-133z^3+195z^2-130z+2}{54(1-z)}\,H_{1,0}\nonumber\\
&-\frac{14z^4-34z^3+51z^2-35z+16}{18(1-z)}\,H_{1,2}+\frac{26z^4-49z^3+73z^2-48z+24}{6(1-z)}\,H_{0,0,0}\nonumber\\
&-\frac{44z^4-91z^3+138z^2-93z+48}{18(1-z)}\,H_{2,0}-\frac{46z^4-98z^3+147z^2-99z+48}{18(1-z)}\,H_{1,0,0}\nonumber\\
&-\frac{108z^4-251z^3+319z^2-176z+64}{18(1-z)}\,H_{0,0}+\frac{1}{36}\left(299z^3-772z^2+889z-260\right)\,H_1\nonumber\\
&-\frac{64z^4-131z^3+198z^2-133z+68}{6(1-z)}\,\zeta_2\,H_0\nonumber\\
&+\frac{301z^4-602z^3+903z^2-602z-20}{18(1-z)}\,\zeta_2\,H_1-\frac{22z^4-50z^3+75z^2-51z+24}{18(1-z)}\,H_3\nonumber\\
&-\frac{68z^4-259z^3+213z^2-22z-64}{54(1-z)}\,H_2\nonumber\\
&-\frac{2\left(94z^4-221z^3+342z^2-224z+103\right)}{81(1-z)}\,H_0\nonumber\\
&+\frac{1}{432}\left(-11707z^3+18121z^2-25601z+3003\right)\,\zeta_2\nonumber\\
&+\frac{-16744z^3-8897z^2-20960z-11313}{1944}+\frac{161z^4-304z^3+453z^2-298z-16}{18(1-z)}\,\zeta_3\,,\nonumber\\
C_{gg}^{(-1)} &= \frac{4z^2}{27}\,H_1+\frac{1}{162}\left(-9z^3+44z^2-3z+12\right)\,,\\
D_{gg}^{(-1)} &= \frac{\left(z^2-z+1\right)^2}{1-z}\,H_0+z\left(z^2-z+2\right)\,H_1-4z\left(z^2-z+2\right)\,\zeta_3\\
&+\frac{1}{24}\left(157z^3-181z^2+325z-11\right)\,,\nonumber
\end{align}
\begin{align}
%
%
A_{gg}^{(0)} &= 
\frac{6917536z^3-14883745z^2+16560434z-3796749}{11664}\\
&-\frac{4\left(z^2-z+1\right)^2}{1-z}\,H_{4,1}-\frac{32\left(z^2-z+1\right)^2}{3(1-z)}\,H_{1,1,3}+\frac{94\left(z^2-z+1\right)^2}{3(1-z)}\,H_{2,2,0}\nonumber\\
&+\frac{32\left(z^2-z+1\right)^2}{1-z}\,H_{2,1,0,0}+\frac{104\left(z^2-z+1\right)^2}{3(1-z)}\,H_{2,3}-\frac{36\left(z^2-z+1\right)^2}{1-z}\,H_{2,0,0,0}\nonumber\\
&+\frac{110\left(z^2-z+1\right)^2}{3(1-z)}\,H_{1,3,1}+\frac{40\left(z^2-z+1\right)^2}{1-z}\,H_{1,4}-\frac{48\left(z^2-z+1\right)^2}{1-z}\,H_{3,0,0}\nonumber\\
&-\frac{54\left(z^2-z+1\right)^2}{1-z}\,H_5+\frac{80\left(z^2-z+1\right)^2}{1-z}\,H_{3,1,0}+\frac{82\left(z^2-z+1\right)^2}{1-z}\,H_{3,2}\nonumber\\
&-\frac{84\left(z^2-z+1\right)^2}{1-z}\,H_{4,0}+\frac{88\left(z^2-z+1\right)^2}{1-z}\,H_{1,3,0}+\frac{278\left(z^2-z+1\right)^2}{3(1-z)}\,H_{3,1,1}\nonumber\\
&-\frac{286\left(z^2-z+1\right)^2}{3(1-z)}\,H_{2,1,2}-\frac{96\left(z^2-z+1\right)^2}{1-z}\,H_{1,2,2}-\frac{292\left(z^2-z+1\right)^2}{3(1-z)}\,H_{2,2,1}\nonumber\\
&-\frac{296\left(z^2-z+1\right)^2}{3(1-z)}\,H_{1,1,1,0,0}-\frac{102\left(z^2-z+1\right)^2}{1-z}\,H_{1,1,2,0}-\frac{308\left(z^2-z+1\right)^2}{3(1-z)}\,H_{2,1,1,0}\nonumber\\
&-\frac{308\left(z^2-z+1\right)^2}{3(1-z)}\,H_{1,2,1,0}-\frac{108\left(z^2-z+1\right)^2}{1-z}\,H_{1,0,0,0,0}+\frac{114\left(z^2-z+1\right)^2}{1-z}\,H_{1,1,0,0,0}\nonumber\\
&+\frac{117\left(z^2-z+1\right)^2}{1-z}\,H_{0,0,0,0,0}-\frac{776\left(z^2-z+1\right)^2}{3(1-z)}\,H_{2,1,1,1}-\frac{776\left(z^2-z+1\right)^2}{3(1-z)}\,H_{1,2,1,1}\nonumber\\
&-\frac{880\left(z^2-z+1\right)^2}{3(1-z)}\,H_{1,1,2,1}-\frac{360\left(z^2-z+1\right)^2}{1-z}\,H_{1,1,1,2}-\frac{1408\left(z^2-z+1\right)^2}{3(1-z)}\,H_{1,1,1,1,0}\nonumber\\
&+2384z\left(z^2-z+2\right)\,H_{1,1,1,1,1}+\frac{2}{3}\left(1757z^3-5035z^2+5153z-1639\right)\,H_{1,1,1,1}\nonumber\\
&+\frac{1}{9}\left(5835z^3-17949z^2+16690z-6991\right)\,H_{1,1,1}+\frac{94\left(z^2-z+1\right)^2}{3(1-z)}\,H_{1,2,0,0}\nonumber\\
&-\frac{67z^4-236z^3+224z^2-66z-21}{1-z}\,H_{0,0,0,0}-\frac{2\left(73z^4-56z^3+84z^2-46z+62\right)}{9(1-z)}\,H_{1,3}\nonumber\\
&+\frac{1}{216}\left(96z^4+88650z^3-222173z^2+221368z-61997\right)\,H_1\nonumber\\
&+\frac{115z^4-512z^3+213z^2+228z-336}{18(1-z)}\,H_{2,0,0}\nonumber\\
&+\frac{170z^4-1834z^3+2751z^2-1824z+192}{18(1-z)}\,H_{1,0,0,0}\nonumber\\
&-\frac{287z^4-831z^3+889z^2-356z+34}{3(1-z)}\,H_{2,2}\nonumber\\
&+\frac{515z^4-1648z^3+1617z^2-528z-90}{9(1-z)}\,H_{3,1}\nonumber\\
&+\frac{569z^4-1654z^3+1671z^2-564z-36}{18(1-z)}\,H_4\nonumber\\
&-\frac{2\left(637z^4-2504z^3+3756z^2-2494z+626\right)}{9(1-z)}\,H_{1,2,1}\nonumber
\end{align}
\begin{align}
\phantom{A_gg^{(0)}}
&-\frac{730z^4-2462z^3+3693z^2-2412z+675}{18(1-z)}\,H_{1,1,0,0}\nonumber\\
&-\frac{797z^4-2674z^3+4011z^2-2684z+775}{18(1-z)}\,H_{1,2,0}\nonumber\\
&+\frac{889z^4-2886z^3+1911z^2-114z-1788}{36(1-z)}\,H_{0,0,0}\nonumber\\
&+\frac{1415z^4-4246z^3+4209z^2-1356z-180}{18(1-z)}\,H_{3,0}\nonumber\\
&+\frac{1586z^4-3172z^3+4758z^2-3172z-5225}{15(1-z)}\,\zeta_5\nonumber\\
&-\frac{2051z^4-5686z^3+5919z^2-2196z+60}{18(1-z)}\,H_{2,1,0}\nonumber\\
&-\frac{2939z^4-16984z^3+27480z^2-21564z+5169}{108(1-z)}\,H_{1,0,0}\nonumber\\
&-\frac{3079z^4-12596z^3+18894z^2-12556z+3035}{18(1-z)}\,H_{1,1,2}\nonumber\\
&-\frac{3199z^4-9368z^3+9597z^2-3428z-68}{9(1-z)}\,H_{2,1,1}\nonumber\\
&-\frac{3461z^4-9748z^3+8103z^2-1416z-1608}{108(1-z)}\,H_3\nonumber\\
&-\frac{3657z^4-16002z^3+24003z^2-15992z+3646}{18(1-z)}\,H_{1,1,1,0}\nonumber\\
&-\frac{4006z^4-16085z^3+23490z^2-16073z+4006}{27(1-z)}\,H_{1,1,0}\nonumber\\
&-\frac{5006z^4-13312z^3+12237z^2-3831z-1680}{54(1-z)}\,H_{2,0}\nonumber\\
&-\frac{12580z^4-32927z^3+35517z^2-15170z-656}{54(1-z)}\,H_{2,1}\nonumber\\
&-\frac{14311z^4-56954z^3+86754z^2-61564z+16541}{108(1-z)}\,H_{1,2}\nonumber\\
&+\frac{117637z^4-412340z^3+574362z^2-369668z+90169}{432(1-z)}\,\zeta_3\nonumber\\
&+\frac{148347z^4-240328z^3+138972z^2+57432z-92695}{384(1-z)}\,\zeta_4\nonumber\\
&+\frac{1}{108}\left(-48z^5-495z^4+48370z^3-137141z^2+126868z-47810\right)\,H_{1,1}\nonumber\\
&-\frac{216z^5-54056z^4+88189z^3-113004z^2+84694z-51428}{486(1-z)}\,H_0\nonumber\\
&+\frac{48z^6+447z^5-11754z^4+25186z^3-28256z^2+17101z-2540}{108(1-z)}\,H_{0,0}\nonumber\\
&+\frac{144z^6+1341z^5-32230z^4+88754z^3-104124z^2+47015z+4340}{324(1-z)}\,H_2\nonumber\\
&+\frac{144z^6+1341z^5-23266z^4+95600z^3-136668z^2+86531z-16354}{324(1-z)}\,H_{1,0}\nonumber
\end{align}
\begin{align}
\phantom{A_gg^{(-1)}}
&-\frac{3456z^6+32184z^5-1450121z^4+4685548z^3-6395130z^2+4498564z-1391205}{2592(1-z)}\,\zeta_2\nonumber\\
&+\frac{70\left(z^2-z+1\right)^2}{3(1-z)}\,\zeta_3\,H_{1,0}+\frac{89\left(z^2-z+1\right)^2}{3(1-z)}\,\zeta_3\,H_{0,0}-\frac{116\left(z^2-z+1\right)^2}{3(1-z)}\,\zeta_2\,H_{2,1}\nonumber\\
&-\frac{116\left(z^2-z+1\right)^2}{3(1-z)}\,\zeta_2\,H_{1,2}+\frac{154\left(z^2-z+1\right)^2}{3(1-z)}\,\zeta_3\,H_2-\frac{310\left(z^2-z+1\right)^2}{3(1-z)}\,\zeta_2\,H_{1,1,0}\nonumber\\
&+\frac{7463\left(z^2-z+1\right)^2}{48(1-z)}\,\zeta_4\,H_0+\frac{168\left(z^2-z+1\right)^2}{1-z}\,\zeta_2\,H_{2,0}+\frac{169\left(z^2-z+1\right)^2}{1-z}\,\zeta_2\,H_3\nonumber\\
&+\frac{318\left(z^2-z+1\right)^2}{1-z}\,\zeta_2\,H_{1,0,0}-\frac{729\left(z^2-z+1\right)^2}{2(1-z)}\,\zeta_2\,H_{0,0,0}\nonumber\\
&+\frac{4\left(171z^4-342z^3+513z^2-342z-13\right)}{1-z}\,\zeta_3\,H_{1,1}\nonumber\\
&+\frac{583z^4-1772z^3+1098z^2+388z-770}{18(1-z)}\,\zeta_3\,H_0\nonumber\\
&+\frac{667z^4-2018z^3+2847z^2-1848z+810}{18(1-z)}\,\zeta_2\,H_2\nonumber\\
&+\frac{1037z^4-1024z^3+1536z^2-1164z+1389}{36(1-z)}\,\zeta_2\,H_{1,0}\nonumber\\
&+\frac{1433z^4-4492z^3+4328z^2-1302z-327}{6(1-z)}\,\zeta_2\,H_{0,0}\nonumber\\
&-\frac{1549z^4-3098z^3+4647z^2-3098z+141}{3(1-z)}\,\zeta_2\,\zeta_3\nonumber\\
&+\frac{2\left(1657z^4-3314z^3+4971z^2-3314z-164\right)}{3(1-z)}\,\zeta_2\,H_{1,1,1}\nonumber\\
&-\frac{3016z^4-8611z^3+4033z^2+1762z-5078}{36(1-z)}\,\zeta_2\,H_0\nonumber\\
&+\frac{6008z^4-23614z^3+35421z^2-23634z+5745}{18(1-z)}\,\zeta_3\,H_1\nonumber\\
&-\frac{11359z^4-22718z^3+34077z^2-22718z-7040}{48(1-z)}\,\zeta_4\,H_1\nonumber\\
&+\frac{17457z^4-70956z^3+106434z^2-71096z+19249}{36(1-z)}\,\zeta_2\,H_{1,1}\nonumber\\
&+\frac{128729z^4-615292z^3+899310z^2-632356z+226457}{432(1-z)}\,\zeta_2\,H_1\,,\nonumber\\
B_{gg}^{(0)}&=\frac{-9697813z^3+6173047z^2-15914939z-2575755}{116640}\\
&-\frac{80\left(z^2-z+1\right)^2}{9(1-z)}\,H_{2,1,1}+\frac{232}{3}z\left(z^2-z+2\right)\,H_{1,1,1,1}\nonumber\\
&+\frac{1}{9}\left(671z^3-1206z^2+1661z-319\right)\,H_{1,1,1}\nonumber\\
&+\frac{-3168z^4-11889z^3-411311z^2+373833z-287025}{6480}\,H_1\nonumber\\
&+\frac{2\left(10z^4-14z^3+21z^2-13z+8\right)}{9(1-z)}\,H_{1,3}-\frac{14z^4-25z^3+37z^2-24z+12}{1-z}\,H_{0,0,0,0}\nonumber
\end{align}
\begin{align}
\phantom{A_gg^{(-1)}}
&+\frac{2\left(14z^4-22z^3+33z^2-21z+12\right)}{9(1-z)}\,H_{3,1}-\frac{28z^4-49z^3+73z^2-48z+24}{6(1-z)}\,H_{2,2}\nonumber\\
&-\frac{2\left(32z^4-70z^3+105z^2-71z+34\right)}{9(1-z)}\,H_{1,2,1}-\frac{58z^4-110z^3+165z^2-109z+62}{18(1-z)}\,H_{1,2,0}\nonumber\\
&-\frac{68z^4-154z^3+231z^2-159z+78}{18(1-z)}\,H_{1,1,0,0}-\frac{70z^4-164z^3+255z^2-177z+96}{18(1-z)}\,H_{2,1,0}\nonumber\\
&+\frac{70z^4-146z^3+219z^2-147z+72}{18(1-z)}\,H_4-\frac{97z^4-200z^3+300z^2-202z+101}{9(1-z)}\,H_{1,1,2}\nonumber\\
&+\frac{110z^4-235z^3+354z^2-237z+120}{18(1-z)}\,H_{2,0,0}+\frac{121z^4-431z^3+222z^2+120z-192}{54(1-z)}\,H_3\nonumber\\
&+\frac{142z^4-290z^3+435z^2-291z+144}{18(1-z)}\,H_{3,0}\nonumber\\
&+\frac{238z^4-482z^3+723z^2-483z+234}{18(1-z)}\,H_{1,0,0,0}\nonumber\\
&-\frac{269z^4-13z^3-1113z^2+1896z-687}{54(1-z)}\,H_{1,0,0}\nonumber\\
&-\frac{277z^4-566z^3-294z^2+1358z-679}{54(1-z)}\,H_{1,2}\nonumber\\
&-\frac{354z^4-714z^3+1071z^2-715z+356}{18(1-z)}\,H_{1,1,1,0}\nonumber\\
&-\frac{392z^4-1285z^3+1926z^2-1297z+392}{27(1-z)}\,H_{1,1,0}\nonumber\\
&-\frac{854z^4-2389z^3+2643z^2-1108z+128}{54(1-z)}\,H_{2,1}\nonumber\\
&-\frac{968z^4-2287z^3+3147z^2-1860z+768}{108(1-z)}\,H_{2,0}\nonumber\\
&+\frac{1010z^4-2427z^3+2829z^2-1380z+384}{36(1-z)}\,H_{0,0,0}\nonumber\\
&-\frac{6405z^4-12286z^3+18219z^2-11946z-400}{96(1-z)}\,\zeta_4\nonumber\\
&+\frac{9409z^4-26048z^3+43110z^2-33572z+6973}{216(1-z)}\,\zeta_3\nonumber\\
&+\frac{1}{540}\left(264z^5+900z^4+23560z^3-60085z^2+68650z-18689\right)\,H_{1,1}\nonumber\\
&+\frac{9504z^5-867377z^4+2107411z^3-3141027z^2+1899229z-749000}{19440(1-z)}\,H_0\nonumber\\
&-\frac{528z^6+1272z^5+11805z^4-30200z^3+3475z^2+15820z-8240}{1080(1-z)}\,H_{0,0}\nonumber\\
&-\frac{1584z^6+3816z^5+48155z^4-184075z^3+275505z^2-177889z+54824}{3240(1-z)}\,H_{1,0}\nonumber\\
&-\frac{1584z^6+3816z^5+63740z^4-159985z^3+122820z^2-26755z+8240}{3240(1-z)}\,H_2\nonumber\\
&+\frac{9504z^6+22896z^5+363505z^4-1459070z^3+1999530z^2-1225646z+272361}{6480(1-z)}\,\zeta_2\nonumber
\end{align}
\begin{align}
\phantom{A_gg^{(-1)}}
&-\frac{47z^4-52z^3+72z^2-40z+20}{9(1-z)}\,\zeta_3\,H_0+\frac{202z^4-395z^3+595z^2-396z+204}{6(1-z)}\,\zeta_2\,H_{0,0}\nonumber\\
&-\frac{268z^4-635z^3+975z^2-672z+348}{18(1-z)}\,\zeta_2\,H_2+\frac{308z^4-616z^3+924z^2-615z-15}{9(1-z)}\,\zeta_3\,H_1\nonumber\\
&-\frac{430z^4-902z^3+1353z^2-909z+462}{18(1-z)}\,\zeta_2\,H_{1,0}\nonumber\\
&-\frac{465z^4-1128z^3+1421z^2-766z+272}{9(1-z)}\,\zeta_2\,H_0\nonumber\\
&+\frac{954z^4-1854z^3+2781z^2-1847z-194}{18(1-z)}\,\zeta_2\,H_{1,1}\nonumber\\
&+\frac{5\left(2857z^4-7088z^3+9834z^2-5756z+25\right)}{216(1-z)}\,\zeta_2\,H_1\,,\nonumber\\
C_{gg}^{(0)} &=\frac{16z^2}{27}\,H_{1,1}+\frac{1}{81}\left(-9z^3+79z^2+3z+15\right)\,H_1-\frac{z^2}{6}\,\zeta_2\\
&+\frac{1}{162}\left(-81z^3+230z^2-67z+112\right)\,,\nonumber\\
D_{gg}^{(0)} &= \frac{1}{45}\left(2z^5-30z^4+165z^3-140z^2+210z-27\right)\,H_{1,1}\\
&-\frac{2z^6-32z^5+15z^4+55z^3-190z^2+123z-63}{45(1-z)}\,H_{1,0}\nonumber\\
&-\frac{4z^6-64z^5+390z^4-525z^3+720z^2-540z+270}{90(1-z)}\,H_{0,0}\nonumber
\end{align}
\begin{align}
\phantom{A_{gg}^{(0)}}
&-\frac{1}{3}z(4z-3)\,H_{1,0,0}-\frac{1}{3}z(4z-3)\,H_{1,2}-\frac{8\left(z^2-z+1\right)^2}{1-z}\,\zeta_3\,H_0-8z\left(z^2-z+2\right)\,\zeta_3\,H_1\nonumber\\
&+\frac{1}{180}\left(-8z^4+2201z^3-2796z^2+5028z-255\right)\,H_1\nonumber\\
&+\frac{8z^5+2501z^4-4843z^3+6921z^2-4632z+2370}{180(1-z)}\,H_0\nonumber\\
&-\frac{4z^6-64z^5+30z^4+195z^3-360z^2+180z-90}{90(1-z)}\,H_2+\frac{1}{3}z(4z-3)\,\zeta_2\,H_1\nonumber\\
&-\frac{20}{3}z\left(z^2-z+2\right)\,\zeta_4+\frac{1}{3}\left(-69z^3+86z^2-146z+11\right)\,\zeta_3\nonumber\\
&+\frac{1}{720}\left(29441z^3-39619z^2+64213z-5225\right)\nonumber\\
&+\frac{24z^6-384z^5+2685z^4-4210z^3+5305z^2-3666z+396}{180(1-z)}\,\zeta_2\,.\nonumber
\end{align}


\subsection{The $gq$ initial state}
The contribution of the two-loop amplitude for $gq\to Hq$ to the inclusive Higgs cross section at N$^3$LO can be written as
\beq
\hat{\sigma}_{gq\to Hq}^{(3)} = 2\,\left(\frac{\alpha_0}{2\pi}\right)^3\,s^{-1-3\eps}\,\sigma_0\,\hat{\sigma}_{gq\to Hq}^{(3)}\,.
\eeq
Note that, unlike for the pure-gluon initial state, the result is regular in the limit $z\to1$, and so we do not need to separate off the contribution from the soft limit. The coefficient $\hat{\sigma}_{gq\to Hq}^{(3)}$ can be written as
\beq\bsp
\hat{\sigma}_{gq\to Hq}^{(3)} = \sum_{k=-5}^0\eps^k\Bigg[&\,N^3\,A_{gq}^{(k)} + N^2\,N_f\,B_{gq}^{(k)} + N\,N_f^2\,C_{gq}^{(k)}+N_f\,D_{gq}^{(k)} + N\,E_{gq}^{(k)}\\
&\,+ \frac{N_f^2}{N}\,F_{gq}^{(k)}+ \frac{1}{N}\,G_{gq}^{(k)} + \frac{N_f}{N^2}\,H_{gq}^{(k)} 
+ \frac{1}{N^3}\,I_{gq}^{(k)}\Bigg]\,,
\esp\eeq
with
\begin{align}
A_{gq}^{(-5)} &= -\frac{1}{4}\left(z^2-2z+2\right)\,,\\
B_{gq}^{(-5)} &= 0\,,\\
C_{gq}^{(-5)} &= 0\,,\\
D_{gq}^{(-5)} &=0\,,\\
E_{gq}^{(-5)} &=\frac{3}{8}\left(z^2-2z+2\right)\,,\\
F_{gq}^{(-5)} &=0\,,\\
G_{gq}^{(-5)} &=-\frac{7}{48}\left(z^2-2z+2\right)\,,\\
H_{gq}^{(-5)} &=0\,,\\
I_{gq}^{(-5)} &=\frac{1}{48}\left(z^2-2z+2\right)\,,
\end{align}
\begin{align}
A_{gq}^{(-4)} &= \frac{1}{4}\left(z^2-2z+2\right)\,H_0-\frac{5}{8}\left(z^2-2z+2\right)\,H_1+\frac{1}{8}(z+1)^2\,,\\
B_{gq}^{(-4)} &= -\frac{1}{8}\left(z^2-2z+2\right)\,,\\
C_{gq}^{(-4)} &= 0\,,\\
D_{gq}^{(-4)} &=\frac{23}{144}\left(z^2-2z+2\right)\,,\\
E_{gq}^{(-4)} &=-\frac{7}{24}\left(z^2-2z+2\right)\,H_0+\frac{25}{24}\left(z^2-2z+2\right)\,H_1\\
&+\frac{1}{288}\left(89z^2-394z+178\right)\,,\nonumber\\
F_{gq}^{(-4)} &=0\,,\\
G_{gq}^{(-4)} &=-\frac{1}{2}\left(z^2-2z+2\right)\,H_1+\frac{1}{24}\left(z^2-2z+2\right)\,H_0\\
&+\frac{1}{288}\left(-170z^2+424z-289\right)\,,\nonumber\\
H_{gq}^{(-4)} &=-\frac{5}{144}\left(z^2-2z+2\right)\,,\\
I_{gq}^{(-4)} &=\frac{1}{12}\left(z^2-2z+2\right)\,H_1+\frac{1}{96}\left(15z^2-34z+25\right)\,,
\end{align}
%
%
\begin{align}
A_{gq}^{(-3)} &= \frac{7}{24}\left(z^2-2z+2\right)\,H_{1,0}-\frac{2}{3}\left(z^2-2z+2\right)\,H_{0,0}-\frac{23}{12}\left(z^2-2z+2\right)\,H_{1,1}\\
&+\frac{1}{48}\left(-19z^2+14z-14\right)\,H_0+\frac{7}{24}\left(z^2-2z+2\right)\,H_2+\frac{1}{16}\left(5z^2+10z+8\right)\,H_1\nonumber\\
&+\frac{27}{8}\left(z^2-2z+2\right)\,\zeta_2+\frac{1471z^2-2942z+2780}{1728}\,,\nonumber\\
B_{gq}^{(-3)} &= \frac{1}{12}\left(z^2-2z+2\right)\,H_0-\frac{3}{8}\left(z^2-2z+2\right)\,H_1\\
&+\frac{1}{432}\left(-164z^2+436z-301\right)\,,\nonumber\\
C_{gq}^{(-3)} &= -\frac{1}{54}\left(z^2-2z+2\right)\,,\\
D_{gq}^{(-3)} &= -\frac{1}{12}\left(z^2-2z+2\right)\,H_0+\frac{37}{72}\left(z^2-2z+2\right)\,H_1\\
&+\frac{1}{288}\left(189z^2-470z+335\right)\,,\nonumber\\
E_{gq}^{(-3)} &=-\frac{1}{3}\left(z^2-2z+2\right)\,H_{1,0}+\frac{19}{24}\left(z^2-2z+2\right)\,H_{0,0}+\frac{41}{12}\left(z^2-2z+2\right)\,H_{1,1}\\
&-\frac{1}{3}\left(z^2-2z+2\right)\,H_2+\frac{1}{12}\left(4z^2-z+1\right)\,H_0+\frac{1}{144}\left(160z^2-620z+269\right)\,H_1\nonumber\\
&-\frac{35}{8}\left(z^2-2z+2\right)\,\zeta_2+\frac{802z^2-3104z+971}{1728}\,,\nonumber\\
F_{gq}^{(-3)} &=\frac{1}{54}\left(z^2-2z+2\right)\,,\\
G_{gq}^{(-3)} &=-\frac{1}{8}\left(z^2-2z+2\right)\,H_{0,0}+\frac{1}{24}\left(z^2-2z+2\right)\,H_{1,0}-\frac{11}{6}\left(z^2-2z+2\right)\,H_{1,1}\\
&+\frac{1}{144}\left(-295z^2+734z-491\right)\,H_1+\frac{1}{24}\left(z^2-2z+2\right)\,H_2\nonumber\\
&+\frac{1}{48}\left(3z^2-10z+10\right)\,H_0+\frac{35}{32}\left(z^2-2z+2\right)\,\zeta_2+\frac{-3830z^2+9664z-6181}{1728}\nonumber\,,\\
H_{gq}^{(-3)} &=\frac{1}{864}\left(-239z^2+538z-403\right)-\frac{5}{36}\left(z^2-2z+2\right)\,H_1\,,\\
I_{gq}^{(-3)} &=\frac{1}{3}\left(z^2-2z+2\right)\,H_{1,1}+\frac{1}{24}\left(15z^2-34z+25\right)\,H_1-\frac{3}{32}\left(z^2-2z+2\right)\,\zeta_2\\
&+\frac{1}{192}\left(173z^2-402z+270\right)\,,\nonumber\\
\nonumber\\
A_{gq}^{(-2)} &= -\frac{1}{48}\left(55z^2-82z+46\right)\,H_{1,0}-\frac{1}{8}\left(z^2-2z+2\right)\,H_{2,1}\\
&+\frac{1}{8}\left(z^2-2z+2\right)\,H_{1,2}-\frac{3}{8}\left(z^2-2z+2\right)\,H_{1,1,0}-\frac{25}{24}\left(z^2-2z+2\right)\,H_{1,0,0}\nonumber\\
&-\frac{31}{24}\left(z^2-2z+2\right)\,H_{2,0}+\frac{7}{4}\left(z^2-2z+2\right)\,H_{0,0,0}-\frac{29}{4}\left(z^2-2z+2\right)\,H_{1,1,1}\nonumber\\
&+\frac{1}{48}\left(3z^2+10z+50\right)\,H_{0,0}+\frac{1}{144}\left(89z^2+374z+247\right)\,H_{1,1}\nonumber\\
&-\frac{127}{24}\left(z^2-2z+2\right)\,\zeta_2\,H_0+\frac{347}{48}\left(z^2-2z+2\right)\,\zeta_2\,H_1+\frac{1}{48}\left(-73z^2+70z-10\right)\,H_2\nonumber\\
&+\frac{1}{72}\left(-67z^2+146z-137\right)\,H_0-\frac{19}{24}\left(z^2-2z+2\right)\,H_3\nonumber\\
&+\frac{1}{864}\left(2357z^2-4930z+4201\right)\,H_1+\frac{1}{16}\left(-16z^2-60z-29\right)\,\zeta_2\nonumber
\end{align}
\begin{align}
&+\frac{23}{8}\left(z^2-2z+2\right)\,\zeta_3+\frac{18356z^2-52852z+29659}{10368}\nonumber\,,\\
B_{gq}^{(-2)} &= -\frac{1}{4}\left(z^2-2z+2\right)\,H_{0,0}+\frac{1}{12}\left(z^2-2z+2\right)\,H_{1,0}-\frac{47}{36}\left(z^2-2z+2\right)\,H_{1,1}\\
&+\frac{1}{432}\left(-485z^2+1294z-835\right)\,H_1+\frac{1}{12}\left(z^2-2z+2\right)\,H_2\nonumber\\
&+\frac{1}{36}\left(5z^2-16z+16\right)\,H_0+\frac{23}{16}\left(z^2-2z+2\right)\,\zeta_2+\frac{-1135z^2+4346z-1862}{2592}\,,\nonumber\\
C_{gq}^{(-2)} &= \frac{1}{324}\left(-47z^2+106z-79\right)-\frac{2}{27}\left(z^2-2z+2\right)\,H_1\,,\\
D_{gq}^{(-2)} &= -\frac{1}{12}\left(z^2-2z+2\right)\,H_{1,0}+\frac{1}{4}\left(z^2-2z+2\right)\,H_{0,0}+\frac{67}{36}\left(z^2-2z+2\right)\,H_{1,1}\\
&+\frac{1}{36}\left(-5z^2+16z-16\right)\,H_0-\frac{1}{12}\left(z^2-2z+2\right)\,H_2\nonumber\\
&+\frac{1}{144}\left(321z^2-790z+547\right)\,H_1-\frac{455}{288}\left(z^2-2z+2\right)\,\zeta_2\nonumber\\
&+\frac{5123z^2-13666z+8023}{2592}\,,\nonumber\\
E_{gq}^{(-2)} &=\frac{5}{24}\left(z^2-2z+2\right)\,H_{1,2}+\frac{1}{3}\left(z^2-2z+2\right)\,H_{2,1}+\frac{2}{3}\left(z^2-2z+2\right)\,H_{1,1,0}\\
&+\frac{37}{24}\left(z^2-2z+2\right)\,H_{1,0,0}+\frac{5}{3}\left(z^2-2z+2\right)\,H_{2,0}-\frac{17}{8}\left(z^2-2z+2\right)\,H_{0,0,0}\nonumber\\
&+\frac{79}{6}\left(z^2-2z+2\right)\,H_{1,1,1}+\frac{1}{24}\left(35z^2-54z+27\right)\,H_{1,0}\nonumber\\
&+\frac{1}{48}\left(51z^2-82z-20\right)\,H_{0,0}+\frac{1}{72}\left(334z^2-1160z+485\right)\,H_{1,1}\nonumber\\
&+\frac{287}{48}\left(z^2-2z+2\right)\,\zeta_2\,H_0-\frac{521}{48}\left(z^2-2z+2\right)\,\zeta_2\,H_1+\frac{11}{12}\left(z^2-2z+2\right)\,H_3\nonumber\\
&+\frac{1}{144}\left(101z^2-205z+196\right)\,H_0+\frac{1}{432}\left(316z^2-1763z+467\right)\,H_1\nonumber\\
&+\frac{1}{48}z(115z-102)\,H_2+\frac{1}{576}\left(-2159z^2+8206z-2824\right)\,\zeta_2-\frac{75}{16}\left(z^2-2z+2\right)\,\zeta_3\nonumber\\
&+\frac{1778z^2-7064z+4481}{3456}\,,\nonumber\\
F_{gq}^{(-2)} &=\frac{2}{27}\left(z^2-2z+2\right)\,H_1+\frac{1}{324}\left(47z^2-106z+79\right)\,,\\
G_{gq}^{(-2)} &=\frac{1}{144}\left(-1117z^2+2762z-1817\right)\,H_{1,1}+\frac{1}{16}\left(-21z^2+26z-10\right)\,H_{0,0}\\
&-\frac{5}{24}\left(z^2-2z+2\right)\,H_{2,1}-\frac{7}{24}\left(z^2-2z+2\right)\,H_{1,1,0}+\frac{3}{8}\left(z^2-2z+2\right)\,H_{0,0,0}\nonumber\\
&-\frac{3}{8}\left(z^2-2z+2\right)\,H_{2,0}-\frac{11}{24}\left(z^2-2z+2\right)\,H_{1,2}-\frac{5}{8}\left(z^2-2z+2\right)\,H_{1,0,0}\nonumber\\
&-\frac{29}{4}\left(z^2-2z+2\right)\,H_{1,1,1}-\frac{1}{48}(3z-4)(5z-2)\,H_{1,0}-\frac{11}{16}\left(z^2-2z+2\right)\,\zeta_2\,H_0\nonumber\\
&+\frac{33}{8}\left(z^2-2z+2\right)\,\zeta_2\,H_1+\frac{1}{432}\left(-2957z^2+7684z-4930\right)\,H_1\nonumber\\
&+\frac{1}{48}\left(-51z^2+38z+10\right)\,H_2+\frac{1}{8}\left(-z^2+2z-2\right)\,H_3+\frac{1}{24}\left(7z^2-16z+13\right)\,H_0\nonumber\\
&+\frac{13}{6}\left(z^2-2z+2\right)\,\zeta_3+\frac{1}{576}\left(3212z^2-6964z+4471\right)\,\zeta_2\nonumber\\
&+\frac{-67970z^2+179128z-109873}{10368}\,,\nonumber
\end{align}
\begin{align}
H_{gq}^{(-2)} &=-\frac{5}{9}\left(z^2-2z+2\right)\,H_{1,1}+\frac{1}{216}\left(-239z^2+538z-403\right)\,H_1\\
&+\frac{41}{288}\left(z^2-2z+2\right)\,\zeta_2+\frac{-3988z^2+9320z-6161}{2592}\,,\nonumber\\
I_{gq}^{(-2)} &=\frac{1}{8}\left(z^2-2z+2\right)\,H_{1,0,0}+\frac{1}{8}\left(z^2-2z+2\right)\,H_{1,2}+\frac{4}{3}\left(z^2-2z+2\right)\,H_{1,1,1}\\
&+\frac{1}{6}\left(15z^2-34z+25\right)\,H_{1,1}+\frac{1}{16}z(3z-2)\,H_{0,0}-\frac{1}{2}\left(z^2-2z+2\right)\,\zeta_2\,H_1\nonumber\\
&+\frac{1}{96}\left(325z^2-768z+525\right)\,H_1+\frac{1}{16}(1-z)z\,H_0+\frac{1}{16}z(3z-2)\,H_2\nonumber\\
&+\frac{1}{64}\left(-53z^2+102z-67\right)\,\zeta_2-\frac{17}{48}\left(z^2-2z+2\right)\,\zeta_3+\frac{1}{384}\left(1640z^2-3892z+2473\right)\,,\nonumber\\
\nonumber\\
A_{gq}^{(-1)} &= \frac{-41678z^2-16840z-79505}{20736}+\frac{1}{384}\left(-3871z^2+6334z-8048\right)\,\zeta_2\\
&+\frac{1}{144}\left(-206z^2+751z-262\right)\,H_2+\frac{1}{48}\left(-163z^2+312z-136\right)\,H_{1,2}\nonumber\\
&+\frac{1}{48}\left(-163z^2+362z-110\right)\,H_{1,1,0}+\frac{1}{72}\left(-127z^2+353z-284\right)\,H_{1,0}\nonumber\\
&+\frac{1}{16}\left(-91z^2+90z+30\right)\,H_{2,1}+\frac{1}{144}\left(-56z^2-560z-421\right)\,\zeta_3\nonumber\\
&+\frac{1}{432}\left(-43z^2+161z+154\right)\,H_0+\frac{1}{48}\left(-11z^2+24z+130\right)\,H_{2,0}\nonumber\\
&+\frac{1}{24}\left(-z^2-6z-87\right)\,H_{0,0,0}+\frac{1}{24}\left(-z^2+2z-2\right)\,H_{3,1}-\frac{13}{12}\left(z^2-2z+2\right)\,H_{1,3}\nonumber\\
&-\frac{41}{24}\left(z^2-2z+2\right)\,H_{1,2,1}+\frac{15}{8}\left(z^2-2z+2\right)\,H_{2,0,0}+\frac{17}{8}\left(z^2-2z+2\right)\,H_4\nonumber\\
&-\frac{55}{24}\left(z^2-2z+2\right)\,H_{1,2,0}-\frac{19}{8}\left(z^2-2z+2\right)\,H_{1,1,0,0}-\frac{71}{24}\left(z^2-2z+2\right)\,H_{2,1,0}\nonumber\\
&+3\left(z^2-2z+2\right)\,H_{1,0,0,0}-\frac{73}{24}\left(z^2-2z+2\right)\,H_{2,2}+\frac{25}{8}\left(z^2-2z+2\right)\,H_{3,0}\nonumber\\
&-\frac{77}{24}\left(z^2-2z+2\right)\,H_{2,1,1}-\frac{7}{2}\left(z^2-2z+2\right)\,H_{1,1,2}-\frac{9}{2}\left(z^2-2z+2\right)\,H_{0,0,0,0}\nonumber\\
&-\frac{23}{4}\left(z^2-2z+2\right)\,H_{1,1,1,0}-\frac{1087}{64}\left(z^2-2z+2\right)\,\zeta_4-\frac{193}{6}\left(z^2-2z+2\right)\,H_{1,1,1,1}\nonumber\\
&+\frac{1}{24}\left(20z^2-9z+19\right)\,H_3+\frac{1}{48}\left(37z^2+622z+323\right)\,H_{1,1,1}\nonumber\\
&+\frac{1}{48}\left(39z^2+20z+30\right)\,H_{1,0,0}+\frac{1}{48}\left(106z^2-139z+282\right)\,H_{0,0}\nonumber\\
&+\frac{1}{432}\left(4181z^2-8590z+6943\right)\,H_{1,1}+\frac{15026z^2-47650z+27967}{2592}\,H_1\nonumber\\
&+\frac{1}{96}\left(-443z^2-502z-704\right)\,\zeta_2\,H_1-\frac{7}{6}\left(z^2-2z+2\right)\,\zeta_3\,H_0\nonumber\\
&-\frac{323}{48}\left(z^2-2z+2\right)\,\zeta_2\,H_2-\frac{335}{48}\left(z^2-2z+2\right)\,\zeta_2\,H_{1,0}+\frac{235}{24}\left(z^2-2z+2\right)\,\zeta_3\,H_1\nonumber\\
&+\frac{29}{2}\left(z^2-2z+2\right)\,\zeta_2\,H_{0,0}+20\left(z^2-2z+2\right)\,\zeta_2\,H_{1,1}\nonumber\\
&+\frac{1}{96}\left(135z^2+62z+766\right)\,\zeta_2\,H_0\,,\nonumber
\end{align}
\begin{align}
B_{gq}^{(-1)} &= \frac{1}{36}\left(-150z^2+394z-241\right)\,H_{1,1}+\frac{1}{36}\left(-13z^2+20z-2\right)\,H_{1,0}\\
&+\frac{1}{6}\left(-7z^2+13z-8\right)\,H_{0,0}+\frac{1}{12}\left(-z^2+2z-2\right)\,H_{1,2}+\frac{1}{4}\left(-z^2+2z-2\right)\,H_{2,1}\nonumber\\
&-\frac{5}{12}\left(z^2-2z+2\right)\,H_{1,0,0}-\frac{5}{12}\left(z^2-2z+2\right)\,H_{1,1,0}-\frac{7}{12}\left(z^2-2z+2\right)\,H_{2,0}\nonumber\\
&+\frac{3}{4}\left(z^2-2z+2\right)\,H_{0,0,0}-\frac{16}{3}\left(z^2-2z+2\right)\,H_{1,1,1}-\frac{15}{8}\left(z^2-2z+2\right)\,\zeta_2\,H_0\nonumber\\
&+\frac{193}{48}\left(z^2-2z+2\right)\,\zeta_2\,H_1+\frac{1}{648}\left(-613z^2+2654z-845\right)\,H_1\nonumber\\
&+\frac{1}{216}\left(-97z^2+80z-62\right)\,H_0+\frac{1}{18}\left(-11z^2+7z+8\right)\,H_2+\frac{1}{4}\left(-z^2+2z-2\right)\,H_3\nonumber\\
&+\frac{89}{36}\left(z^2-2z+2\right)\,\zeta_3+\frac{1}{96}\left(534z^2-1280z+903\right)\,\zeta_2+\frac{15476z^2-25644z+27525}{5184}\,,\nonumber\\
C_{gq}^{(-1)} &= -\frac{8}{27}\left(z^2-2z+2\right)\,H_{1,1}+\frac{1}{81}\left(-47z^2+106z-79\right)\,H_1\\
&+\frac{1}{12}\left(z^2-2z+2\right)\,\zeta_2+\frac{1}{324}\left(-261z^2+610z-403\right)\,,\nonumber\\
D_{gq}^{(-1)} &=\frac{1}{4}\left(z^2-2z+2\right)\,H_{2,1}+\frac{1}{4}\left(z^2-2z+2\right)\,H_{1,2}+\frac{5}{12}\left(z^2-2z+2\right)\,H_{1,1,0}\\
&+\frac{7}{12}\left(z^2-2z+2\right)\,H_{2,0}+\frac{7}{12}\left(z^2-2z+2\right)\,H_{1,0,0}-\frac{3}{4}\left(z^2-2z+2\right)\,H_{0,0,0}\nonumber\\
&+\frac{68}{9}\left(z^2-2z+2\right)\,H_{1,1,1}+\frac{1}{36}\left(13z^2-20z+2\right)\,H_{1,0}+\frac{1}{12}\left(17z^2-28z+16\right)\,H_{0,0}\nonumber\\
&+\frac{1}{108}\left(928z^2-2258z+1529\right)\,H_{1,1}+\frac{15}{8}\left(z^2-2z+2\right)\,\zeta_2\,H_0\nonumber\\
&-\frac{685}{144}\left(z^2-2z+2\right)\,\zeta_2\,H_1+\frac{1}{4}\left(z^2-2z+2\right)\,H_3+\frac{1}{36}\left(31z^2-20z-16\right)\,H_2\nonumber\\
&+\frac{1}{108}\left(53z^2-58z+58\right)\,H_0+\frac{1}{648}\left(4493z^2-11812z+7033\right)\,H_1\nonumber\\
&+\frac{-11975z^2+27682z-19501}{1728}\,\zeta_2-\frac{325}{144}\left(z^2-2z+2\right)\,\zeta_3\nonumber\\
&+\frac{30122z^2-89308z+35497}{7776}\,,\\
E_{gq}^{(-1)} &=\frac{-18400z^2+44972z+190765}{62208}+\frac{-32990z^2+103120z-18259}{3456}\,\zeta_2\\
&+\frac{-4079z^2+9530z-8549}{1728}\,H_1+\frac{1}{288}\left(-2428z^2+6980z-2759\right)\,\zeta_3\nonumber\\
&+\frac{1}{16}\left(-61z^2+86z+28\right)\,H_{0,0,0}+\frac{1}{48}\left(-28z^2-86z+11\right)\,H_{1,0,0}\nonumber\\
&+\frac{1}{48}\left(-21z^2-16z-8\right)\,H_3-\frac{7}{12}\left(z^2-2z+2\right)\,H_{3,1}+\frac{31}{24}\left(z^2-2z+2\right)\,H_{1,3}\nonumber\\
&-\frac{9}{4}\left(z^2-2z+2\right)\,H_{2,0,0}-\frac{5}{2}\left(z^2-2z+2\right)\,H_4+\frac{29}{8}\left(z^2-2z+2\right)\,H_{1,1,0,0}\nonumber\\
&+\frac{89}{24}\left(z^2-2z+2\right)\,H_{1,2,0}+\frac{49}{12}\left(z^2-2z+2\right)\,H_{2,1,0}+\frac{25}{6}\left(z^2-2z+2\right)\,H_{2,2}\nonumber\\
&-\frac{17}{4}\left(z^2-2z+2\right)\,H_{3,0}+\frac{103}{24}\left(z^2-2z+2\right)\,H_{1,2,1}-\frac{37}{8}\left(z^2-2z+2\right)\,H_{1,0,0,0}\nonumber\\
&+\frac{31}{6}\left(z^2-2z+2\right)\,H_{2,1,1}+\frac{45}{8}\left(z^2-2z+2\right)\,H_{0,0,0,0}+\frac{143}{24}\left(z^2-2z+2\right)\,H_{1,1,2}\nonumber\\
&+\frac{26}{3}\left(z^2-2z+2\right)\,H_{1,1,1,0}+\frac{2681}{128}\left(z^2-2z+2\right)\,\zeta_4+\frac{173}{3}\left(z^2-2z+2\right)\,H_{1,1,1,1}\nonumber
\end{align}
\begin{align}
&+\frac{1}{48}\left(52z^2-177z-204\right)\,H_{0,0}+\frac{1}{12}\left(82z^2-180z+81\right)\,H_{1,1,0}\nonumber\\
&+\frac{1}{432}\left(151z^2+379z-1000\right)\,H_0+\frac{1}{48}\left(167z^2-236z-40\right)\,H_{2,0}\nonumber\\
&+\frac{1}{48}\left(234z^2-482z+209\right)\,H_{1,2}+\frac{1}{288}\left(511z^2-3952z+1117\right)\,H_{1,1}\nonumber\\
&+\frac{1}{144}\left(533z^2-1363z+922\right)\,H_{1,0}+\frac{1}{144}\left(560z^2-1363z+184\right)\,H_2\nonumber\\
&+\frac{1}{48}\left(585z^2-602z-40\right)\,H_{2,1}+\frac{1}{36}\left(772z^2-2492z+1007\right)\,H_{1,1,1}\nonumber\\
&+\frac{1}{288}\left(-2554z^2+11360z-4421\right)\,\zeta_2\,H_1+\frac{11}{8}\left(z^2-2z+2\right)\,\zeta_3\,H_0\nonumber\\
&+\frac{23}{3}\left(z^2-2z+2\right)\,\zeta_2\,H_2+\frac{241}{24}\left(z^2-2z+2\right)\,\zeta_2\,H_{1,0}-\frac{265}{16}\left(z^2-2z+2\right)\,\zeta_2\,H_{0,0}\nonumber\\
&-\frac{141}{8}\left(z^2-2z+2\right)\,\zeta_3\,H_1-\frac{401}{12}\left(z^2-2z+2\right)\,\zeta_2\,H_{1,1}\nonumber\\
&+\frac{1}{48}\left(243z^2-430z-218\right)\,\zeta_2\,H_0\,,\nonumber\\
F_{gq}^{(-1)} &=\frac{8}{27}\left(z^2-2z+2\right)\,H_{1,1}+\frac{1}{81}\left(47z^2-106z+79\right)\,H_1\\
&+\frac{1}{12}\left(-z^2+2z-2\right)\,\zeta_2+\frac{1}{324}\left(261z^2-610z+403\right)\,,\nonumber\\
G_{gq}^{(-1)} &=\frac{-980846z^2+2724232z-1639885}{62208}+\frac{-49097z^2+130744z-78310}{2592}\,H_1\\
&+\frac{1}{216}\left(-5257z^2+13928z-8933\right)\,H_{1,1}+\frac{1}{144}\left(-4639z^2+11366z-7397\right)\,H_{1,1,1}\nonumber\\
&+\frac{1}{48}\left(-375z^2+374z-50\right)\,H_{2,1}+\frac{1}{48}\left(-200z^2+355z-78\right)\,H_{0,0}\nonumber\\
&+\frac{1}{48}\left(-183z^2+230z-90\right)\,H_{2,0}+\frac{1}{48}\left(-165z^2+358z-214\right)\,H_{1,1,0}\nonumber\\
&+\frac{1}{48}\left(-148z^2+231z+26\right)\,H_2+\frac{1}{48}\left(-89z^2+218z-112\right)\,H_{1,2}\nonumber\\
&+\frac{1}{24}\left(-42z^2+105z-59\right)\,H_{1,0}+\frac{1}{48}\left(-29z^2+114z-80\right)\,H_{1,0,0}\nonumber\\
&+\frac{1}{24}\left(-14z^2+20z-15\right)\,H_3+\frac{1}{3}\left(-z^2+2z-2\right)\,H_{1,3}+\frac{3}{8}\left(z^2-2z+2\right)\,H_4\nonumber\\
&+\frac{3}{8}\left(z^2-2z+2\right)\,H_{2,0,0}+\frac{5}{8}\left(z^2-2z+2\right)\,H_{3,1}-\frac{9}{8}\left(z^2-2z+2\right)\,H_{0,0,0,0}\nonumber\\
&+\frac{9}{8}\left(z^2-2z+2\right)\,H_{3,0}-\frac{9}{8}\left(z^2-2z+2\right)\,H_{2,2}-\frac{9}{8}\left(z^2-2z+2\right)\,H_{2,1,0}\nonumber\\
&-\frac{13}{8}\left(z^2-2z+2\right)\,H_{1,1,0,0}-\frac{43}{24}\left(z^2-2z+2\right)\,H_{1,2,0}-\frac{47}{24}\left(z^2-2z+2\right)\,H_{2,1,1}\nonumber\\
&+2\left(z^2-2z+2\right)\,H_{1,0,0,0}-\frac{17}{6}\left(z^2-2z+2\right)\,H_{1,1,2}-\frac{35}{12}\left(z^2-2z+2\right)\,H_{1,1,1,0}\nonumber\\
&-\frac{83}{24}\left(z^2-2z+2\right)\,H_{1,2,1}-\frac{3739}{768}\left(z^2-2z+2\right)\,\zeta_4-\frac{185}{6}\left(z^2-2z+2\right)\,H_{1,1,1,1}\nonumber\\
&+\frac{1}{48}\left(3z^2-75z+94\right)\,H_0+\frac{1}{24}\left(106z^2-132z+45\right)\,H_{0,0,0}\nonumber\\
&+\frac{1}{288}\left(3557z^2-7954z+5056\right)\,\zeta_3+\frac{83390z^2-192112z+109609}{3456}\,\zeta_2\nonumber\\
&+\frac{1}{32}\left(-225z^2+278z-110\right)\,\zeta_2\,H_0-\frac{5}{24}\left(z^2-2z+2\right)\,\zeta_3\,H_0\nonumber
\end{align}
\begin{align}
&-\frac{15}{16}\left(z^2-2z+2\right)\,\zeta_2\,H_2+\frac{33}{16}\left(z^2-2z+2\right)\,\zeta_2\,H_{0,0}-\frac{55}{16}\left(z^2-2z+2\right)\,\zeta_2\,H_{1,0}\nonumber\\
&+\frac{77}{8}\left(z^2-2z+2\right)\,\zeta_3\,H_1+\frac{367}{24}\left(z^2-2z+2\right)\,\zeta_2\,H_{1,1}\nonumber\\
&+\frac{1}{288}\left(4729z^2-11834z+7973\right)\,\zeta_2\,H_1\,,\nonumber\\
H_{gq}^{(-1)} &=\frac{1}{54}\left(-239z^2+538z-403\right)\,H_{1,1}+\frac{1}{6}\left(-z^2+2z-2\right)\,H_{1,0,0}\\
&+\frac{1}{6}\left(-z^2+2z-2\right)\,H_{1,2}-\frac{20}{9}\left(z^2-2z+2\right)\,H_{1,1,1}-\frac{1}{12}z(3z-2)\,H_{0,0}\nonumber\\
&+\frac{53}{72}\left(z^2-2z+2\right)\,\zeta_2\,H_1+\frac{1}{324}\left(-1940z^2+4579z-3094\right)\,H_1\nonumber\\
&+\frac{1}{24}\left(-z^2+4z-6\right)\,H_0-\frac{1}{12}z(3z-2)\,H_2-\frac{31}{144}\left(z^2-2z+2\right)\,\zeta_3\nonumber\\
&+\frac{2363z^2-4642z+3247}{1728}\,\zeta_2+\frac{-106672z^2+255548z-153569}{15552}\,,\nonumber\\
I_{gq}^{(-1)} &=\frac{3}{8}\left(z^2-2z+2\right)\,\zeta_2\,H_{1,0}-\frac{15}{8}\left(z^2-2z+2\right)\,\zeta_2\,H_{1,1}\\
&-\frac{3}{8}\left(z^2-2z+2\right)\,H_{1,0,0,0}+\frac{3}{8}\left(z^2-2z+2\right)\,H_{1,2,0}+\frac{3}{8}\left(z^2-2z+2\right)\,H_{1,1,0,0}\nonumber\\
&+\frac{3}{8}\left(z^2-2z+2\right)\,H_{1,1,2}+\frac{7}{8}\left(z^2-2z+2\right)\,H_{1,2,1}+\frac{16}{3}\left(z^2-2z+2\right)\,H_{1,1,1,1}\nonumber\\
&+\frac{1}{16}\left(6z^2-16z+13\right)\,H_{1,0,0}+\frac{1}{16}\left(6z^2-16z+13\right)\,H_{1,2}\nonumber\\
&+\frac{2}{3}\left(15z^2-34z+25\right)\,H_{1,1,1}+\frac{1}{96}\left(1237z^2-2964z+2055\right)\,H_{1,1}\nonumber\\
&+\frac{1}{8}\left(z^2-2z+2\right)\,H_{1,3}+\frac{3}{16}(1-z)z\,H_{1,0}-\frac{3}{16}z(3z-2)\,H_{0,0,0}\nonumber\\
&+\frac{3}{16}z(3z-2)\,H_{2,0}+\frac{7}{16}z(3z-2)\,H_{2,1}+\frac{1}{16}z(14z-13)\,H_{0,0}\nonumber\\
&+\frac{1}{16}\left(-47z^2+110z-80\right)\,\zeta_2\,H_1-\frac{43}{24}\left(z^2-2z+2\right)\,\zeta_3\,H_1\nonumber\\
&+\frac{1}{192}\left(2977z^2-7214z+4679\right)\,H_1+\frac{3}{16}z(3z-2)\,\zeta_2\,H_0+\frac{5}{16}(1-z)z\,H_0\nonumber\\
&+\frac{1}{16}z(3z-2)\,H_3+\frac{1}{16}z(10z-9)\,H_2+\frac{1}{384}\left(-1729z^2+3554z-2102\right)\,\zeta_2\nonumber\\
&+\frac{1}{96}\left(-339z^2+698z-485\right)\,\zeta_3+\frac{697}{768}\left(z^2-2z+2\right)\,\zeta_4\nonumber\\
&+\frac{1}{768}\left(13880z^2-33564z+20835\right)\,,\nonumber\\
\nonumber\\
A_{gq}^{(0)} &= \frac{-16397336z^2+30450460z-25185649}{373248}\\
&+\frac{1}{576}\left(-5541z^2+10102z-8938\right)\,\zeta_3+\frac{1}{288}\left(-1286z^2+4288z-3079\right)\,H_{1,2}\nonumber\\
&+\frac{1}{48}\left(-1177z^2+1174z+686\right)\,H_{2,1,1}+\frac{1}{48}\left(-677z^2+833z-870\right)\,H_{0,0,0}\nonumber\\
&+\frac{1}{48}\left(-621z^2+1398z-550\right)\,H_{1,2,1}+\frac{1}{48}\left(-575z^2+1478z-387\right)\,H_{1,1,2}\nonumber\\
&+\frac{1}{48}\left(-546z^2+1642z-143\right)\,H_{1,1,1,0}+\frac{1}{48}\left(-219z^2+276z+302\right)\,H_{2,1,0}\nonumber
\end{align}
\begin{align}
&+\frac{1}{144}\left(-187z^2+9638z+4753\right)\,H_{1,1,1,1}+\frac{1}{48}\left(-175z^2+244z-338\right)\,H_{2,0,0}\nonumber\\
&+\frac{1}{48}\left(-122z^2-42z-197\right)\,H_{1,0,0,0}+\frac{1}{24}\left(-86z^2+128z+103\right)\,H_{2,2}\nonumber\\
&+\frac{1}{48}\left(-55z^2-8z-138\right)\,H_4+\frac{1}{24}\left(-54z^2+20z+135\right)\,H_3\nonumber\\
&+\frac{1}{48}\left(-31z^2+88z-462\right)\,H_{3,0}+\frac{1}{48}\left(-27z^2+276z-74\right)\,H_{1,2,0}\nonumber\\
&+\frac{1}{24}\left(-18z^2+145z-47\right)\,H_{1,1,0,0}+\frac{13}{24}\left(z^2-2z+2\right)\,H_{1,3,1}\nonumber\\
&+\frac{11}{8}\left(z^2-2z+2\right)\,H_{4,1}-\frac{13}{8}\left(z^2-2z+2\right)\,H_{1,1,3}+\frac{19}{8}\left(z^2-2z+2\right)\,H_{1,2,0,0}\nonumber\\
&+\frac{61}{24}\left(z^2-2z+2\right)\,H_{1,4}+\frac{67}{24}\left(z^2-2z+2\right)\,H_{2,2,0}+\frac{23}{8}\left(z^2-2z+2\right)\,H_{2,1,0,0}\nonumber\\
&+\frac{77}{24}\left(z^2-2z+2\right)\,H_{2,3}-\frac{33}{8}\left(z^2-2z+2\right)\,H_{2,0,0,0}-\frac{39}{8}\left(z^2-2z+2\right)\,H_{3,0,0}\nonumber\\
&-\frac{45}{8}\left(z^2-2z+2\right)\,H_5+\frac{137}{24}\left(z^2-2z+2\right)\,H_{3,1,1}+\frac{139}{24}\left(z^2-2z+2\right)\,H_{1,3,0}\nonumber\\
&+\frac{53}{8}\left(z^2-2z+2\right)\,H_{3,1,0}-\frac{53}{8}\left(z^2-2z+2\right)\,H_{1,2,1,0}+\frac{55}{8}\left(z^2-2z+2\right)\,H_{3,2}\nonumber\\
&-7\left(z^2-2z+2\right)\,H_{1,2,2}-\frac{57}{8}\left(z^2-2z+2\right)\,H_{4,0}-\frac{15}{2}\left(z^2-2z+2\right)\,H_{1,1,2,0}\nonumber\\
&-\frac{97}{12}\left(z^2-2z+2\right)\,H_{1,1,1,0,0}+\frac{49}{6}\left(z^2-2z+2\right)\,H_{1,1,0,0,0}\nonumber\\
&-\frac{33}{4}\left(z^2-2z+2\right)\,H_{1,0,0,0,0}+\frac{42}{5}\left(z^2-2z+2\right)\,\zeta_5-\frac{205}{24}\left(z^2-2z+2\right)\,H_{2,1,2}\nonumber\\
&-\frac{211}{24}\left(z^2-2z+2\right)\,H_{2,2,1}-\frac{227}{24}\left(z^2-2z+2\right)\,H_{2,1,1,0}+\frac{45}{4}\left(z^2-2z+2\right)\,H_{0,0,0,0,0}\nonumber\\
&-\frac{107}{8}\left(z^2-2z+2\right)\,H_{1,2,1,1}-\frac{497}{24}\left(z^2-2z+2\right)\,H_{1,1,2,1}-\frac{169}{8}\left(z^2-2z+2\right)\,H_{2,1,1,1}\nonumber\\
&-\frac{229}{8}\left(z^2-2z+2\right)\,H_{1,1,1,2}-\frac{479}{12}\left(z^2-2z+2\right)\,H_{1,1,1,1,0}\nonumber\\
&-\frac{635}{4}\left(z^2-2z+2\right)\,H_{1,1,1,1,1}+\frac{1}{24}\left(4z^2+269z+149\right)\,H_{2,1}\nonumber\\
&+\frac{1}{8}\left(14z^2-22z+99\right)\,H_{0,0,0,0}+\frac{1}{48}\left(38z^2+20z-63\right)\,H_{1,3}\nonumber\\
&+\frac{1}{288}\left(74z^2+98z+673\right)\,H_{1,0,0}+\frac{1}{48}\left(85z^2+54z-310\right)\,H_{3,1}\nonumber\\
&+\frac{1}{288}\left(89z^2+1358z-365\right)\,H_{1,1,0}+\frac{1}{72}\left(190z^2-341z+971\right)\,H_{2,0}\nonumber\\
&+\frac{32851z^2+8134z-5641}{1152}\,\zeta_4+\frac{1}{864}\left(36545z^2-72694z+57709\right)\,H_{1,1,1}\nonumber\\
&+\frac{-540z^3-91615z^2-20078z-85789}{10368}\,H_1+\frac{-270z^3+47581z^2-97775z+97568}{5184}\,H_0\nonumber\\
&+\frac{-360z^4-1584z^3-82700z^2+147900z-72535}{2304}\,\zeta_2\nonumber\\
&+\frac{1}{288}\left(15z^4+66z^3+2498z^2-2009z+172\right)\,H_{0,0}\nonumber\\
&+\frac{1}{864}\left(45z^4+198z^3+7672z^2-4667z+1028\right)\,H_2\nonumber\\
&+\frac{90z^4+396z^3+7973z^2-12622z+8959}{1728}\,H_{1,0}\nonumber
\end{align}
\begin{align}
&+\frac{90z^4+396z^3+42239z^2-128522z+75557}{1728}\,H_{1,1}\nonumber\\
&+\frac{1}{576}\left(-21455z^2+43270z-38449\right)\,\zeta_2\,H_1+\frac{1}{288}\left(-3095z^2-5414z-8389\right)\,\zeta_2\,H_{1,1}\nonumber\\
&+\frac{1}{144}\left(-659z^2-1514z-2344\right)\,\zeta_3\,H_1+\frac{1}{8}\left(z^2-2z+2\right)\,\zeta_3\,H_{1,0}\nonumber\\
&+\frac{37}{12}\left(z^2-2z+2\right)\,\zeta_3\,H_{0,0}-\frac{55}{16}\left(z^2-2z+2\right)\,\zeta_2\,H_{1,2}-\frac{169}{48}\left(z^2-2z+2\right)\,\zeta_2\,H_{2,1}\nonumber\\
&-\frac{215}{48}\left(z^2-2z+2\right)\,\zeta_2\,H_{1,1,0}+\frac{39}{8}\left(z^2-2z+2\right)\,\zeta_3\,H_2+\frac{293}{16}\left(z^2-2z+2\right)\,\zeta_2\,H_3\nonumber\\
&+\frac{297}{16}\left(z^2-2z+2\right)\,\zeta_2\,H_{2,0}+\frac{435}{16}\left(z^2-2z+2\right)\,\zeta_2\,H_{1,0,0}+\frac{5539}{192}\left(z^2-2z+2\right)\,\zeta_4\,H_0\nonumber\\
&-\frac{11843}{384}\left(z^2-2z+2\right)\,\zeta_4\,H_1-\frac{315}{8}\left(z^2-2z+2\right)\,\zeta_2\,H_{0,0,0}-\frac{1927}{48}\left(z^2-2z+2\right)\,\zeta_2\,\zeta_3\nonumber\\
&+\frac{177}{4}\left(z^2-2z+2\right)\,\zeta_3\,H_{1,1}+\frac{611}{8}\left(z^2-2z+2\right)\,\zeta_2\,H_{1,1,1}\nonumber\\
&+\frac{3}{16}\left(17z^2+10z-50\right)\,\zeta_3\,H_0+\frac{1}{96}\left(881z^2-1266z-2562\right)\,\zeta_2\,H_{0,0}\nonumber\\
&+\frac{1}{48}\left(911z^2-1288z+2195\right)\,\zeta_2\,H_0+\frac{1}{96}\left(1599z^2-1886z+1794\right)\,\zeta_2\,H_{1,0}\nonumber\\
&+\frac{1}{96}\left(1791z^2-1994z+1650\right)\,\zeta_2\,H_2\,,\nonumber\\
B_{gq}^{(0)} &= \frac{7\left(380164z^2-827612z+637397\right)}{93312}\\
&+\frac{1}{216}\left(-4207z^2+10718z-6479\right)\,H_{1,1,1}+\frac{1}{72}\left(-187z^2+362z-224\right)\,H_{2,0}\nonumber\\
&+\frac{1}{72}\left(-172z^2+338z-167\right)\,H_{1,2}+\frac{1}{36}\left(-169z^2+368z-224\right)\,H_{1,1,0}\nonumber\\
&+\frac{1}{24}\left(-121z^2+158z-32\right)\,H_{2,1}+\frac{1}{72}\left(-68z^2+178z-151\right)\,H_{1,0,0}\nonumber\\
&+\frac{1}{12}\left(-z^2+2z-2\right)\,H_{1,3}+\frac{3}{4}\left(z^2-2z+2\right)\,H_4+\frac{3}{4}\left(z^2-2z+2\right)\,H_{3,1}\nonumber\\
&-\frac{5}{6}\left(z^2-2z+2\right)\,H_{1,2,0}-\frac{5}{6}\left(z^2-2z+2\right)\,H_{1,1,0,0}+\frac{13}{12}\left(z^2-2z+2\right)\,H_{2,0,0}\nonumber\\
&-\frac{7}{6}\left(z^2-2z+2\right)\,H_{1,2,1}-\frac{5}{4}\left(z^2-2z+2\right)\,H_{2,2}-\frac{17}{12}\left(z^2-2z+2\right)\,H_{2,1,0}\nonumber\\
&+\frac{19}{12}\left(z^2-2z+2\right)\,H_{1,0,0,0}+\frac{7}{4}\left(z^2-2z+2\right)\,H_{3,0}-\frac{9}{4}\left(z^2-2z+2\right)\,H_{0,0,0,0}\nonumber\\
&-\frac{29}{12}\left(z^2-2z+2\right)\,H_{2,1,1}-\frac{5}{2}\left(z^2-2z+2\right)\,H_{1,1,2}-\frac{53}{12}\left(z^2-2z+2\right)\,H_{1,1,1,0}\nonumber\\
&-\frac{5269\left(z^2-2z+2\right)}{1152}\,\zeta_4-\frac{224}{9}\left(z^2-2z+2\right)\,H_{1,1,1,1}+\frac{1}{6}\left(2z^2+3z-8\right)\,H_3\nonumber\\
&+\frac{1}{6}\left(30z^2-49z+24\right)\,H_{0,0,0}+\frac{1}{432}\left(3923z^2-9802z+6310\right)\,\zeta_3\nonumber\\
&+\frac{27z^3-6409z^2+12065z-10778}{1296}\,H_0+\frac{54z^3+31507z^2-58986z+52929}{2592}\,H_1\nonumber\\
&+\frac{-27z^4-6604z^2+22910z-7547}{1296}\,H_{1,1}+\frac{1}{432}\left(-9z^4-2552z^2+2938z-124\right)\,H_2\nonumber\\
&+\frac{1}{432}\left(-9z^4-2012z^2+3994z-2401\right)\,H_{1,0}\nonumber\\
&+\frac{1}{144}\left(-3z^4-592z^2+766z+124\right)\,H_{0,0}+\frac{1}{576}\left(36z^4+9511z^2-22666z+10514\right)\,\zeta_2\nonumber
\end{align}
\begin{align}
&+\frac{1}{12}\left(-108z^2+197z-120\right)\,\zeta_2\,H_0-\frac{5}{12}\left(z^2-2z+2\right)\,\zeta_3\,H_0\nonumber\\
&-\frac{73}{24}\left(z^2-2z+2\right)\,\zeta_2\,H_2-\frac{79}{24}\left(z^2-2z+2\right)\,\zeta_2\,H_{1,0}+\frac{45}{8}\left(z^2-2z+2\right)\,\zeta_2\,H_{0,0}\nonumber\\
&+\frac{691}{72}\left(z^2-2z+2\right)\,\zeta_3\,H_1+\frac{1007}{72}\left(z^2-2z+2\right)\,\zeta_2\,H_{1,1}\nonumber\\
&+\frac{1}{288}\left(4399z^2-11042z+7757\right)\,\zeta_2\,H_1\,,\nonumber\\
C_{gq}^{(0)} &= -\frac{32}{27}\left(z^2-2z+2\right)\,H_{1,1,1}-\frac{4}{81}\left(47z^2-106z+79\right)\,H_{1,1}\\
&+\frac{1}{3}\left(z^2-2z+2\right)\,\zeta_2\,H_1+\frac{1}{81}\left(-261z^2+610z-403\right)\,H_1+\frac{17}{54}\left(z^2-2z+2\right)\,\zeta_3\nonumber\\
&+\frac{1}{72}\left(47z^2-106z+79\right)\,\zeta_2+\frac{-11287z^2+26714z-16643}{2916}\,,\nonumber\\
D_{gq}^{(0)} &=\frac{-7127z^2-90310z-142088}{15552}+\frac{1}{288}\left(-2899z^2+6834z-4065\right)\,\zeta_3\\
&+\frac{1}{4}\left(-25z^2+36z-16\right)\,H_{0,0,0}+\frac{1}{12}\left(-7z^2-4z+16\right)\,H_3+\frac{1}{12}\left(-z^2+2z-2\right)\,H_{1,3}\nonumber\\
&-\frac{3}{4}\left(z^2-2z+2\right)\,H_4-\frac{3}{4}\left(z^2-2z+2\right)\,H_{3,1}-\frac{13}{12}\left(z^2-2z+2\right)\,H_{2,0,0}\nonumber\\
&+\frac{13}{12}\left(z^2-2z+2\right)\,H_{1,2,0}+\frac{5}{4}\left(z^2-2z+2\right)\,H_{2,2}+\frac{5}{4}\left(z^2-2z+2\right)\,H_{1,1,0,0}\nonumber\\
&+\frac{17}{12}\left(z^2-2z+2\right)\,H_{2,1,0}-\frac{7}{4}\left(z^2-2z+2\right)\,H_{3,0}+\frac{25}{12}\left(z^2-2z+2\right)\,H_{1,2,1}\nonumber\\
&+\frac{9}{4}\left(z^2-2z+2\right)\,H_{0,0,0,0}+\frac{29}{12}\left(z^2-2z+2\right)\,H_{2,1,1}-\frac{29}{12}\left(z^2-2z+2\right)\,H_{1,0,0,0}\nonumber\\
&+\frac{35}{12}\left(z^2-2z+2\right)\,H_{1,1,2}+\frac{53}{12}\left(z^2-2z+2\right)\,H_{1,1,1,0}+\frac{19127\left(z^2-2z+2\right)}{2304}\,\zeta_4\nonumber\\
&+\frac{304}{9}\left(z^2-2z+2\right)\,H_{1,1,1,1}+\frac{1}{24}\left(67z^2-140z+83\right)\,H_{1,2}\nonumber\\
&+\frac{1}{12}\left(77z^2-90z+16\right)\,H_{2,1}+\frac{1}{72}\left(97z^2-260z+233\right)\,H_{1,0,0}\nonumber\\
&+\frac{1}{36}\left(107z^2-190z+112\right)\,H_{2,0}+\frac{1}{36}\left(169z^2-368z+224\right)\,H_{1,1,0}\nonumber\\
&+\frac{1}{72}\left(2677z^2-6442z+4309\right)\,H_{1,1,1}+\frac{-162z^3+114881z^2-332356z+161365}{7776}\,H_1\nonumber\\
&+\frac{-27z^3+7660z^2-15773z+15044}{1296}\,H_0+\frac{-324z^4-127855z^2+288266z-139847}{5184}\,\zeta_2\nonumber\\
&+\frac{1}{144}\left(3z^4+763z^2-842z-232\right)\,H_{0,0}+\frac{1}{432}\left(9z^4+2156z^2-4336z+2707\right)\,H_{1,0}\nonumber\\
&+\frac{1}{432}\left(9z^4+3137z^2-3454z+232\right)\,H_2\nonumber\\
&+\frac{1}{432}\left(9z^4+12467z^2-31950z+19098\right)\,H_{1,1}\nonumber\\
&+\frac{1}{864}\left(-17407z^2+42818z-30749\right)\,\zeta_2\,H_1-\frac{7}{12}\left(z^2-2z+2\right)\,\zeta_3\,H_0\nonumber\\
&+\frac{73}{24}\left(z^2-2z+2\right)\,\zeta_2\,H_2+\frac{113}{24}\left(z^2-2z+2\right)\,\zeta_2\,H_{1,0}-\frac{45}{8}\left(z^2-2z+2\right)\,\zeta_2\,H_{0,0}\nonumber\\
&-\frac{737}{72}\left(z^2-2z+2\right)\,\zeta_3\,H_1-\frac{1201}{72}\left(z^2-2z+2\right)\,\zeta_2\,H_{1,1}\nonumber\\
&+\frac{1}{24}\left(267z^2-428z+240\right)\,\zeta_2\,H_0\,,\nonumber
\end{align}
\begin{align}
E_{gq}^{(0)} &=\frac{-556382z^2+5813728z+1746191}{373248}+\frac{1}{48}(31z-36)(91z+6)\,H_{2,1,1}\\
&+\frac{-365539z^2+164798z-12626}{4608}\,\zeta_4+\frac{-15752z^2+63484z-25099}{1728}\,\zeta_3\nonumber\\
&+\frac{1}{48}\left(-549z^2+530z+160\right)\,H_{3,1}+\frac{1}{16}\left(-167z^2+208z+64\right)\,H_{3,0}\nonumber\\
&+\frac{1}{8}\left(-21z^2+39z+8\right)\,H_4+\frac{1}{48}\left(-20z^2-60z+147\right)\,H_{1,3}\nonumber\\
&+\frac{1}{48}\left(z^2-32z+248\right)\,H_{2,0,0}+\frac{1}{2}\left(z^2-2z+2\right)\,H_{4,1}+\frac{13}{12}\left(z^2-2z+2\right)\,H_{1,1,3}\nonumber\\
&-\frac{47}{12}\left(z^2-2z+2\right)\,H_{2,2,0}-\frac{47}{12}\left(z^2-2z+2\right)\,H_{1,2,0,0}-4\left(z^2-2z+2\right)\,H_{2,1,0,0}\nonumber\\
&-\frac{101}{24}\left(z^2-2z+2\right)\,H_{1,3,1}-\frac{17}{4}\left(z^2-2z+2\right)\,H_{1,4}-\frac{13}{3}\left(z^2-2z+2\right)\,H_{2,3}\nonumber\\
&+\frac{9}{2}\left(z^2-2z+2\right)\,H_{2,0,0,0}+6\left(z^2-2z+2\right)\,H_{3,0,0}+\frac{27}{4}\left(z^2-2z+2\right)\,H_5\nonumber\\
&-10\left(z^2-2z+2\right)\,H_{3,1,0}-\frac{81}{8}\left(z^2-2z+2\right)\,H_{1,3,0}-\frac{41}{4}\left(z^2-2z+2\right)\,H_{3,2}\nonumber\\
&+\frac{21}{2}\left(z^2-2z+2\right)\,H_{4,0}+\frac{85}{8}\left(z^2-2z+2\right)\,H_{1,2,2}+\frac{65}{6}\left(z^2-2z+2\right)\,H_{1,2,1,0}\nonumber\\
&+\frac{133}{12}\left(z^2-2z+2\right)\,H_{1,1,1,0,0}-\frac{139}{12}\left(z^2-2z+2\right)\,H_{3,1,1}+\frac{93}{8}\left(z^2-2z+2\right)\,H_{1,1,2,0}\nonumber\\
&+\frac{143}{12}\left(z^2-2z+2\right)\,H_{2,1,2}+\frac{73}{6}\left(z^2-2z+2\right)\,H_{2,2,1}+\frac{51}{4}\left(z^2-2z+2\right)\,H_{1,0,0,0,0}\nonumber\\
&+\frac{77}{6}\left(z^2-2z+2\right)\,H_{2,1,1,0}-\frac{105}{8}\left(z^2-2z+2\right)\,H_{1,1,0,0,0}\nonumber\\
&-\frac{117}{8}\left(z^2-2z+2\right)\,H_{0,0,0,0,0}-\frac{1937}{120}\left(z^2-2z+2\right)\,\zeta_5+\frac{647}{24}\left(z^2-2z+2\right)\,H_{1,2,1,1}\nonumber\\
&+\frac{97}{3}\left(z^2-2z+2\right)\,H_{2,1,1,1}+\frac{205}{6}\left(z^2-2z+2\right)\,H_{1,1,2,1}+\frac{175}{4}\left(z^2-2z+2\right)\,H_{1,1,1,2}\nonumber\\
&+\frac{176}{3}\left(z^2-2z+2\right)\,H_{1,1,1,1,0}+\frac{830}{3}\left(z^2-2z+2\right)\,H_{1,1,1,1,1}\nonumber\\
&+\frac{1}{16}\left(19z^2-144z+64\right)\,H_{1,1,0,0}+\frac{3}{8}\left(25z^2-34z-18\right)\,H_{0,0,0,0}\nonumber\\
&+\frac{1}{24}\left(58z^2-297z+150\right)\,H_{1,2,0}+\frac{1}{48}\left(125z^2+188z+166\right)\,H_{1,0,0,0}\nonumber\\
&+\frac{1}{48}\left(247z^2-196z-192\right)\,H_3+\frac{1}{96}\left(517z^2+426z+1272\right)\,H_{0,0,0}\nonumber\\
&+\frac{1}{48}\left(591z^2-862z+64\right)\,H_{2,2}+\frac{1}{48}\left(679z^2-916z-32\right)\,H_{2,1,0}\nonumber\\
&+\frac{1}{24}\left(815z^2-2043z+858\right)\,H_{1,1,1,0}+\frac{1}{48}\left(923z^2-2293z-168\right)\,H_{2,1}\nonumber\\
&+\frac{1}{288}\left(950z^2-2746z+823\right)\,H_{1,0,0}+\frac{1}{48}\left(965z^2-2318z+1019\right)\,H_{1,2,1}\nonumber\\
&+\frac{1}{48}\left(1267z^2-3082z+1360\right)\,H_{1,1,2}+\frac{1}{96}\left(1507z^2-4052z+2466\right)\,H_{1,2}\nonumber\\
&+\frac{1}{18}\left(1912z^2-5900z+2309\right)\,H_{1,1,1,1}+\frac{1}{288}\left(2375z^2-4576z-2480\right)\,H_{2,0}\nonumber\\
&+\frac{1}{144}\left(2702z^2-7429z+4351\right)\,H_{1,1,0}+\frac{1}{432}\left(4384z^2-29357z+10241\right)\,H_{1,1,1}\nonumber\\
&+\frac{243z^3-29689z^2+105344z-132128}{5184}\,H_0\nonumber
\end{align}
\begin{align}
&+\frac{729z^3-239143z^2+815777z-506546}{15552}\,H_1\nonumber\\
&+\frac{-243z^4-1836z^3-66500z^2+169756z-153745}{5184}\,H_{1,1}\nonumber\\
&+\frac{-81z^4-612z^3+8155z^2-32108z-5440}{1728}\,H_2\nonumber\\
&+\frac{-81z^4-612z^3+16075z^2-39092z+17726}{1728}\,H_{1,0}\nonumber\\
&+\frac{1}{576}\left(-27z^4-204z^3+2075z^2-12478z+3040\right)\,H_{0,0}\nonumber\\
&+\frac{2916z^4+22032z^3-706466z^2+2354032z-1361215}{20736}\,\zeta_2\nonumber\\
&+\frac{1}{864}\left(-17312z^2+61927z-22048\right)\,\zeta_2\,H_1+\frac{1}{144}\left(-6283z^2+22316z-8663\right)\,\zeta_2\,H_{1,1}\nonumber\\
&+\frac{1}{144}\left(-3971z^2+13054z-4810\right)\,\zeta_3\,H_1+\frac{1}{96}\left(-1291z^2+1254z-1200\right)\,\zeta_2\,H_2\nonumber\\
&+\frac{1}{32}\left(-1049z^2+1374z+524\right)\,\zeta_2\,H_{0,0}+\frac{1}{48}\left(-847z^2+772z-865\right)\,\zeta_2\,H_{1,0}\nonumber\\
&+\frac{1}{96}\left(-96z^2-783z-3532\right)\,\zeta_2\,H_0+\frac{1}{12}\left(-92z^2+4z+125\right)\,\zeta_3\,H_0\nonumber\\
&-\frac{43}{24}\left(z^2-2z+2\right)\,\zeta_3\,H_{1,0}-\frac{89}{24}\left(z^2-2z+2\right)\,\zeta_3\,H_{0,0}+\frac{29}{6}\left(z^2-2z+2\right)\,\zeta_2\,H_{2,1}\nonumber\\
&+\frac{307}{48}\left(z^2-2z+2\right)\,\zeta_2\,H_{1,2}-\frac{77}{12}\left(z^2-2z+2\right)\,\zeta_3\,H_2+\frac{265}{24}\left(z^2-2z+2\right)\,\zeta_2\,H_{1,1,0}\nonumber\\
&-\frac{7463}{384}\left(z^2-2z+2\right)\,\zeta_4\,H_0-21\left(z^2-2z+2\right)\,\zeta_2\,H_{2,0}-\frac{169}{8}\left(z^2-2z+2\right)\,\zeta_2\,H_3\nonumber\\
&+\frac{10901}{384}\left(z^2-2z+2\right)\,\zeta_4\,H_1-\frac{607}{16}\left(z^2-2z+2\right)\,\zeta_2\,H_{1,0,0}\nonumber\\
&+\frac{729}{16}\left(z^2-2z+2\right)\,\zeta_2\,H_{0,0,0}+\frac{2057}{32}\left(z^2-2z+2\right)\,\zeta_2\,\zeta_3-\frac{943}{12}\left(z^2-2z+2\right)\,\zeta_3\,H_{1,1}\nonumber\\
&-\frac{785}{6}\left(z^2-2z+2\right)\,\zeta_2\,H_{1,1,1}\,,\nonumber\\
F_{gq}^{(0)} &=\frac{32}{27}\left(z^2-2z+2\right)\,H_{1,1,1}+\frac{4}{81}\left(47z^2-106z+79\right)\,H_{1,1}\\
&+\frac{1}{3}\left(-z^2+2z-2\right)\,\zeta_2\,H_1+\frac{1}{81}\left(261z^2-610z+403\right)\,H_1\nonumber\\
&+\frac{1}{72}\left(-47z^2+106z-79\right)\,\zeta_2-\frac{17}{54}\left(z^2-2z+2\right)\,\zeta_3+\frac{11287z^2-26714z+16643}{2916}\,,\nonumber\\
G_{gq}^{(0)} &=\frac{-9858416z^2+29314708z-16722367}{373248}\\
&+\frac{1}{864}\left(-86929z^2+232766z-149741\right)\,H_{1,1,1}\nonumber\\
&+\frac{1}{144}\left(-20869z^2+50618z-32825\right)\,H_{1,1,1,1}+\frac{1}{288}\left(-3136z^2+8192z-4571\right)\,H_{1,2}\nonumber\\
&+\frac{1}{48}\left(-2031z^2+2174z-470\right)\,H_{2,1,1}+\frac{1}{96}\left(-1603z^2+4134z-2641\right)\,H_{1,1,0}\nonumber\\
&+\frac{1}{48}\left(-1084z^2+2444z-1573\right)\,H_{1,1,1,0}+\frac{1}{288}\left(-1024z^2+3062z-1739\right)\,H_{1,0,0}\nonumber\\
&+\frac{1}{48}\left(-755z^2+1766z-1099\right)\,H_{1,1,2}+\frac{1}{48}\left(-613z^2+730z-270\right)\,H_{2,1,0}\nonumber
\end{align}
\begin{align}
&+\frac{1}{24}\left(-566z^2+969z-65\right)\,H_{2,1}+\frac{1}{48}\left(-443z^2+1202z-706\right)\,H_{1,2,1}\nonumber\\
&+\frac{1}{24}\left(-307z^2+537z-117\right)\,H_{2,0}+\frac{1}{24}\left(-259z^2+336z-135\right)\,H_{2,2}\nonumber\\
&+\frac{1}{48}\left(-175z^2+198z-78\right)\,H_3+\frac{1}{48}\left(-125z^2+426z-316\right)\,H_{1,2,0}\nonumber\\
&+\frac{1}{8}\left(-98z^2+130z-45\right)\,H_{0,0,0,0}+\frac{1}{12}\left(-21z^2+76z-56\right)\,H_{1,1,0,0}\nonumber\\
&+\frac{1}{48}\left(-18z^2+52z-105\right)\,H_{1,3}+\frac{7}{24}\left(z^2-2z+2\right)\,H_{1,1,3}-\frac{3}{8}\left(z^2-2z+2\right)\,H_{2,0,0,0}\nonumber\\
&-\frac{9}{8}\left(z^2-2z+2\right)\,H_5-\frac{9}{8}\left(z^2-2z+2\right)\,H_{3,0,0}+\frac{9}{8}\left(z^2-2z+2\right)\,H_{2,3}\nonumber\\
&+\frac{9}{8}\left(z^2-2z+2\right)\,H_{2,2,0}+\frac{9}{8}\left(z^2-2z+2\right)\,H_{2,1,0,0}+\frac{37}{24}\left(z^2-2z+2\right)\,H_{1,2,0,0}\nonumber\\
&-\frac{15}{8}\left(z^2-2z+2\right)\,H_{4,1}+\frac{59}{24}\left(z^2-2z+2\right)\,H_{1,4}+\frac{27}{8}\left(z^2-2z+2\right)\,H_{0,0,0,0,0}\nonumber\\
&-\frac{27}{8}\left(z^2-2z+2\right)\,H_{4,0}+\frac{27}{8}\left(z^2-2z+2\right)\,H_{3,2}+\frac{27}{8}\left(z^2-2z+2\right)\,H_{3,1,0}\nonumber\\
&-\frac{27}{8}\left(z^2-2z+2\right)\,H_{2,2,1}-\frac{27}{8}\left(z^2-2z+2\right)\,H_{2,1,2}-\frac{27}{8}\left(z^2-2z+2\right)\,H_{2,1,1,0}\nonumber\\
&+\frac{97}{24}\left(z^2-2z+2\right)\,H_{1,3,1}-\frac{17}{4}\left(z^2-2z+2\right)\,H_{1,1,1,0,0}+\frac{289}{60}\left(z^2-2z+2\right)\,\zeta_5\nonumber\\
&-5\left(z^2-2z+2\right)\,H_{1,2,2}+\frac{125}{24}\left(z^2-2z+2\right)\,H_{1,3,0}-\frac{21}{4}\left(z^2-2z+2\right)\,H_{1,0,0,0,0}\nonumber\\
&-\frac{21}{4}\left(z^2-2z+2\right)\,H_{1,1,2,0}+\frac{47}{8}\left(z^2-2z+2\right)\,H_{3,1,1}+\frac{73}{12}\left(z^2-2z+2\right)\,H_{1,1,0,0,0}\nonumber\\
&-\frac{149}{24}\left(z^2-2z+2\right)\,H_{1,2,1,0}-\frac{269}{24}\left(z^2-2z+2\right)\,H_{2,1,1,1}-\frac{383}{24}\left(z^2-2z+2\right)\,H_{1,1,2,1}\nonumber\\
&-\frac{131}{8}\left(z^2-2z+2\right)\,H_{1,1,1,2}-\frac{75}{4}\left(z^2-2z+2\right)\,H_{1,1,1,1,0}-\frac{455}{24}\left(z^2-2z+2\right)\,H_{1,2,1,1}\nonumber\\
&-\frac{557}{4}\left(z^2-2z+2\right)\,H_{1,1,1,1,1}+\frac{1}{48}\left(60z^2-308z+157\right)\,H_{1,0,0,0}\nonumber\\
&+\frac{1}{48}\left(165z^2-218z+90\right)\,H_{2,0,0}+\frac{1}{48}\left(235z^2-262z+90\right)\,H_4\nonumber\\
&+\frac{1}{48}\left(491z^2-602z+150\right)\,H_{3,1}+\frac{1}{48}\left(540z^2-1151z+234\right)\,H_{0,0,0}\nonumber\\
&+\frac{1}{48}\left(595z^2-754z+270\right)\,H_{3,0}+\frac{65720z^2-166888z+102169}{1728}\,\zeta_3\nonumber\\
&+\frac{81918z^2-60080z+6689}{1536}\,\zeta_4+\frac{1}{576}\left(18z^3-1349z^2-1405z+3840\right)\,H_0\nonumber\\
&+\frac{972z^3-1261663z^2+3374378z-1847717}{31104}\,H_1\nonumber\\
&+\frac{-162z^4+108z^3-352897z^2+934550z-546047}{5184}\,H_{1,1}\nonumber\\
&+\frac{1}{288}\left(-9z^4+6z^3-4272z^2+7231z+564\right)\,H_2\nonumber\\
&+\frac{1}{288}\left(-9z^4+6z^3-4134z^2+8815z-1692\right)\,H_{0,0}\nonumber\\
&+\frac{1}{192}\left(-6z^4+4z^3-2201z^2+4854z-2519\right)\,H_{1,0}\nonumber\\
&+\frac{1944z^4-1296z^3+1847882z^2-4492864z+2451781}{20736}\,\zeta_2\nonumber
\end{align}
\begin{align}
&+\frac{1}{48}\left(-1109z^2+1906z-429\right)\,\zeta_2\,H_0+\frac{1}{96}\left(-311z^2+566z-450\right)\,\zeta_2\,H_2\nonumber\\
&+\frac{1}{96}\left(-193z^2+990z-496\right)\,\zeta_2\,H_{1,0}+\frac{5}{8}\left(z^2-2z+2\right)\,\zeta_3\,H_{0,0}\nonumber\\
&+\frac{121}{96}\left(z^2-2z+2\right)\,\zeta_4\,H_1-\frac{21}{16}\left(z^2-2z+2\right)\,\zeta_2\,H_{2,1}-\frac{67}{48}\left(z^2-2z+2\right)\,\zeta_2\,H_{1,2}\nonumber\\
&+\frac{37}{24}\left(z^2-2z+2\right)\,\zeta_3\,H_2+\frac{39}{16}\left(z^2-2z+2\right)\,\zeta_2\,H_{2,0}+\frac{67}{24}\left(z^2-2z+2\right)\,\zeta_3\,H_{1,0}\nonumber\\
&+\frac{45}{16}\left(z^2-2z+2\right)\,\zeta_2\,H_3-\frac{99}{16}\left(z^2-2z+2\right)\,\zeta_2\,H_{0,0,0}-\frac{135}{16}\left(z^2-2z+2\right)\,\zeta_2\,H_{1,1,0}\nonumber\\
&-\frac{1205}{128}\left(z^2-2z+2\right)\,\zeta_4\,H_0+\frac{201}{16}\left(z^2-2z+2\right)\,\zeta_2\,H_{1,0,0}-\frac{1171}{48}\left(z^2-2z+2\right)\,\zeta_2\,\zeta_3\nonumber\\
&+\frac{165}{4}\left(z^2-2z+2\right)\,\zeta_3\,H_{1,1}+\frac{1481}{24}\left(z^2-2z+2\right)\,\zeta_2\,H_{1,1,1}\nonumber\\
&+\frac{1}{48}\left(287z^2-166z-50\right)\,\zeta_3\,H_0+\frac{1}{36}\left(1639z^2-3977z+2597\right)\,\zeta_3\,H_1\nonumber\\
&+\frac{1}{96}\left(2527z^2-3030z+990\right)\,\zeta_2\,H_{0,0}+\frac{1}{288}\left(18991z^2-46958z+31295\right)\,\zeta_2\,H_{1,1}\nonumber\\
&+\frac{1}{864}\left(59993z^2-152392z+96547\right)\,\zeta_2\,H_1\,,\nonumber\\
H_{gq}^{(0)} &=-\frac{17}{12}\left(z^2-2z+2\right)\,\zeta_2\,H_{1,0}+\frac{97}{36}\left(z^2-2z+2\right)\,\zeta_2\,H_{1,1}\\
&+\frac{-30797z^2+72940z-49747}{1296}\,H_{1,1}+\frac{1}{144}\left(-171z^2+76z+108\right)\,H_{0,0}\nonumber\\
&+\frac{1}{72}\left(-29z^2+82z-82\right)\,H_{1,0,0}+\frac{1}{72}\left(-29z^2+82z-82\right)\,H_{1,2}\nonumber\\
&+\frac{1}{24}\left(-8z^2+19z-17\right)\,H_{1,0}+\frac{1}{4}\left(-z^2+2z-2\right)\,H_{1,2,0}+\frac{1}{6}\left(z^2-2z+2\right)\,H_{1,3}\nonumber\\
&-\frac{5}{12}\left(z^2-2z+2\right)\,H_{1,1,0,0}-\frac{5}{12}\left(z^2-2z+2\right)\,H_{1,1,2}+\frac{5}{6}\left(z^2-2z+2\right)\,H_{1,0,0,0}\nonumber\\
&-\frac{11}{12}\left(z^2-2z+2\right)\,H_{1,2,1}-\frac{80}{9}\left(z^2-2z+2\right)\,H_{1,1,1,1}\nonumber\\
&-\frac{2}{27}\left(239z^2-538z+403\right)\,H_{1,1,1}-\frac{1}{8}z(3z-2)\,H_{2,0}+\frac{5}{12}z(3z-2)\,H_{0,0,0}\nonumber\\
&-\frac{11}{24}z(3z-2)\,H_{2,1}+\frac{23}{36}\left(z^2-2z+2\right)\,\zeta_3\,H_1+\left(z^2-2z+2\right)\,\zeta_3\,H_0\nonumber\\
&+\frac{1}{432}\left(2105z^2-4846z+3739\right)\,\zeta_2\,H_1+\frac{-104701z^2+254657z-160076}{3888}\,H_1\nonumber\\
&+\frac{1}{144}\left(-195z^2+172z-36\right)\,H_2+\frac{1}{144}\left(-139z^2+412z-474\right)\,H_0\nonumber\\
&-\frac{17}{24}z(3z-2)\,\zeta_2\,H_0+\frac{1}{12}z(3z-2)\,H_3-\frac{2863}{768}\left(z^2-2z+2\right)\,\zeta_4\nonumber\\
&+\frac{1}{864}\left(851z^2-898z-425\right)\,\zeta_3+\frac{42256z^2-84272z+45221}{5184}\,\zeta_2\nonumber\\
&+\frac{-2618386z^2+6335144z-3609251}{93312}\,,\nonumber
\end{align}
\begin{align}
I_{gq}^{(0)} &=\frac{110338z^2-269872z+165275}{1536}+\frac{1}{32}(27-46z)\,H_{1,0,0}+\frac{1}{16}(7-4z)\,H_{1,3}\\
&-\frac{3}{16}z(3z-2)\,H_{3,1}+\frac{3}{8}z(3z-2)\,H_{0,0,0,0}-\frac{3}{8}z(3z-2)\,H_4-\frac{7}{16}z(3z-2)\,H_{3,0}\nonumber\\
&+\frac{11}{16}z(3z-2)\,H_{2,2}+\frac{43}{16}z(3z-2)\,H_{2,1,1}+\frac{1}{16}z(3z+2)\,H_{2,0,0}+\frac{1}{8}z(6z-7)\,H_3\nonumber\\
&+\frac{3}{16}z(17z-10)\,H_{2,1,0}+\frac{1}{16}(1-z)(38z-23)\,H_{1,1,0}+\frac{1}{32}z(61z-56)\,H_{2,0}\nonumber\\
&+\frac{1}{16}z(67z-61)\,H_{2,1}-\frac{1}{32}z(81z-70)\,H_{0,0,0}+\frac{-3873z^2-5698z+5041}{1536}\,\zeta_4\nonumber\\
&+\frac{1}{192}\left(-3705z^2+8122z-5584\right)\,\zeta_3+\frac{1}{32}\left(-11z^2-36z+28\right)\,H_{1,2}\nonumber\\
&+\frac{1}{4}\left(z^2-2z+2\right)\,H_{1,1,3}-\frac{3}{8}\left(z^2-2z+2\right)\,H_{1,3,1}+\frac{3}{4}\left(z^2-2z+2\right)\,H_{1,0,0,0,0}\nonumber\\
&-\frac{3}{4}\left(z^2-2z+2\right)\,H_{1,4}-\frac{7}{8}\left(z^2-2z+2\right)\,H_{1,3,0}-\frac{9}{8}\left(z^2-2z+2\right)\,H_{1,1,0,0,0}\nonumber\\
&+\frac{9}{8}\left(z^2-2z+2\right)\,H_{1,1,2,0}+\frac{5}{4}\left(z^2-2z+2\right)\,H_{1,1,1,0,0}+\frac{5}{4}\left(z^2-2z+2\right)\,H_{1,1,1,2}\nonumber\\
&+\frac{11}{8}\left(z^2-2z+2\right)\,H_{1,2,2}+2\left(z^2-2z+2\right)\,H_{1,2,1,0}+\frac{5}{2}\left(z^2-2z+2\right)\,H_{1,1,2,1}\nonumber\\
&+\frac{117}{40}\left(z^2-2z+2\right)\,\zeta_5+\frac{43}{8}\left(z^2-2z+2\right)\,H_{1,2,1,1}+\frac{64}{3}\left(z^2-2z+2\right)\,H_{1,1,1,1,1}\nonumber\\
&+\frac{3}{8}\left(2z^2-6z+5\right)\,H_{1,2,0}-\frac{1}{192}z\left(5z^2+213z-188\right)\,H_0\nonumber\\
&-\frac{3}{16}\left(7z^2-18z+14\right)\,H_{1,0,0,0}+\frac{3}{16}\left(7z^2-18z+14\right)\,H_{1,1,0,0}\nonumber\\
&+\frac{3}{16}\left(7z^2-18z+14\right)\,H_{1,1,2}+\frac{8}{3}\left(15z^2-34z+25\right)\,H_{1,1,1,1}\nonumber\\
&+\frac{1}{16}\left(33z^2-94z+79\right)\,H_{1,2,1}+\frac{1}{48}\left(2312z^2-5631z+3975\right)\,H_{1,1,1}\nonumber\\
&+\frac{1}{192}\left(-5z^3+12437z^2-30529z+19248\right)\,H_1+\frac{1}{192}z\left(5z^3+20z^2+237z-216\right)\,H_2\nonumber\\
&+\frac{1}{192}z\left(5z^3+20z^2+399z-378\right)\,H_{0,0}-\frac{1}{192}(1-z)\left(5z^3+25z^2-446z+446\right)\,H_{1,0}\nonumber\\
&+\frac{1}{768}\left(-60z^4-240z^3-14708z^2+29916z-16213\right)\,\zeta_2\nonumber\\
&+\frac{1}{192}\left(5z^4+20z^3+10840z^2-26620z+17523\right)\,H_{1,1}-\frac{29}{32}z(3z-2)\,\zeta_2\,H_{0,0}\nonumber\\
&-\frac{1}{4}z(6z-5)\,\zeta_3\,H_0-\frac{1}{32}z(63z-58)\,\zeta_2\,H_2+\frac{1}{32}z(164z-151)\,\zeta_2\,H_0\nonumber\\
&+\frac{1}{192}\left(-2333z^2+5680z-3739\right)\,\zeta_2\,H_1+\frac{1}{24}\left(-321z^2+728z-539\right)\,\zeta_3\,H_1\nonumber\\
&+\frac{25}{96}\left(z^2-2z+2\right)\,\zeta_2\,\zeta_3-\frac{9}{8}\left(z^2-2z+2\right)\,\zeta_3\,H_{1,0}+\frac{229}{192}\left(z^2-2z+2\right)\,\zeta_4\,H_1\nonumber\\
&-\frac{25}{16}\left(z^2-2z+2\right)\,\zeta_2\,H_{1,2}-\frac{29}{16}\left(z^2-2z+2\right)\,\zeta_2\,H_{1,0,0}+\frac{15}{8}\left(z^2-2z+2\right)\,\zeta_2\,H_{1,1,0}\nonumber\\
&-\frac{83}{12}\left(z^2-2z+2\right)\,\zeta_3\,H_{1,1}-\frac{29}{4}\left(z^2-2z+2\right)\,\zeta_2\,H_{1,1,1}+\frac{3}{4}\left(4z^2-9z+6\right)\,\zeta_2\,H_{1,0}\nonumber\\
&-\frac{5}{16}\left(37z^2-86z+62\right)\,\zeta_2\,H_{1,1}\,.\nonumber
\end{align}


\subsection{The $q\bar{q}$ initial state}
The contribution of the two-loop amplitude for $q\bar{q}\to Hg$ to the inclusive Higgs cross section at N$^3$LO can be written as
\beq
\hat{\sigma}_{q\bar{q}\to Hg}^{(3)} = 2\,\left(\frac{\alpha_0}{2\pi}\right)^3\,s^{-1-3\eps}\,\sigma_0\,\hat{\sigma}_{q\bar{q}\to Hg}^{(3)}\,.
\eeq
The result is regular in the limit $z\to1$, and so we do not need to separate off the contribution from the soft limit. The coefficient $\hat{\sigma}_{q\bar{q}\to Hg}^{(3)}$ can be written as
\beq\bsp
\hat{\sigma}_{q\bar{q}\to Hg}^{(3)} = \sum_{k=-4}^0\eps^k\,\frac{V}{N}\,\Bigg[&\,N^3\,A_{q\bar{q}}^{(k)} + N^2\,N_f\,B_{q\bar{q}}^{(k)} + N\,N_f^2\,C_{q\bar{q}}^{(k)}+N_f\,D_{q\bar{q}}^{(k)} + N\,E_{q\bar{q}}^{(k)}\\
&\,+ \frac{N_f^2}{N}\,F_{q\bar{q}}^{(k)}+ \frac{1}{N}\,G_{q\bar{q}}^{(k)} + \frac{N_f}{N^2}\,H_{q\bar{q}}^{(k)} 
+ \frac{1}{N^3}\,I_{q\bar{q}}^{(k)}\Bigg]\,,
\esp\eeq
with
\begin{align}
A_{q\bar{q}}^{(-4)} &= \frac{1}{6}(1-z)^3\,,\\
B_{q\bar{q}}^{(-4)} &= 0\,,\\
C_{q\bar{q}}^{(-4)} &= 0\,,\\
D_{q\bar{q}}^{(-4)} &= 0\,,\\
E_{q\bar{q}}^{(-4)} &= -\frac{1}{3}(1-z)^3\,,\\
F_{q\bar{q}}^{(-4)} &= 0\,,\\
G_{q\bar{q}}^{(-4)} &=\frac{5}{24}(1-z)^3\,,\\
H_{q\bar{q}}^{(-4)} &= 0\,,\\
I_{q\bar{q}}^{(-4)} &= -\frac{1}{24}(1-z)^3\,,\\
\nonumber\\
A_{q\bar{q}}^{(-3)} &= \frac{2}{3}(1-z)^3\,H_1-\frac{1}{24}(1-z)^3\,,\\
B_{q\bar{q}}^{(-3)} &= \frac{5}{36}(1-z)^3\,,\\
C_{q\bar{q}}^{(-3)} &= 0\,,\\
D_{q\bar{q}}^{(-3)} &= -\frac{5}{24}(1-z)^3\,,\\
E_{q\bar{q}}^{(-3)} &= -\frac{7}{6}(1-z)^3\,H_1-\frac{35}{144}(1-z)^3\,,\\
F_{q\bar{q}}^{(-3)} &= 0\,,\\
G_{q\bar{q}}^{(-3)} &= \frac{7}{12}(1-z)^3\,H_1+\frac{7}{16}(1-z)^3\,,
\end{align}
%
%
\begin{align}
H_{q\bar{q}}^{(-3)} &= \frac{5}{72}(1-z)^3\,,\\
I_{q\bar{q}}^{(-3)} &= -\frac{1}{12}(1-z)^3\,H_1-\frac{11}{72}(1-z)^3\,,\\
\nonumber\\
A_{q\bar{q}}^{(-2)} &= -\frac{1}{6}z\left(2z^2-3z+3\right)\,H_{0,0}+\frac{8}{3}(1-z)^3\,H_{1,1}-\frac{1}{6}(1-z)\left(z^2+1\right)\,H_1\\
&-\frac{1}{6}z\left(2z^2-3z+3\right)\,H_2-\frac{1}{3}(1-z)z\,H_0+\frac{1}{288}(1-z)\left(25z^2-86z-35\right)\nonumber\\
&+\frac{1}{12}\left(13z^3-33z^2+33z-9\right)\,\zeta_2\,,\nonumber\\
B_{q\bar{q}}^{(-2)} &= \frac{4}{9}(1-z)^3\,H_1+\frac{2}{9}(1-z)^3\,,\\
C_{q\bar{q}}^{(-2)} &= \frac{1}{27}(1-z)^3\,,\\
D_{q\bar{q}}^{(-2)} &= -\frac{7}{12}(1-z)^3\,H_1-\frac{55}{108}(1-z)^3\,,\\
%
E_{q\bar{q}}^{(-2)} &= \frac{5}{12}z\left(2z^2-3z+3\right)\,H_{0,0}-\frac{1}{3}(1-z)^3\,H_{1,0}-\frac{9}{2}(1-z)^3\,H_{1,1}\\
&+\frac{5}{12}z\left(2z^2-3z+3\right)\,H_2-\frac{5}{72}(1-z)\left(11z^2-34z+11\right)\,H_1+\frac{5}{6}(1-z)z\,H_0\nonumber\\
&-\frac{1}{288}(1-z)\left(221z^2-604z+143\right)+\frac{1}{24}\left(-79z^3+207z^2-207z+59\right)\,\zeta_2\,,\nonumber\\
F_{q\bar{q}}^{(-2)} &= -\frac{1}{27}(1-z)^3\,,\\
G_{q\bar{q}}^{(-2)} &= -\frac{1}{3}z\left(2z^2-3z+3\right)\,H_{0,0}+\frac{1}{2}(1-z)^3\,H_{1,0}+\frac{13}{6}(1-z)^3\,H_{1,1}\\
&-\frac{1}{3}z\left(2z^2-3z+3\right)\,H_2+\frac{1}{72}(1-z)\left(89z^2-226z+89\right)\,H_1-\frac{2}{3}(1-z)z\,H_0\nonumber\\
&+\frac{1}{48}\left(159z^3-429z^2+429z-127\right)\,\zeta_2+\frac{1}{864}(1-z)(23z-11)(47z-95)\,,\nonumber\\
H_{q\bar{q}}^{(-2)} &=\frac{5}{36}(1-z)^3\,H_1+\frac{31}{108}(1-z)^3\,,\\
I_{q\bar{q}}^{(-2)} &=\frac{1}{12}z\left(2z^2-3z+3\right)\,H_{0,0}-\frac{1}{6}(1-z)^3\,H_{1,0}-\frac{1}{3}(1-z)^3\,H_{1,1}\\
&+\frac{1}{12}z\left(2z^2-3z+3\right)\,H_2-\frac{1}{36}(1-z)\left(11z^2-28z+11\right)\,H_1+\frac{1}{6}(1-z)z\,H_0\nonumber\\
&-\frac{1}{864}(1-z)\left(493z^2-1148z+511\right)+\frac{1}{48}\left(-53z^3+147z^2-147z+45\right)\,\zeta_2\,,\nonumber\\
\nonumber\\
A_{q\bar{q}}^{(-1)} &= -\frac{1}{2}z\left(2z^2-3z+3\right)\,H_{2,0}+\frac{7}{12}z\left(2z^2-3z+3\right)\,H_{0,0,0}\\
&-\frac{7}{6}z\left(2z^2-3z+3\right)\,H_{2,1}-\frac{1}{3}(1-z)\left(2z^2+3z+2\right)\,H_{1,1}\nonumber\\
&+\frac{1}{72}z\left(28z^2-81z+27\right)\,H_{0,0}+\frac{1}{6}(1-z)^3\,H_{1,0,0}+\frac{1}{6}(1-z)^3\,H_{1,2}\nonumber\\
&+\frac{32}{3}(1-z)^3\,H_{1,1,1}-((1-z)z)\,H_{1,0}-\frac{2}{3}z\left(2z^2-3z+3\right)\,\zeta_2\,H_0\nonumber\\
&-\frac{1}{12}z\left(2z^2-3z+3\right)\,H_3+\frac{1}{72}z\left(28z^2+15z-69\right)\,H_2\nonumber
\end{align}
\begin{align}
&+\frac{1}{432}(1-z)\left(35z^2-802z-397\right)\,H_1-\frac{19}{6}(1-z)^3\,\zeta_2\,H_1+\frac{1}{36}(1-z)z(6z-49)\,H_0\nonumber\\
&+\frac{(1-z)\left(9481z^2-28196z+4747\right)}{5184}+\frac{1}{144}\left(-127z^3+399z^2-291z+71\right)\,\zeta_2\nonumber\\
&+\frac{1}{12}\left(38z^3-105z^2+105z-32\right)\,\zeta_3\,,\nonumber\\
B_{q\bar{q}}^{(-1)} &= -\frac{1}{9}z\left(2z^2-3z+3\right)\,H_{0,0}+\frac{13}{9}(1-z)^3\,H_{1,1}-\frac{1}{9}z\left(2z^2-3z+3\right)\,H_2\\
&-\frac{2}{9}(1-z)z\,H_0+\frac{1}{18}(1-z)(3z-5)(5z-3)\,H_1+\frac{1}{432}(1-z)\left(19z^2-74z-41\right)\nonumber\\
&+\frac{1}{72}\left(77z^3-207z^2+207z-61\right)\,\zeta_2\,,\nonumber\\
C_{q\bar{q}}^{(-1)} &= \frac{2}{27}(1-z)^3\,H_1+\frac{4}{27}(1-z)^3\,,\\
%
D_{q\bar{q}}^{(-1)} &= \frac{2}{9}z\left(2z^2-3z+3\right)\,H_{0,0}-\frac{2}{9}(1-z)^3\,H_{1,0}-\frac{35}{18}(1-z)^3\,H_{1,1}\\
&+\frac{2}{9}z\left(2z^2-3z+3\right)\,H_2-\frac{2}{27}(1-z)\left(19z^2-44z+19\right)\,H_1+\frac{4}{9}(1-z)z\,H_0\\
&-\frac{1}{36}(1-z)\left(37z^2-86z+33\right)+\frac{1}{144}\left(-391z^3+1077z^2-1077z+327\right)\,\zeta_2\,,\nonumber\\
E_{q\bar{q}}^{(-1)} &=-\frac{1}{3}(1-z)\left(z^2-9z+1\right)\,H_{1,0}+\frac{7}{6}z\left(2z^2-3z+3\right)\,H_{2,0}\\
&-\frac{17}{12}z\left(2z^2-3z+3\right)\,H_{0,0,0}+\frac{11}{4}z\left(2z^2-3z+3\right)\,H_{2,1}\nonumber\\
&+\frac{1}{72}z\left(22z^2+27z+45\right)\,H_{0,0}-\frac{1}{18}(1-z)\left(47z^2-193z+47\right)\,H_{1,1}\nonumber\\
&-\frac{2}{3}(1-z)^3\,H_{1,0,0}-2(1-z)^3\,H_{1,2}-\frac{8}{3}(1-z)^3\,H_{1,1,0}-\frac{115}{6}(1-z)^3\,H_{1,1,1}\nonumber\\
&+\frac{13}{6}z\left(2z^2-3z+3\right)\,\zeta_2\,H_0+\frac{1}{6}z\left(2z^2-3z+3\right)\,H_3+\frac{1}{72}z\left(22z^2-201z+273\right)\,H_2\nonumber\\
&-\frac{1}{432}(1-z)\left(1123z^2-4655z+592\right)\,H_1+10(1-z)^3\,\zeta_2\,H_1\nonumber\\
&-\frac{1}{36}(1-z)z(9z-131)\,H_0-\frac{(1-z)\left(15149z^2-55138z+4709\right)}{5184}\nonumber\\
&+\frac{1}{288}\left(-459z^3+573z^2-861z+371\right)\,\zeta_2+\frac{1}{24}\left(-155z^3+423z^2-423z+127\right)\,\zeta_3\,,\nonumber\\
F_{q\bar{q}}^{(-1)} &=-\frac{2}{27}(1-z)^3\,H_1-\frac{4}{27}(1-z)^3\,,\\
G_{q\bar{q}}^{(-1)} &=-\frac{5}{6}z\left(2z^2-3z+3\right)\,H_{2,0}+\frac{13}{12}z\left(2z^2-3z+3\right)\,H_{0,0,0}\\
&-2z\left(2z^2-3z+3\right)\,H_{2,1}+\frac{1}{18}(1-z)\left(19z^2-68z+19\right)\,H_{1,0}\nonumber\\
&-\frac{1}{24}z\left(34z^2-43z+49\right)\,H_{0,0}+\frac{1}{18}(1-z)\left(83z^2-238z+83\right)\,H_{1,1}\nonumber\\
&+\frac{2}{3}(1-z)^3\,H_{1,0,0}+\frac{5}{2}(1-z)^3\,H_{1,2}+\frac{11}{3}(1-z)^3\,H_{1,1,0}+\frac{31}{3}(1-z)^3\,H_{1,1,1}\nonumber\\
&-\frac{7}{3}z\left(2z^2-3z+3\right)\,\zeta_2\,H_0-\frac{1}{12}z\left(2z^2-3z+3\right)\,H_3-\frac{1}{24}z\left(34z^2-99z+105\right)\,H_2\nonumber\\
&+\frac{1}{144}(1-z)\left(527z^2-1846z+503\right)\,H_1-\frac{67}{8}(1-z)^3\,\zeta_2\,H_1\nonumber
\end{align}
\begin{align}
&+\frac{1}{12}(1-z)(z-40)z\,H_0+\frac{(1-z)\left(14687z^2-52756z+8693\right)}{5184}\nonumber\\
&+\frac{1}{12}\left(52z^3-141z^2+141z-42\right)\,\zeta_3+\frac{1}{288}\left(1939z^3-4389z^2+4461z-1531\right)\,\zeta_2\,,\nonumber\\
H_{q\bar{q}}^{(-1)} &=-\frac{1}{9}z\left(2z^2-3z+3\right)\,H_{0,0}+\frac{2}{9}(1-z)^3\,H_{1,0}+\frac{1}{2}(1-z)^3\,H_{1,1}\\
&-\frac{1}{9}z\left(2z^2-3z+3\right)\,H_2+\frac{1}{54}(1-z)\left(31z^2-74z+31\right)\,H_1-\frac{2}{9}(1-z)z\,H_0\nonumber\\
&+\frac{1}{432}(1-z)\left(425z^2-958z+437\right)+\frac{1}{144}\left(237z^3-663z^2+663z-205\right)\,\zeta_2\,,\nonumber\\
I_{q\bar{q}}^{(-1)} &=\frac{1}{6}z\left(2z^2-3z+3\right)\,H_{2,0}-\frac{1}{4}z\left(2z^2-3z+3\right)\,H_{0,0,0}\\
&+\frac{5}{12}z\left(2z^2-3z+3\right)\,H_{2,1}-\frac{1}{6}(1-z)\left(8z^2-21z+8\right)\,H_{1,1}\nonumber\\
&-\frac{1}{18}(1-z)\left(13z^2-32z+13\right)\,H_{1,0}+\frac{1}{72}z\left(52z^2-75z+75\right)\,H_{0,0}\nonumber\\
&-\frac{1}{6}(1-z)^3\,H_{1,0,0}-\frac{2}{3}(1-z)^3\,H_{1,2}-(1-z)^3\,H_{1,1,0}-\frac{11}{6}(1-z)^3\,H_{1,1,1}\nonumber\\
&+\frac{5}{6}z\left(2z^2-3z+3\right)\,\zeta_2\,H_0+\frac{1}{72}z\left(52z^2-111z+111\right)\,H_2\nonumber\\
&-\frac{1}{432}(1-z)\left(493z^2-1685z+520\right)\,H_1+\frac{37}{24}(1-z)^3\,\zeta_2\,H_1+\frac{19}{18}(1-z)z\,H_0\nonumber\\
&-\frac{(1-z)\left(9019z^2-25814z+8731\right)}{5184}+\frac{1}{144}\left(-613z^3+1509z^2-1509z+509\right)\,\zeta_2\nonumber\\
&+\frac{1}{24}\left(-25z^3+69z^2-69z+21\right)\,\zeta_3\,,\nonumber\\
\nonumber\\
A_{q\bar{q}}^{(0)} &= \frac{(1-z)\left(104741z^2-281554z+36581\right)}{10368}-\frac{1}{6}(1-z)^3\,H_{1,3}\\
&+\frac{1}{6}(1-z)^3\,H_{1,1,0,0}+\frac{1}{3}(1-z)^3\,H_{1,2,1}+\frac{1}{3}(1-z)^3\,H_{1,1,2}-\frac{1}{2}(1-z)^3\,H_{1,0,0,0}\nonumber\\
&-\frac{2}{3}(1-z)^3\,H_{1,1,1,0}+\frac{253}{6}(1-z)^3\,H_{1,1,1,1}-\frac{1}{18}(1-z)^2(13z-94)\,H_{1,2}\nonumber\\
&-\frac{1}{864}(1-z)z(765z-2927)\,H_0+\frac{1}{8}z\left(2z^2-14z-3\right)\,H_{2,0}\nonumber\\
&+\frac{5}{12}z\left(2z^2-3z+3\right)\,H_{2,0,0}+\frac{3}{4}z\left(2z^2-3z+3\right)\,H_4-\frac{3}{2}z\left(2z^2-3z+3\right)\,H_{2,2}\nonumber\\
&+\frac{19}{12}z\left(2z^2-3z+3\right)\,H_{3,1}-\frac{19}{12}z\left(2z^2-3z+3\right)\,H_{2,1,0}-\frac{7}{4}z\left(2z^2-3z+3\right)\,H_{0,0,0,0}\nonumber\\
&+\frac{7}{4}z\left(2z^2-3z+3\right)\,H_{3,0}-\frac{37}{6}z\left(2z^2-3z+3\right)\,H_{2,1,1}\nonumber\\
&-\frac{1}{12}(1-z)\left(11z^2+16z+11\right)\,H_{1,1,0}+\frac{1}{18}(1-z)\left(13z^2-38z+94\right)\,H_{1,0,0}\nonumber\\
&-\frac{1}{12}z\left(16z^2-11z+17\right)\,H_3-\frac{1}{12}(1-z)\left(43z^2+62z+43\right)\,H_{1,1,1}\nonumber\\
&+\frac{1}{24}z\left(58z^2-47z-85\right)\,H_{2,1}-\frac{1}{432}(1-z)\left(61z^2+1507z+394\right)\,H_{1,0}\nonumber\\
&-\frac{1}{24}z\left(84z^2-155z+86\right)\,H_{0,0,0}-\frac{1}{216}(1-z)\left(107z^2+2027z+1196\right)\,H_{1,1}\nonumber\\
&+\frac{1}{432}z\left(340z^2+2694z-2217\right)\,H_2+\frac{1}{432}z\left(745z^2-6z+78\right)\,H_{0,0}\nonumber
\end{align}
\begin{align}
&+\frac{(1-z)\left(4885z^2-13745z-8669\right)}{1296}\,H_1+\frac{-585z^3-4785z^2-615z+2717}{1728}\,\zeta_2\nonumber\\
&+\frac{1}{72}\left(-167z^3+378z^2-576z+449\right)\,\zeta_3+\frac{1}{96}\left(1447z^3-1929z^2+1929z+161\right)\,\zeta_4\nonumber\\
&+3(1-z)^3\,\zeta_2\,H_{1,0}-\frac{65}{6}(1-z)^3\,\zeta_3\,H_1-\frac{71}{6}(1-z)^3\,\zeta_2\,H_{1,1}\nonumber\\
&-\frac{1}{12}z\left(2z^2-3z+3\right)\,\zeta_2\,H_2+\frac{2}{3}z\left(2z^2-3z+3\right)\,\zeta_3\,H_0+\frac{13}{6}z\left(2z^2-3z+3\right)\,\zeta_2\,H_{0,0}\nonumber\\
&+\frac{1}{18}(1-z)\left(50z^2-76z-31\right)\,\zeta_2\,H_1+\frac{1}{24}z\left(74z^2-186z+153\right)\,\zeta_2\,H_0\,,\nonumber\\
%
B_{q\bar{q}}^{(0)} &= -\frac{1}{6}z\left(2z^2-3z+3\right)\,H_{2,0}+\frac{5}{9}z\left(2z^2-3z+3\right)\,H_{0,0,0}\\
&-\frac{11}{18}z\left(2z^2-3z+3\right)\,H_{2,1}-\frac{1}{18}(1-z)\left(4z^2-17z+31\right)\,H_{1,0,0}\nonumber\\
&-\frac{1}{18}(1-z)\left(4z^2-17z+31\right)\,H_{1,2}+\frac{1}{27}(1-z)\left(80z^2-193z+80\right)\,H_{1,1}\nonumber\\
&-\frac{1}{216}z\left(112z^2+177z-96\right)\,H_{0,0}+\frac{43}{9}(1-z)^3\,H_{1,1,1}-\frac{1}{3}(1-z)z\,H_{1,0}\nonumber\\
&-\frac{17}{18}z\left(2z^2-3z+3\right)\,\zeta_2\,H_0-\frac{1}{18}(1-z)\left(46z^2-83z+19\right)\,\zeta_2\,H_1\nonumber\\
&+\frac{1}{9}z\left(2z^2-3z+3\right)\,H_3-\frac{1}{216}z\left(112z^2-15z+96\right)\,H_2\nonumber\\
&+\frac{1}{432}(1-z)\left(518z^2-2565z+1133\right)\,H_1+\frac{1}{216}(1-z)z(126z-703)\,H_0\nonumber\\
&-\frac{(1-z)\left(13387z^2-8171z+11236\right)}{7776}+\frac{1}{12}\left(33z^3-99z^2+117z-47\right)\,\zeta_3\nonumber\\
&+\frac{1}{216}\left(202z^3+75z^2+6z-90\right)\,\zeta_2\,,\nonumber\\
C_{q\bar{q}}^{(0)} &= \frac{4}{27}(1-z)^3\,H_{1,1}+\frac{8}{27}(1-z)^3\,H_1-\frac{29}{54}(1-z)^3\,\zeta_2+\frac{38}{81}(1-z)^3\,,\\
D_{q\bar{q}}^{(0)} &=\frac{1}{3}z\left(2z^2-3z+3\right)\,H_{2,0}-\frac{10}{9}z\left(2z^2-3z+3\right)\,H_{0,0,0}\\
&+\frac{11}{9}z\left(2z^2-3z+3\right)\,H_{2,1}+\frac{1}{18}(1-z)\left(4z^2-17z+31\right)\,H_{1,0,0}\nonumber\\
&-\frac{1}{6}(1-z)\left(4z^2-5z-5\right)\,H_{1,2}-\frac{2}{27}(1-z)\left(19z^2-47z+19\right)\,H_{1,0}\nonumber\\
&-\frac{1}{27}(1-z)\left(149z^2-364z+149\right)\,H_{1,1}+\frac{1}{216}z\left(416z^2-213z+276\right)\,H_{0,0}\nonumber\\
&-\frac{13}{9}(1-z)^3\,H_{1,1,0}-\frac{23}{3}(1-z)^3\,H_{1,1,1}+\frac{17}{9}z\left(2z^2-3z+3\right)\,\zeta_2\,H_0\nonumber\\
&+\frac{1}{24}(1-z)\left(143z^2-274z+107\right)\,\zeta_2\,H_1-\frac{2}{9}z\left(2z^2-3z+3\right)\,H_3\nonumber\\
&+\frac{1}{216}z\left(416z^2-597z+660\right)\,H_2-\frac{1}{432}(1-z)\left(1368z^2-5413z+2031\right)\,H_1\nonumber\\
&-\frac{1}{216}(1-z)z(144z-1061)\,H_0-\frac{(1-z)\left(2198z^2-17401z+3119\right)}{2592}\nonumber\\
&+\frac{1}{216}\left(-1981z^3+4572z^2-4635z+1565\right)\,\zeta_2\nonumber\\
&+\frac{1}{72}\left(-253z^3+723z^2-831z+313\right)\,\zeta_3\,,\nonumber
\end{align}
%
%
\begin{align}
E_{q\bar{q}}^{(0)} &=-\frac{(1-z)\left(338279z^2-1388872z-39343\right)}{31104}+\frac{1}{6}(1-z)^3\,H_{1,3}\\
&-\frac{13}{6}(1-z)^3\,H_{1,1,0,0}-\frac{7}{3}(1-z)^3\,H_{1,2,0}+3(1-z)^3\,H_{1,0,0,0}-\frac{21}{2}(1-z)^3\,H_{1,2,1}\nonumber\\
&-\frac{27}{2}(1-z)^3\,H_{1,1,2}-\frac{103}{6}(1-z)^3\,H_{1,1,1,0}-90(1-z)^3\,H_{1,1,1,1}\nonumber\\
&+\frac{31}{288}(1-z)z(5z+31)\,H_0-\frac{4}{3}z\left(2z^2-3z+3\right)\,H_4-\frac{17}{6}z\left(2z^2-3z+3\right)\,H_{3,1}\nonumber\\
&-\frac{43}{12}z\left(2z^2-3z+3\right)\,H_{3,0}+\frac{23}{6}z\left(2z^2-3z+3\right)\,H_{2,2}+4z\left(2z^2-3z+3\right)\,H_{0,0,0,0}\nonumber\\
&+\frac{179}{12}z\left(2z^2-3z+3\right)\,H_{2,1,1}-\frac{1}{12}z\left(20z^2-33z+33\right)\,H_{2,0,0}\nonumber\\
&-\frac{1}{12}(1-z)\left(40z^2-183z+43\right)\,H_{1,1,0}+\frac{1}{24}z\left(71z^2-130z+46\right)\,H_{0,0,0}\nonumber\\
&+\frac{1}{24}z\left(83z^2-79z+88\right)\,H_3+\frac{1}{12}z\left(100z^2-147z+147\right)\,H_{2,1,0}\nonumber\\
&-\frac{1}{72}(1-z)\left(193z^2-437z+562\right)\,H_{1,0,0}-\frac{1}{18}(1-z)\left(205z^2-947z+205\right)\,H_{1,1,1}\nonumber\\
&+\frac{1}{72}z\left(238z^2-279z+351\right)\,H_{2,0}+\frac{1}{72}z\left(274z^2-1002z+1353\right)\,H_{2,1}\nonumber\\
&-\frac{1}{72}(1-z)\left(307z^2-1319z+658\right)\,H_{1,2}-\frac{(1-z)\left(23594z^2-101827z-12595\right)}{2592}\,H_1\nonumber\\
&+\frac{1}{96}\left(-1193z^3-465z^2+465z-1395\right)\,\zeta_4+\frac{1}{144}\left(-249z^3+1269z^2-999z-583\right)\,\zeta_3\nonumber\\
&+\frac{1}{144}z\left(9z^3-131z^2-388z+10\right)\,H_{0,0}+\frac{1}{144}z\left(9z^3+58z^2-2608z+2041\right)\,H_2\nonumber\\
&-\frac{1}{432}(1-z)\left(27z^3-74z^2-5357z+232\right)\,H_{1,0}\nonumber\\
&-\frac{1}{432}(1-z)\left(27z^3+3431z^2-21709z+1613\right)\,H_{1,1}\nonumber\\
&+\frac{1}{576}\left(-108z^4-385z^3+6669z^2-4755z+579\right)\,\zeta_2+\frac{1}{6}(1-z)^3\,\zeta_2\,H_{1,0}\nonumber\\
&+\frac{103}{4}(1-z)^3\,\zeta_3\,H_1+\frac{183}{4}(1-z)^3\,\zeta_2\,H_{1,1}-\frac{239}{24}z\left(2z^2-3z+3\right)\,\zeta_2\,H_{0,0}\nonumber\\
&-\frac{1}{2}z\left(9z^2-14z+14\right)\,\zeta_3\,H_0-\frac{1}{8}z\left(34z^2-55z+55\right)\,\zeta_2\,H_2\nonumber\\
&-\frac{1}{72}z\left(352z^2-1299z+813\right)\,\zeta_2\,H_0+\frac{1}{144}(1-z)\left(1651z^2-4148z+2281\right)\,\zeta_2\,H_1\,,\nonumber\\
F_{q\bar{q}}^{(0)} &=-\frac{4}{27}(1-z)^3\,H_{1,1}-\frac{8}{27}(1-z)^3\,H_1+\frac{29}{54}(1-z)^3\,\zeta_2-\frac{38}{81}(1-z)^3\,,\\
G_{q\bar{q}}^{(0)} &=\frac{(1-z)\left(59869z^2-380522z+22933\right)}{10368}-\frac{1}{2}(1-z)^3\,H_{1,3}\\
&+\frac{5}{2}(1-z)^3\,H_{1,1,0,0}+\frac{8}{3}(1-z)^3\,H_{1,2,0}-\frac{10}{3}(1-z)^3\,H_{1,0,0,0}+\frac{37}{3}(1-z)^3\,H_{1,2,1}\nonumber\\
&+16(1-z)^3\,H_{1,1,2}+\frac{68}{3}(1-z)^3\,H_{1,1,1,0}+\frac{343}{6}(1-z)^3\,H_{1,1,1,1}\nonumber\\
&+\frac{3}{4}(1-z)(2z-5)(4z-1)\,H_{1,2}+\frac{1}{144}(1-z)z(41z-1694)\,H_0\nonumber\\
&+\frac{7}{12}z\left(2z^2-3z+3\right)\,H_{2,0,0}+\frac{3}{4}z\left(2z^2-3z+3\right)\,H_4+\frac{19}{12}z\left(2z^2-3z+3\right)\,H_{3,1}\nonumber\\
&+\frac{9}{4}z\left(2z^2-3z+3\right)\,H_{3,0}-\frac{5}{2}z\left(2z^2-3z+3\right)\,H_{2,2}-\frac{11}{4}z\left(2z^2-3z+3\right)\,H_{0,0,0,0}\nonumber
\end{align}
\begin{align}
&-\frac{37}{12}z\left(2z^2-3z+3\right)\,H_{2,1,0}-\frac{31}{3}z\left(2z^2-3z+3\right)\,H_{2,1,1}-\frac{7}{36}z\left(10z^2-9z+9\right)\,H_3\nonumber\\
&+\frac{1}{36}z\left(91z^2-153z+189\right)\,H_{0,0,0}+\frac{1}{36}(1-z)\left(94z^2-128z+13\right)\,H_{1,0,0}\nonumber\\
&+\frac{2}{9}(1-z)\left(98z^2-289z+98\right)\,H_{1,1,1}-\frac{1}{24}z\left(220z^2-523z+511\right)\,H_{2,1}\nonumber\\
&+\frac{1}{36}(1-z)\left(289z^2-800z+289\right)\,H_{1,1,0}-\frac{1}{72}z\left(338z^2-537z+429\right)\,H_{2,0}\nonumber\\
&+\frac{(1-z)\left(22681z^2-118154z+21043\right)}{2592}\,H_1\nonumber\\
&+\frac{1}{384}\left(-289z^3+12531z^2-12531z+8065\right)\,\zeta_4\nonumber\\
&-\frac{1}{144}z\left(18z^3+442z^2-655z+106\right)\,H_{0,0}-\frac{1}{144}z\left(18z^3+496z^2-2403z+1800\right)\,H_2\nonumber\\
&+\frac{1}{216}(1-z)\left(27z^3+494z^2-3379z+872\right)\,H_{1,0}\nonumber\\
&+\frac{1}{216}(1-z)\left(27z^3+2816z^2-12274z+3095\right)\,H_{1,1}\nonumber\\
&+\frac{1}{144}\left(1111z^3-3855z^2+4539z-1187\right)\,\zeta_3\nonumber\\
&+\frac{648z^4+26537z^3-69903z^2+68139z-26489}{1728}\,\zeta_2-\frac{5}{4}(1-z)^3\,\zeta_2\,H_{1,0}\nonumber\\
&-\frac{37}{2}(1-z)^3\,\zeta_3\,H_1-\frac{397}{12}(1-z)^3\,\zeta_2\,H_{1,1}+\frac{5}{6}z\left(2z^2-3z+3\right)\,\zeta_2\,H_2\nonumber\\
&+2z\left(2z^2-3z+3\right)\,\zeta_3\,H_0+\frac{133}{12}z\left(2z^2-3z+3\right)\,\zeta_2\,H_{0,0}\nonumber\\
&-\frac{1}{72}z\left(542z^2-171z+423\right)\,\zeta_2\,H_0-\frac{1}{144}(1-z)\left(2449z^2-5066z+2125\right)\,\zeta_2\,H_1\,,\nonumber\\
H_{q\bar{q}}^{(0)} &=-\frac{1}{6}z\left(2z^2-3z+3\right)\,H_{2,0}+\frac{5}{9}z\left(2z^2-3z+3\right)\,H_{0,0,0}\\
&-\frac{11}{18}z\left(2z^2-3z+3\right)\,H_{2,1}+\frac{1}{9}(1-z)\left(23z^2-57z+23\right)\,H_{1,1}\nonumber\\
&+\frac{1}{27}(1-z)\left(38z^2-85z+38\right)\,H_{1,0}-\frac{1}{108}z\left(152z^2-195z+186\right)\,H_{0,0}\nonumber\\
&+\frac{8}{9}(1-z)^3\,H_{1,2}+\frac{13}{9}(1-z)^3\,H_{1,1,0}+\frac{26}{9}(1-z)^3\,H_{1,1,1}-\frac{17}{18}z\left(2z^2-3z+3\right)\,\zeta_2\,H_0\nonumber\\
&+\frac{1}{9}z\left(2z^2-3z+3\right)\,H_3-\frac{1}{108}z\left(152z^2-291z+282\right)\,H_2\nonumber\\
&+\frac{1}{216}(1-z)\left(425z^2-1424z+449\right)\,H_1-\frac{245}{72}(1-z)^3\,\zeta_2\,H_1\nonumber\\
&+\frac{1}{108}(1-z)z(9z-179)\,H_0+\frac{(1-z)\left(19981z^2-60374z+20593\right)}{7776}\nonumber\\
&+\frac{1}{72}\left(55z^3-129z^2+129z-31\right)\,\zeta_3+\frac{1}{216}\left(1779z^3-4647z^2+4629z-1475\right)\,\zeta_2\,,
\nonumber\\
I_{q\bar{q}}^{(0)} &=-\frac{(1-z)\left(155551z^2-597356z+217885\right)}{31104}-\frac{1}{3}(1-z)^3\,H_{1,2,0}\\
&+\frac{1}{2}(1-z)^3\,H_{1,3}-\frac{1}{2}(1-z)^3\,H_{1,1,0,0}+\frac{5}{6}(1-z)^3\,H_{1,0,0,0}-\frac{13}{6}(1-z)^3\,H_{1,2,1}\nonumber\\
&-\frac{17}{6}(1-z)^3\,H_{1,1,2}-\frac{29}{6}(1-z)^3\,H_{1,1,1,0}-\frac{28}{3}(1-z)^3\,H_{1,1,1,1}\nonumber\\
&+\frac{1}{432}(1-z)z(27z+2177)\,H_0-\frac{1}{6}z\left(2z^2-3z+3\right)\,H_4+\frac{1}{6}z\left(2z^2-3z+3\right)\,H_{2,2}\nonumber
\end{align}
\begin{align}
&-\frac{1}{3}z\left(2z^2-3z+3\right)\,H_{3,1}-\frac{5}{12}z\left(2z^2-3z+3\right)\,H_{3,0}+\frac{1}{2}z\left(2z^2-3z+3\right)\,H_{0,0,0,0}\nonumber\\
&+\frac{19}{12}z\left(2z^2-3z+3\right)\,H_{2,1,1}+\frac{1}{4}z\left(4z^2-7z+7\right)\,H_{2,1,0}-\frac{1}{12}z\left(4z^2-3z+3\right)\,H_{2,0,0}\nonumber\\
&-\frac{1}{72}z\left(13z^2-45z+36\right)\,H_3-\frac{1}{72}(1-z)\left(47z^2+29z-160\right)\,H_{1,0,0}\nonumber\\
&-\frac{1}{24}(1-z)\left(59z^2-99z-4\right)\,H_{1,2}+\frac{1}{72}z\left(82z^2-132z+105\right)\,H_{2,0}\nonumber\\
&-\frac{1}{36}(1-z)\left(136z^2-299z+127\right)\,H_{1,1,0}-\frac{1}{72}z\left(143z^2-231z+258\right)\,H_{0,0,0}\nonumber\\
&+\frac{1}{72}z\left(212z^2-426z+435\right)\,H_{2,1}-\frac{1}{36}(1-z)\left(245z^2-604z+245\right)\,H_{1,1,1}\nonumber\\
&-\frac{(1-z)\left(8857z^2-43817z+16300\right)}{2592}\,H_1+\frac{1}{384}\left(-727z^3-2955z^2+2955z-3129\right)\,\zeta_4\nonumber\\
&+\frac{1}{72}\left(-264z^3+915z^2-1194z+436\right)\,\zeta_3+\frac{1}{432}z\left(27z^3+974z^2-2079z+1494\right)\,H_2\nonumber\\
&+\frac{1}{432}z\left(27z^3+974z^2-795z+210\right)\,H_{0,0}\nonumber\\
&-\frac{1}{432}(1-z)\left(27z^3+1001z^2-2908z+1118\right)\,H_{1,0}\nonumber\\
&-\frac{1}{432}(1-z)\left(27z^3+1987z^2-6893z+2185\right)\,H_{1,1}\nonumber\\
&+\frac{-324z^4-24797z^3+54681z^2-53259z+22035}{1728}\,\zeta_2-\frac{5}{6}(1-z)^3\,\zeta_2\,H_{1,1}\nonumber\\
&-\frac{23}{12}(1-z)^3\,\zeta_2\,H_{1,0}+\frac{43}{12}(1-z)^3\,\zeta_3\,H_1-\frac{79}{24}z\left(2z^2-3z+3\right)\,\zeta_2\,H_{0,0}\nonumber\\
&-\frac{1}{6}z\left(5z^2-6z+6\right)\,\zeta_3\,H_0+\frac{1}{8}z\left(22z^2-37z+37\right)\,\zeta_2\,H_2\nonumber\\
&+\frac{1}{72}(1-z)\left(199z^2-155z+46\right)\,\zeta_2\,H_1+\frac{1}{24}z\left(224z^2-304z+259\right)\,\zeta_2\,H_0\,.\nonumber
\end{align}


\end{document}